\documentclass[useAMS,usenatbib,usegraphicx]{mn2e}
\usepackage[fleqn]{amsmath}
\usepackage{amssymb}
\DeclareMathAlphabet{\mathitbf}{OML}{cmm}{b}{it}
\usepackage{url}
\usepackage{fixltx2e}
\usepackage[T1]{fontenc}
\usepackage{times}

\newcommand{\msun}{\ensuremath{\mathrm{M}_{\odot}}}
\newcommand{\lsun}{\ensuremath{\mathrm{L}_{\odot}}}
\usepackage{color}
\def\changed{}
\title[H\,\textsc{ii} regions around exiled O stars]
{Dynamics of H\,\textsc{ii} regions around exiled O stars}
\author[J.~Mackey, N.~Langer, V.~V.~Gvaramadze]
{Jonathan Mackey$^{1}\thanks{email: \texttt{JMackey@astro.uni-bonn.de}}$,
 Norbert~Langer$^{1}$\thanks{Alexander von Humboldt Professor},
 Vasilii~V.~Gvaramadze$^{2,3}$\\
$^1$Argelander-Institut f\"ur Astronomie, Auf dem H\"ugel 71, 53121 Bonn, Germany.\\
$^2$Sternberg Astronomical Institute, Lomonosov Moscow State University, Universitetskij Pr.~13, Moscow 119992, Russia. \\
$^{3}$Isaac Newton Institute of Chile, Moscow Branch, Universitetskij Pr.~13, Moscow 119992, Russia.}

\begin{document}

\date{Submitted 30 June 2013; Accepted 23 August 2013}
\pagerange{\pageref{firstpage}--\pageref{lastpage}} \pubyear{2013}

\maketitle
\label{firstpage}

\begin{abstract}
At least 25 per cent of massive stars are ejected from their parent cluster, becoming \emph{runaways} or \emph{exiles}, travelling with often-supersonic space velocities through the interstellar medium (ISM).
{\changed Their overpressurised H\,\textsc{ii} regions impart kinetic energy and momentum to the ISM, compress and/or evaporate dense clouds, and can constrain properties of both the star and the ISM.}
Here we present {\changed one-, two-, and (the first)} three-dimensional simulations of the H\,\textsc{ii} region around a massive star moving supersonically through a uniform, magnetised ISM, with properties appropriate for the nearby O star $\zeta$ Oph.
The H\,\textsc{ii} region leaves an expanding overdense shell behind the star and, inside this, an underdense wake that should be filled with hot gas from the shocked stellar wind.
The gas column density in the shell is strongly influenced by the ISM magnetic field strength and orientation.
{\changed H$\alpha$ emission maps show the H\,\textsc{ii} region remains roughly circular, although the star is displaced somewhat from the centre of emission.}
For our model parameters, the kinetic energy feedback from the H\,\textsc{ii} region is comparable to the mechanical luminosity of the stellar wind, and the momentum feedback rate is $>100\times$ larger than that from the wind and $\approx10\times$ larger than the total momentum input rate available from radiation pressure.
Compared to the star's eventual supernova explosion, the kinetic energy feedback from the H\,\textsc{ii} region over the star's main sequence lifetime is $>100\times$ less, but the momentum feedback is up to $4\times$ larger.
{\changed H\,\textsc{ii} region dynamics are found to have only a small effect on the ISM conditions that a bow shock close to the star would encounter.}
\end{abstract}


%%%%%%%%%%%%%%%%%%%%%%%%%%%%%%%%%%%%%%%%%%%%%%%%%%%%%%%%%%%%%%%%%%%%%
\section{Introduction}
\label{sec:intro}
%%%%%%%%%%%%%%%%%%%%%%%%%%%%%%%%%%%%%%%%%%%%%%%%%%%%%%%%%%%%%%%%%%%%%

About 25 per cent of massive stars are classified as isolated \citep{Gie87,Bla93}, meaning they are not currently part of a star cluster or association.
A number of recent studies \citep[e.g.][]{DeWTesPalEA05, SchRoe08, GvaBom08, GvaKniKroEA11, GvaWeiKroEA12} have shown that all but a handful of nearby O stars are very likely to be either classical runaway stars (with space velocity $v_\star\geq30\,\ensuremath{\mathrm{km}\,\mathrm{s}^{-1}}$) or to at least have a significant peculiar velocity directed away from a star cluster that could plausibly have been their birth place.
We will refer to all such stars as \emph{exiles}: stars that have been ejected from their place of birth.
Nearby stars such as $\zeta$ Oph and $\alpha$ Ori (Betelgeuse) belong in this category; indeed by definition all massive stars closer to us than the nearest region of massive star formation must be exiles.

Bow shocks from these stars have been studied for many years and in some depth; see \citet{VilManGar12,MohMacLan12} and references therein for bow shock modelling around cool stars, and e.g.~\citet{vMarLanAchEA06,ArtHoa06,ComKap98} and references therein for bow shocks around hot stars.
H\,\textsc{ii} regions around exiled massive stars have received comparatively less attention, probably because at increasingly hypersonic velocities the H\,\textsc{ii} region produces ever-weaker shocks whereas the bow shock gets stronger.

%Analytic predictions for the shape of H\,\textsc{ii} regions around moving stars were made by \citet{Ras69} and \citet{Thu75} with simplifying assumptions, refined somewhat by \citet{Rag86}.
{\changed
  Analytic and one-dimensional numerical predictions were made by \citet{Ras69} for the shapes and properties of H\,\textsc{ii} regions around moving stars, with the simplifying assumption that recombinations do not occur.
  More detailed numerical models (including recombinations but no hydrodynamics) by \citet{Thu75} predicted a somewhat cometary shape for the  H\,\textsc{ii} region with a broad recombination front in the wake behind the star, and noted that the H\,\textsc{ii} region reaches a steady state in a time shorter than the lifetime of a massive star.
  An analytic model by \citet{Rag86} showed that the H\,\textsc{ii} region should become more cometary as the stellar space velocity increases.
}
\citet{RagNorCanEA97} made the first two-dimensional axisymmetric {\changed radiation-hydrodynamics} simulations of an H\,\textsc{ii} region around a hypersonically-moving star.
The H\,\textsc{ii} region shape was found to agree broadly with predictions, being measurably (but not strongly) aspherical for the parameters chosen in the simulation.
In addition, the hydrodynamic expansion of the H\,\textsc{ii} region was found to drive a weak, outward-moving, conical shell of overdense gas trailing behind the star, which could be observable in infrared dust emission.
Other research has focused on the H\,\textsc{ii} regions produced by stars encountering a dense molecular interstellar medium (ISM)
{\changed \citep{TenYorBod79,MacvBurWooEA91,ArtHoa06},}
or stars moving from a dense cloud into lower density ISM at velocities $v_\star\leq12\,\ensuremath{\mathrm{km}\,\mathrm{s}^{-1}}$ \citep{FraGarKurEA07}, or slowly-moving young stars in proto-star-clusters {\changed \citep{PetBanKleEA10,DalBon11}}.

\citet{ChiRap96} studied H\,\textsc{ii} regions around stationary and moving supersoft X-ray sources.
They assumed isothermal ionized gas, and also ignored hydrodynamics, but included a non-equilibrium ionization calculation with an approximate treatment of the relative motion between the ISM and the radiation source.
They found that recombination-line emission-maps of H\,\textsc{ii} regions are slightly aspherical for moving stars with $v_\star=30\,\ensuremath{\mathrm{km}\,\mathrm{s}^{-1}}$, having a sharper upstream edge and more extended downstream edge.
This distortion becomes more dramatic with $v_\star=100$ and $300\,\mathrm{km}\,\mathrm{s}^{-1}$.
The general features of their results should also be found for H\,\textsc{ii} regions around moving main sequence O stars, although the softer radiation spectrum of O stars means that 
{\changed ionization fronts (I-fronts)}
will be much thinner.
The distortion they find for $v_\star=100\,\ensuremath{\mathrm{km}\,\mathrm{s}^{-1}}$ is much larger than was found by \citet{RagNorCanEA97}, a consequence of the different radiation spectra as well as the very different Str\"omgren radii of the modelled H\,\textsc{ii} regions.

The velocity range between transsonic and hypersonic stellar motion ($10\,\ensuremath{\mathrm{km}\,\mathrm{s}^{-1}} <v_\star<50\,\ensuremath{\mathrm{km}\,\mathrm{s}^{-1}}$) has not yet been explored for exiled stars moving through the diffuse ISM.
The distribution of stellar space velocities for exiles is, however, weighted strongly towards this range \citep{EldLanTou11}.
In addition, many O stars currently in clusters will become exiles later in life \citep[e.g.~because of a binary supernova explosion;][]{Bla61}, so the observed 25 per cent fraction of isolated O stars is a lower limit to the total fraction that become exiles \citep[cf.][]{EldLanTou11}.
It is shown here that in such cases the energy and momentum imparted to the ISM from the H\,\textsc{ii} region can be comparable to, or larger than, that from stellar winds, at least for moderate-velocity stars such as $\zeta$ Oph with $v_\star=26.5\,\ensuremath{\mathrm{km}\,\mathrm{s}^{-1}}$ \citep*{GvaLanMac12}.

The aims of this work are:
\begin{enumerate}
\item to investigate the gas dynamics of an H\,\textsc{ii} region produced by an exiled star moving with supersonic (but not hypersonic) velocities through the warm neutral medium (WNM), taking parameters similar to $\zeta$ Oph as an example case;
\item to predict the effects of the H\,\textsc{ii} region on the ISM, in terms of kinetic energy and momentum feedback; and
\item to assess the effects of the H\,\textsc{ii} region gas dynamics on the ISM near the star, to deduce the properties of the ISM that a stellar wind bow shock encounters.
\end{enumerate}

We address these questions by modelling H\,\textsc{ii} regions with one-, two-, and three-dimensional (1D, 2D, 3D) magnetohydrodynamic (MHD) simulations including non-equilibrium photoionization.
These are the first 3D simulations of H\,\textsc{ii} regions around supersonically-moving exiled massive stars, and also the first to include an ISM magnetic field.
We make simulated observations to investigate the shape of the H\,\textsc{ii} region, the structure of the shell around it, its emission properties, the effects of an ISM magnetic field, and the stability of the  I-front to perturbations.
In addition we investigate how the density and velocity of the ISM at the star are changed by the presence of the H\,\textsc{ii} region, and discuss the potential consequences for the stellar wind bow shock (not modelled here).
Our motivation is partly to test the analytic model of \citet{GvaLanMac12} for constraining the mass-loss rates of massive stars by simultaneous observation of their H\,\textsc{ii} region and bow shock.
This model ignored possible (magneto)hydrodynamic complications, for example the dynamic response of gas to photoionization and density inhomogeneity in the ISM.

Section~\ref{sec:methods} describes the simulation code and suite of simulations we have run.
Section~\ref{sec:1D} describes 1D simulations of H\,\textsc{ii} regions around stars moving with different velocities from sonic to hypersonic, comparing the position of the upstream I-front to analytic predictions.
Results from 3D simulations are presented in Section~\ref{sec:3D}, where the general morphology of the H\,\textsc{ii} regions is discussed, as well as the kinetic energy and momentum feedback, and the properties of the ISM near the star.
The results are discussed further in Section~\ref{sec:discussion}, and our conclusions are presented in Section~\ref{sec:conclusions}.
Effects of spatial and temporal numerical resolution are discussed for a range of stellar space velocities in Appendix~\ref{sec:app:res}.
Effects of limited spatial resolution in 3D simulations are studied in Appendix~\ref{sec:2D} with higher-resolution 2D simulations.
{\changed Equations for heating and cooling rates are listed in Appendix~\ref{app:HeatCool}.
}

%%%%%%%%%%%%%%%%%%%%%%%%%%%%%%%%%%%%%%%%%%%%%%%%%%%%%%%%%%%%%%%%%%%%%
\section{Numerical methods and simulation setup}
\label{sec:methods}
%%%%%%%%%%%%%%%%%%%%%%%%%%%%%%%%%%%%%%%%%%%%%%%%%%%%%%%%%%%%%%%%%%%%%
We use the radiation-MHD code \texttt{pion} \citep{MacLim10,MacLim11,Mac12} for the simulations presented here, solving either the Euler or ideal MHD equations on a uniform rectilinear grid, coupled to a microphysics integrator to solve for the non-equilibrium neutral fraction of hydrogen, $y_n$, and the ionization-dependent heating/cooling rates.
Simulations are run on 1D, 2D, and 3D Cartesian grids, with spatial derivatives set to zero for the dimensions not calculated (slab symmetry).
The finite-volume integration scheme {\changed(including heating and cooling source terms)} is algorithm A3 in \citet{Mac12}  \citep*[based on][]{FalKomJoa98} and is second-order-accurate in space and time and dimensionally unsplit.
Radiation transfer is solved in the on-the-spot approximation, solving only for direct radiation from point sources with a raytracer that calculates the column density of neutral hydrogen and total gas density from the source to every cell.
The radiation flux obeys an inverse-square law regardless of the grid dimensionality.
For the one-dimensional simulations we model a line along the star's direction of motion, passing through the star.
In two dimensions this is a plane containing the star and its velocity vector, and in three dimensions the full space is modelled.
The motivation for this in 1D is that it allows us to model the flow through the upstream I-front realistically (i.e.\ with correct boundary conditions).

\begin{figure}
\centering
\includegraphics[width=0.5\textwidth]{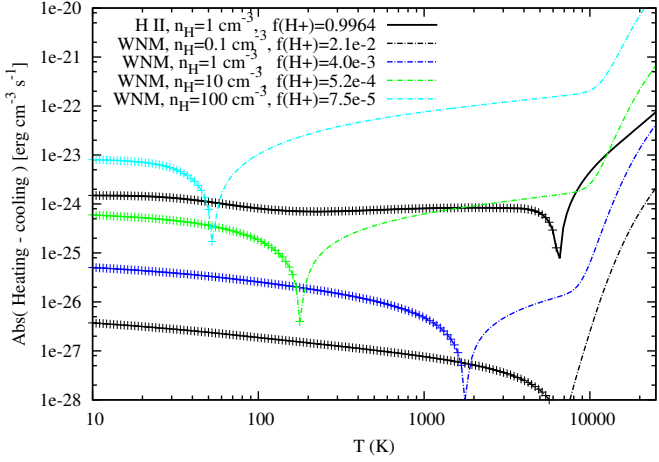}
\caption{
  Volumetric heating and cooling rates obtained for photoionized and neutral gas at different hydrogen number densities (labelled, in $\ensuremath{\mathrm{cm}^{-3}}$), and ionization fractions (set to the equilibrium value for each density).
  For each curve, when there is net heating or cooling the line is plotted with or without crosses, respectively.
  The heavy solid black line shows rates for photoionized gas exposed to an ionizing source (labelled H\,\textsc{ii}), whereas the dot-dashed lines show rates for WNM that is fully shielded from any point sources.
  \label{fig:heatingcooling}
  }
\end{figure}

%%%%%%%%%%%%%%%%%%%%%%%%%%%%%%%%%%%%%%%%%%%%%%%%%%%%%%%%%%%%%%%%%%%%%
\subsection{Microphysics}
%%%%%%%%%%%%%%%%%%%%%%%%%%%%%%%%%%%%%%%%%%%%%%%%%%%%%%%%%%%%%%%%%%%%%
We are considering a hot star moving through the WNM phase of the ISM with hydrogen number density $\ensuremath{n_{\textsc{h}}}\approx0.1-3\,\ensuremath{\mathrm{cm}^{-3}}$, where there is very little attenuation of the background far-ultraviolet heating radiation (photon energies 6-13.6 eV, hereafter FUV).
The molecular fraction is low because of the FUV background, so we ignore molecules altogether.
The main coolants are then fine-structure lines from carbon and oxygen, and polycyclic aromatic hydrocarbons (PAHs) \citep[e.g.][]{WolMcKHolEA03}.
We assume that ISM dust is everywhere on the simulation domain with standard properties and abundance, allowing us to use thermal physics prescriptions from \citet{WolMcKHolEA03}.
With these assumptions we only need to consider dust (including PAH), electrons, and atomic H, He, C, and O in the thermal physics.
We further assume that He is everywhere singly ionized to the same level as H, and never doubly ionized, and that its abundance is $0.1\ensuremath{n_{\textsc{h}}}$.
Gas-phase C is everywhere at least singly ionized because of the lack of shielding, setting a minimum electron fraction 
$ x_e\equiv n_e/\ensuremath{n_{\textsc{h}}} \geq 1.5\times10^{-4}$, where $n_e$ is the electron number density, and we take the gas-phase abundance of C from \citet{SofCarGueEA97}.
With our assumptions about H and He, the electron fraction is then
\begin{equation}
x_e = 1.5\times10^{-4} + 1.1(1-y_n) \;.
\end{equation}

Near the star there is strong photoionizing radiation, and also an elevated FUV radiation field that we model following \citet{HenArtDeCEA09} for the extra FUV heating rate.
In photoionized gas there is strong cooling from forbidden-line emission from ionized C and O.
We do not include stellar winds so the maximum gas temperature is $T\approx12\,000\,$K.

With these assumptions the rate equation for $y_n$ is
\begin{equation}
\dot{y}_n = \alpha_{\mathrm{b}}(1-y_n)x_e\ensuremath{n_{\textsc{h}}} -A_{\mathrm{ci}}y_nx_e\ensuremath{n_{\textsc{h}}} -1.8\times10^{-17}y_n  -A_{\mathrm{pi}}y_n \;.
\label{eqn:yrate}
\end{equation}
The terms on the right-hand side represent, respectively, radiative recombination, collisional ionization, cosmic ray ionization, and photoionization.
Here $\alpha_{\mathrm{b}}$ is the case B recombination coefficient of H \citep{Hum94},
$A_{\mathrm{ci}}$ is the collisional ionization coefficient of H \citep{Vor97},
$1.8\times10^{-17}$ is the cosmic ray ionization rate of H per neutral H atom \citep[e.g.][]{WolMcKHolEA03},
and $A_{\mathrm{pi}}$ is the photoionization rate of H per neutral H atom.
For this we use the photon-conserving discrete form \citep{AbeNorMad99,MelIliAlvEA06}.

{\changed
For most simulations we use multifrequency radiation, but some one-dimensional test calculations in Section~\ref{sec:1D} also use monochromatic radiation.
The photon-conserving form of $A_{\mathrm{pi}}$ for multifrequency radiation is \citep{MelIliAlvEA06}
\begin{equation}
A_{\mathrm{pi}}y_n = \int_{\nu_{\mathrm{th}}}^{\infty}
    \frac{L_\nu \mathrm{e}^{-\tau_\nu}}{h\nu} 
    \frac{1-\mathrm{e}^{-\Delta\tau_\nu}}{n_{\textsc{h}}V_{\mathrm{shell}}}  d\nu \;,
\label{eqn:pion_FVrate}
\end{equation}
integrated over frequency, $\nu$, from the threshold frequency for H photoionization, $\nu_{\textrm{th}}$, for a given source luminosity $L_\nu$ (with units erg\,cm$^{-3}$\,s$^{-1}$\,Hz$^{-1}$).
Here $\tau_\nu(r)\equiv \int_0^r n_\textsc{h}(r^\prime)y_n(r^\prime) \sigma_\nu dr^\prime$ is the optical depth along a ray connecting a grid zone to the source, integrated from the source to the point where the ray enters the zone ($\sigma_\nu$ is the frequency-dependent photoionization cross-section of H$^0$).
The optical depth along the ray section $\Delta s$ within a cell is $\Delta\tau_\nu = n_\textsc{h}y_n \sigma_\nu \Delta s$.
The quantity $V_{\mathrm{shell}}=4\pi[(r+\Delta s)^3-r^3]/3$ is the volume of a spherical shell with inner and outer radii corresponding to the intersection of the ray with the grid zone boundaries.
Following \citet{FraMel94} the integration over frequency is pre-calculated and tabulated for a wide range of optical depths, here for the simpler case where the opacity of helium is ignored.
A blackbody spectrum has been assumed, but in principle a more realistic spectrum could also be used.

To calculate $\tau_\nu$ rays are traced using the short characteristics method with the interpolation scheme proposed in \citet{MelIliAlvEA06}.
Two raytracing are performed per timestep, one of which uses time-centred values of $n_\textsc{h}$ and $y_n$ to ensure the overall scheme is second-order-accurate in time \citep{Mac12}.
Algorithm A3 in \citet{Mac12} is an explicit finite-volume integration scheme, so the timestep must be limited by the velocity of any I-fronts in the simulation (in addition to the hydrodynamic timestep limit).
It has been shown that a sufficient criterion to use with A3 is that the timestep, $\Delta t$, should satisfy $\Delta t \leq 0.25/\dot{y}_n$ \citep[criterion dt02 in][]{Mac12}, and this is used throughout this paper (for further discussion on timestep criteria see Appendix~\ref{sec:app:res}).

For monochromatic radiation, Eq.~(\ref{eqn:pion_FVrate}) reduces to the simpler expression
\begin{equation}
A_{\mathrm{pi}}y_n = \frac{Q_0 \mathrm{e}^{-\tau}(1-\mathrm{e}^{-\Delta\tau})}{n_{\textsc{h}}V_{\mathrm{shell}}} \;,
\label{eqn:pion_FVrate_mono}
\end{equation}
where $Q_0$ is the ionizing photon luminosity, and $\tau$ is measured at the photon frequency which is set to $h\nu=18.6$\,eV (appropriate for a late O star).

Eq.~(\ref{eqn:yrate})}
is integrated together with the rate equation for the change of internal energy density $E\equiv p_g/(\gamma-1)$ (where $p_g$ is the gas thermal pressure), given by

\begin{align}
\frac{\dot{E}}{\ensuremath{n_{\textsc{h}}}}& = 
  H_{\mathrm{pi}}y_n 
 +H_{\mathrm{fuv}}
 +5\times10^{-28}y_n
 +H_{\mathrm{pah}} \nonumber\\
& -A_{\mathrm{ci}}\psi_{\mathrm{H}}y_nx_e\ensuremath{n_{\textsc{h}}}
 -C_{\mathrm{rr}}x_e(1-y_n)\ensuremath{n_{\textsc{h}}}
 -C_{\mathrm{ff}}x_e(1-y_n)\ensuremath{n_{\textsc{h}}} \nonumber\\
&  -C_{\mathrm{cx}}y_nx_e\ensuremath{n_{\textsc{h}}} 
 -C_{\mathrm{nt}}
 -C_{\mathrm{m}}x_e(1-y_n)\ensuremath{n_{\textsc{h}}}  \;,
\label{eqn:energyrate}
\end{align}
{\changed 
where the terms correspond to, respectively,
photoionization heating with $H_{\mathrm{pi}}$ the heating per neutral H atom;
FUV heating from the ionizing star;
cosmic ray heating;
FUV photoelectric heating from the background radiation field;
collisional ionization cooling ($\psi_{\mathrm{H}}$ is the ionization potential of H);
recombination cooling;
Bremsstrahlung;
cooling from collisional excitation of H$^0$;
cooling from grains, C$^+$, and O$^0$ in mostly neutral gas;
and in hotter photoionized gas the forbidden line emission from photoionized O and C.
Most of the equations are taken from \citet{HenArtDeCEA09}, \citet{WolMcKHolEA03}, and \citet{Hum94};
the form of the equations used and references are given in Appendix~\ref{app:HeatCool}.
}

These two equations are coupled in that most of the rates depend on $T$ and $x_e$.
They are integrated together using high-order backwards differencing and adaptive substepping to fixed relative and absolute error tolerances with the \textsc{cvode} numerical integration library \citep{CohHin96}.
{\changed
It has been demonstrated \citep{BovGraLatEA13} that high-order integration methods which integrate the energy equation together with the rate equations are significantly better than uncoupled schemes.
}

Volumetric heating/cooling rates as a function of $T$ are plotted in Fig.~\ref{fig:heatingcooling} for different gas densities, with and without an ionizing radiation field.
The curve for photoionized gas is largely independent of density and distance from the star, except that the equilibrium temperature increases with source attenuation because of spectral hardening.
For neutral gas far from any ionizing sources, the heating rate scales with density whereas cooling scales with density squared.
This means the equilibrium temperature decreases from about 7000\,K at $\ensuremath{n_{\textsc{h}}}=0.1\,\ensuremath{\mathrm{cm}^{-3}}$ to 55\,K at $\ensuremath{n_{\textsc{h}}}=100\,\ensuremath{\mathrm{cm}^{-3}}$.
The main coolants at these temperatures are far-infrared fine-structure metal lines.
At $T\gtrsim10^4$\,K, collisional excitation of neutral H becomes dominant in neutral gas, and C and O optical forbidden lines in ionized gas.

%%%%%%%%%%%%%%%%%%%%%%%%%%%%%%%%%%%%%%%%%%%%%%%%%%%%%%%%%%%%%%%%%%%%%
\subsection{Simulation setup}
%%%%%%%%%%%%%%%%%%%%%%%%%%%%%%%%%%%%%%%%%%%%%%%%%%%%%%%%%%%%%%%%%%%%%
The simulations here are designed to be relevant for the nearby runaway O9.5\,V star $\zeta$ Oph \citep*{MarSchHil05}, discussed in \citet{GvaLanMac12}.
A mean number density of $n \approx 1.1 \ensuremath{n_{\textsc{h}}} = 3\,\ensuremath{\mathrm{cm}^{-3}}$ was derived for the ISM in its vicinity; here we use a similar value $\ensuremath{n_{\textsc{h}}}=2.5\,\ensuremath{\mathrm{cm}^{-3}}$, or $\rho_0=5.845\times10^{-24} \,\mathrm{g}\,\mathrm{cm}^{-3}$.
The initial gas pressure is $p_g=3.795\times10^{-13}\,\mathrm{dyne}\,\mathrm{cm}^{-2}$, corresponding to a temperature of $T=1000\,$K, and the initial H$^+$ fraction $(1-y_n)=0.0021$ is the equilibrium value at this temperature and density.
A random adiabatic perturbation with maximum amplitude of 25 per cent in pressure is applied to each cell in multidimensional simulations.
A point source of ionizing and FUV photons (hereafter `the star') is placed at the origin, and the ISM flows past this at a velocity of $v_\star=26.5\,\ensuremath{\mathrm{km}\,\mathrm{s}^{-1}}$, again motivated by the case of $\zeta$ Oph.

The star has an ionizing photon luminosity $Q_0=3.63\times10^{47}\,\mathrm{s}^{-1}$, distributed according to a blackbody spectrum for an effective temperature $T_\star=30\,500\,$K, appropriate for an O9.5\,V star.
In addition it has a FUV photon luminosity $L_{\mathrm{fuv}}=3\times10^{47}\,\mathrm{s}^{-1}$, implemented as in \citet{HenArtDeCEA09}; this acts in addition to the background interstellar radiation field, which is assumed to have the standard parameters \citep[taken from][]{WolMcKHolEA03}.

Simulations have been run at a number of different spatial resolutions, with zone-size decreasing by factors of 2 from $\Delta x = 0.32\,$pc to $0.005\,$pc (identified by name as r1 to r7, with r1 having the coarsest resolution).
In 3D only r1 and r2 were possible with available computing resources.
To estimate resolution effects, models with resolution r3 and r4 were run in 2D (see Appendix~\ref{sec:2D}), and up to resolution r7 in 1D (see Section~\ref{sec:1D}).
The highest resolutions (r6 and r7) have grid zones that are optically thin to ionizing photons even when fully neutral (the mean free path for a 13.6 eV photon in neutral gas at $\ensuremath{n_{\textsc{h}}}=2.5\,\ensuremath{\mathrm{cm}^{-3}}$ is $\approx 0.021\,$pc, comparable to the zone size for r5).
Spatially resolving the I-front in 3D will be difficult to achieve without adaptive spatial resolution (cf.~\citealt{CanPor11}), but the results presented here show this is not necessary to obtain meaningful results, at least for $v_\star<100\,\ensuremath{\mathrm{km}\,\mathrm{s}^{-1}}$.

For 2D and 3D calculations, the supersonic motion of gas across the grid generates anisotropic numerical diffusivity in grid-aligned flows, and this can produce spurious numerical instability \citep{Qui94,SutBicDop03}.
If the flow is grid-aligned, the I-front instability is much stronger along the grid-axis and this eventually destroys the solution.
A number of corrections were attempted but none completely removes this effect, so the best solution was to rotate the bulk flow by 40\ensuremath{^\circ} so that the numerical diffusivity in the $x$ and $y$ directions are comparable.
The flow velocity is rotated in 3D simulations for the same reason, and is set at 50\ensuremath{^\circ} to the $z$-axis and 40\ensuremath{^\circ} to the $x$-axis.

\begin{figure}
\centering
\includegraphics[width=0.42\textwidth]{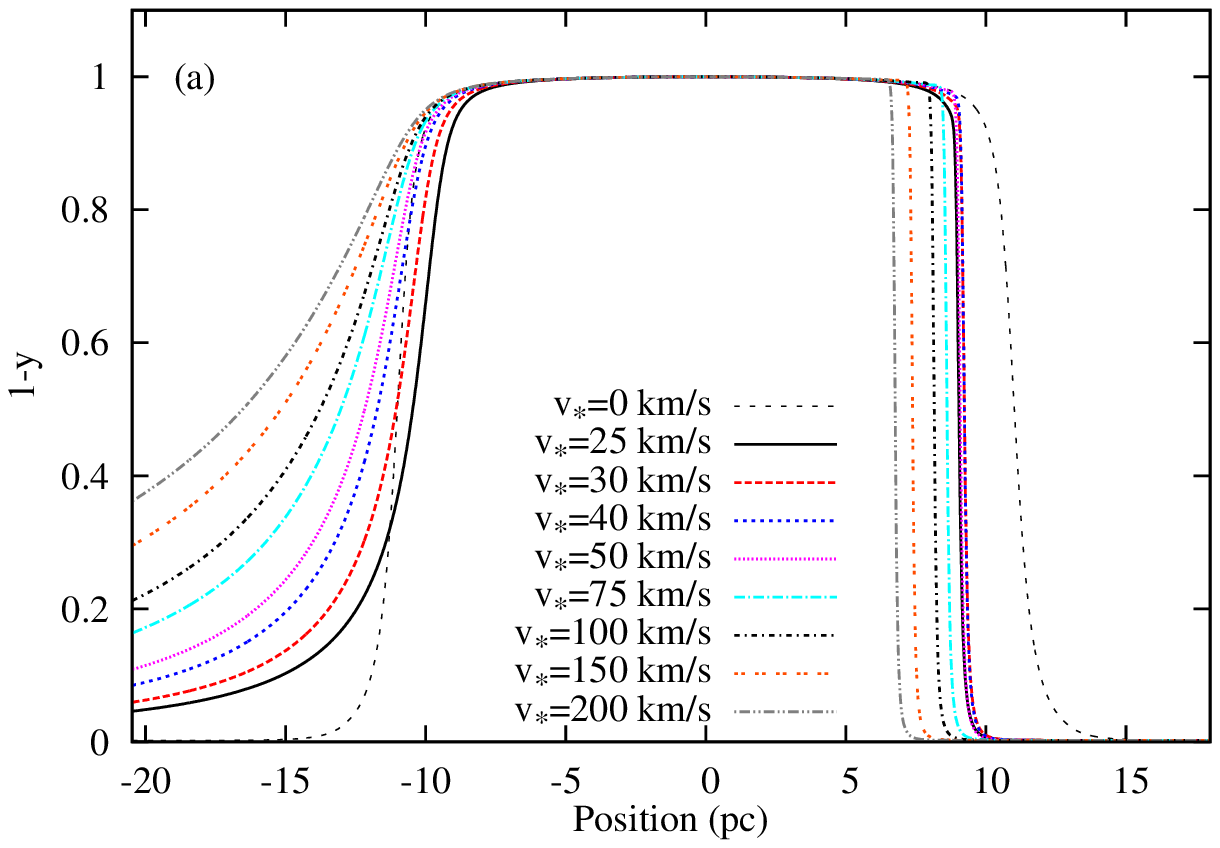}
\includegraphics[width=0.42\textwidth]{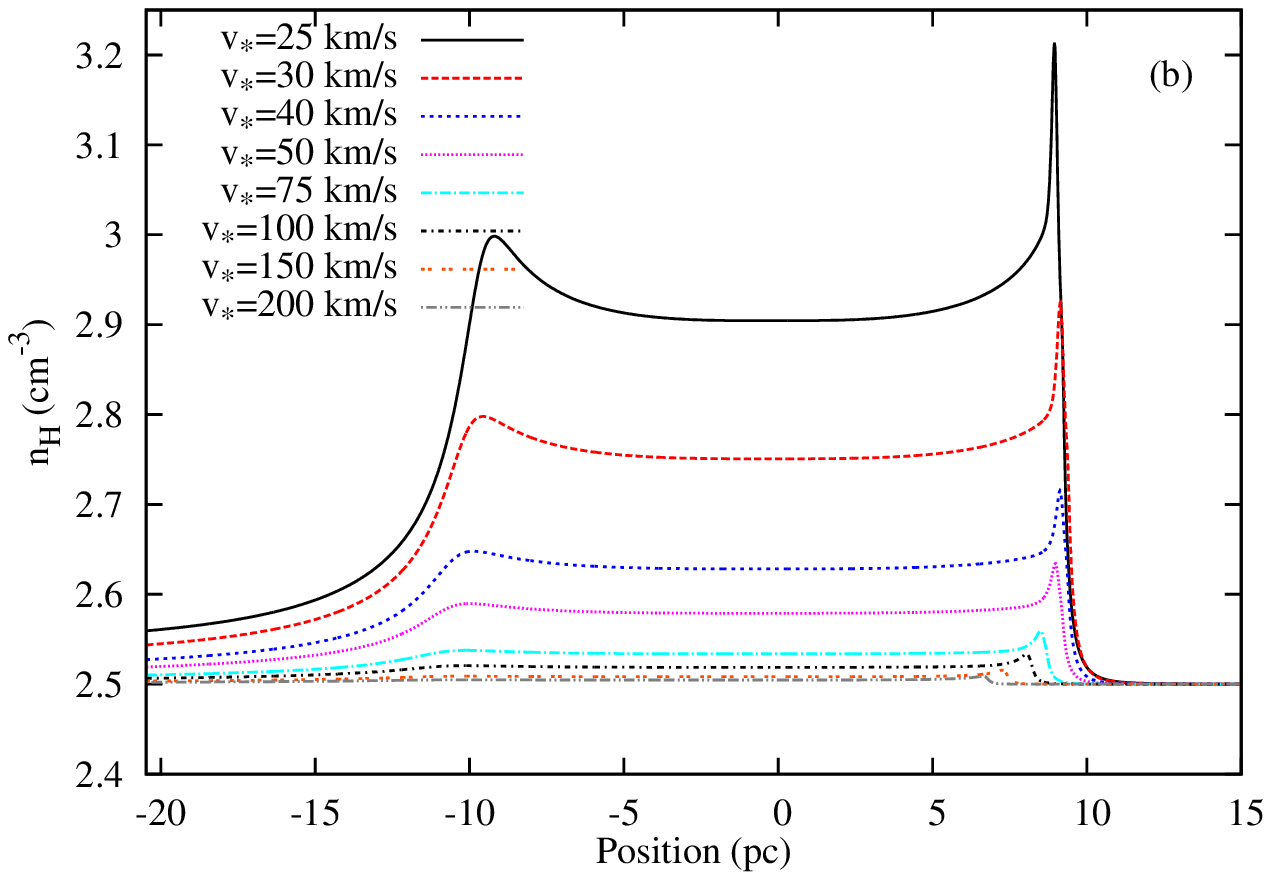}
\includegraphics[width=0.42\textwidth]{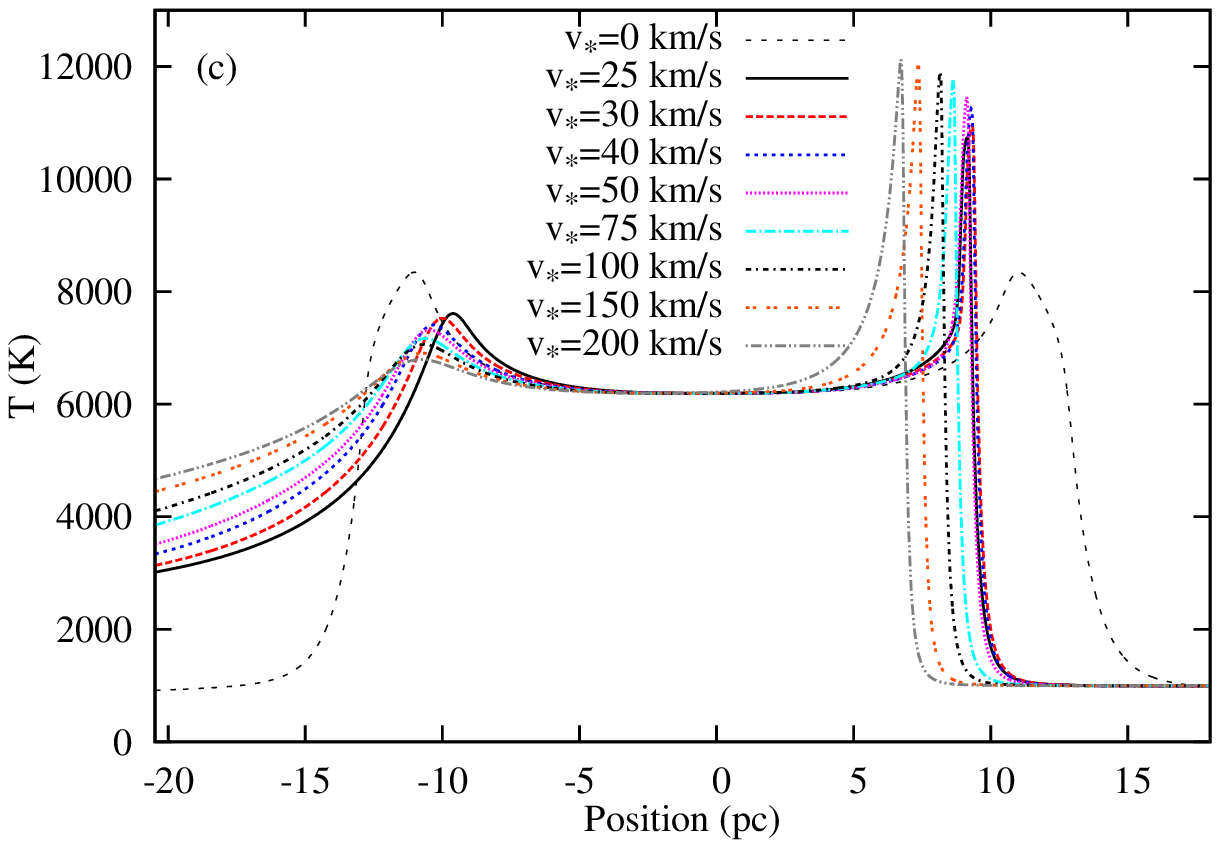}
\includegraphics[width=0.42\textwidth]{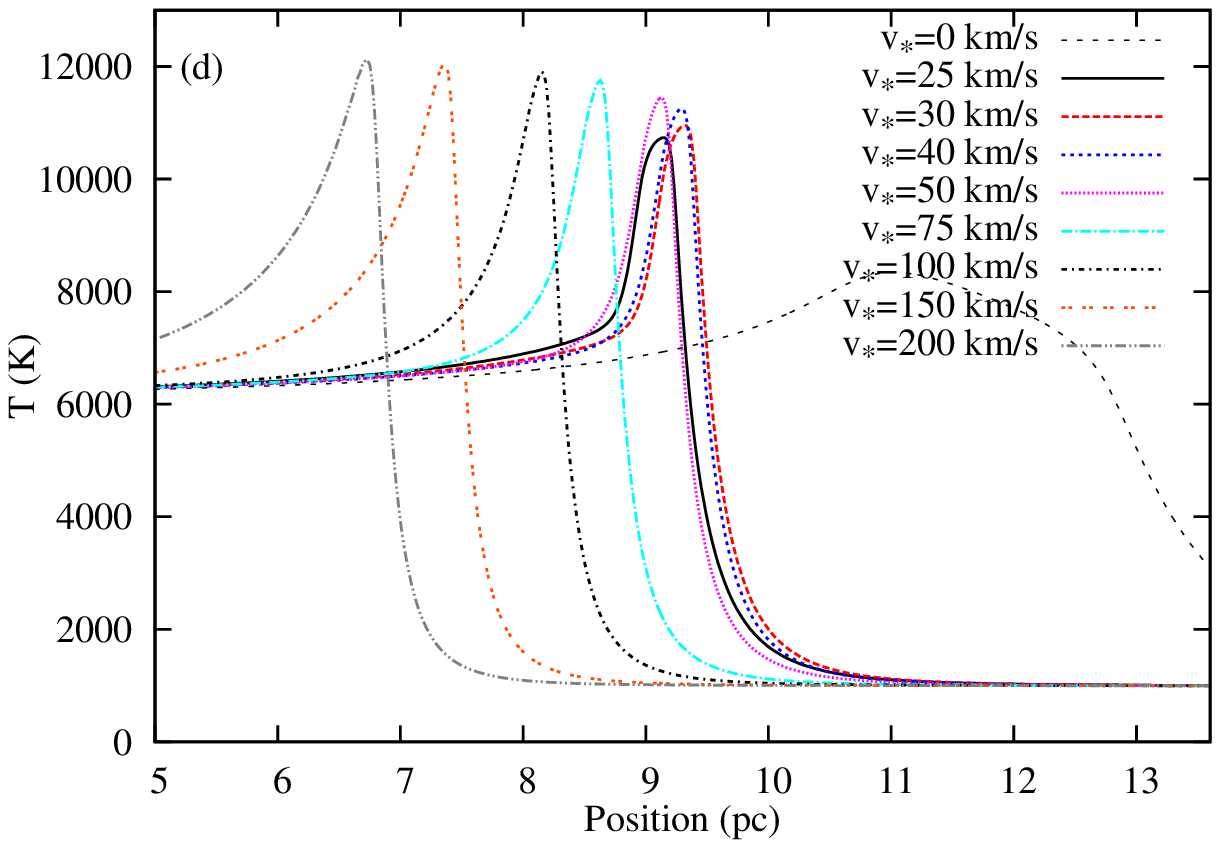}
\caption{
  1D simulations of H\,\textsc{ii} regions from runaway stars with space velocities through the ISM, $v_\star$, as indicated, shown after many advection times when the solution is no longer changing.
  The top panel (a) shows the H$^+$ fraction, $(1-y_n)$, as a function of position relative to the star (in parsecs); (b) shows gas number density; (c) shows gas temperature, and (d) is a zoom-in of the gas temperature at the upstream I-front.
  The thin black dashed line shows results for a static star with no gas dynamics.
  \label{fig:1Dprofiles}
  }
\end{figure}

%%%%%%%%%%%%%%%%%%%%%%%%%%%%%%%%%%%%%%%%%%%%%%%%%%%%%%%%%%%%%%%%%%%%%
\section{One-Dimensional simulations}
\label{sec:1D}
%%%%%%%%%%%%%%%%%%%%%%%%%%%%%%%%%%%%%%%%%%%%%%%%%%%%%%%%%%%%%%%%%%%%%
The primary parameter determining the properties of an I-front is its Mach number relative to the isothermal sound speed in photoionized gas $\mathcal{M}\equiv v/c_i$, and this depends only weakly on the stellar radiation field, ISM number density, and ISM metallicity.
For this reason, only the star's space velocity is varied here; the other properties of the star and the ISM are kept constant.
Stellar space velocities from 10-200 \ensuremath{\mathrm{km}\,\mathrm{s}^{-1}} have been simulated, specifically  $v_{\star} \in \{10,\,20,\,25,\,30,\,40,\,50,\,75,\,100,\,150,\,200\} \,\ensuremath{\mathrm{km}\,\mathrm{s}^{-1}}$.
The models have been run for spatial resolutions r1 to r7 and for different timestep criteria.

One-dimensional (1D) simulations allow us to model both the upstream I-front and the downstream recombination front, although multidimensional effects (not to mention the shocked stellar wind) are expected to alter the recombination front.
As well as studying the physical properties of the fronts, this also allows a very clean test of the numerical error of our solution as a function of spatial and temporal resolution.
The simulations relax to a stationary state in all cases where $v_\star$ is greater than Mach 2 in ionized gas.
The effects of spatial and temporal resolution on the 1D simulations are discussed in Appendix~\ref{sec:app:res}; here we note that it is not necessary to spatially resolve the I-front (cell optical depths could be $\tau>1$ without loss of accuracy) at least for $v_\star<100\,\ensuremath{\mathrm{km}\,\mathrm{s}^{-1}}$.
It is necessary, however, to use a smaller-than-usual Courant-Friedrichs-Lewy (CFL) number of $C_{\mathrm{cfl}}=0.1$ for $v_\star<100\,\ensuremath{\mathrm{km}\,\mathrm{s}^{-1}}$, and even smaller for larger $v_\star$.
When the I-front is spatially resolved the timestep criterion is not so important.

Profiles of the ISM density, H$^{+}$ fraction $(1-y_n)$, and temperature are shown in Fig.~\ref{fig:1Dprofiles} for all models with $v_\star\geq25\,\ensuremath{\mathrm{km}\,\mathrm{s}^{-1}}$, with a profile for a static star in a static medium (i.e.~the original case considered by \citealt{Str39}) overplotted for reference.
We have not plotted lower velocities where the I-front is D-type because the H\,\textsc{ii} region expands further than the Str\"omgren radius $R_{\mathrm{s}}$, and the dense shell that forms remains unstable for many grid advection times (the time for the background flow to cross the grid).

None of the models reaches the Str\"omgren radius in the upstream direction.
A careful look at the plots also shows that the $v_\star=25\,\ensuremath{\mathrm{km}\,\mathrm{s}^{-1}}$ I-front is less advanced than the $v_\star=30\,\ensuremath{\mathrm{km}\,\mathrm{s}^{-1}}$ I-front, which can be attributed to the higher mean density in the H\,\textsc{ii} region in the lower velocity model, and also to the formation of a weak shell at the upstream I-front.
This is somewhat artificial, in that multidimensional expansion of the H\,\textsc{ii} region will be more important at lower velocities and will reduce the mean density as ionized gas flows downstream.
Nevertheless, it does faithfully reflect the 1D jump conditions for a weak R-type I-front \citep[][chapter 106]{MihMih84} according to
\begin{equation}
\rho_1/\rho_0=v_0/v_1\approx 1+\mathcal{M}^{-2} \;,
\label{eqn:WeakR}
\end{equation}
where neutral gas has subscript 0 and ionized subscript 1.
This is derived for isothermal gas, but should apply approximately here because far from the I-front gas relaxes to an equilibrium temperature both upstream and downstream.

\begin{figure}
\centering
\includegraphics[width=0.49\textwidth]{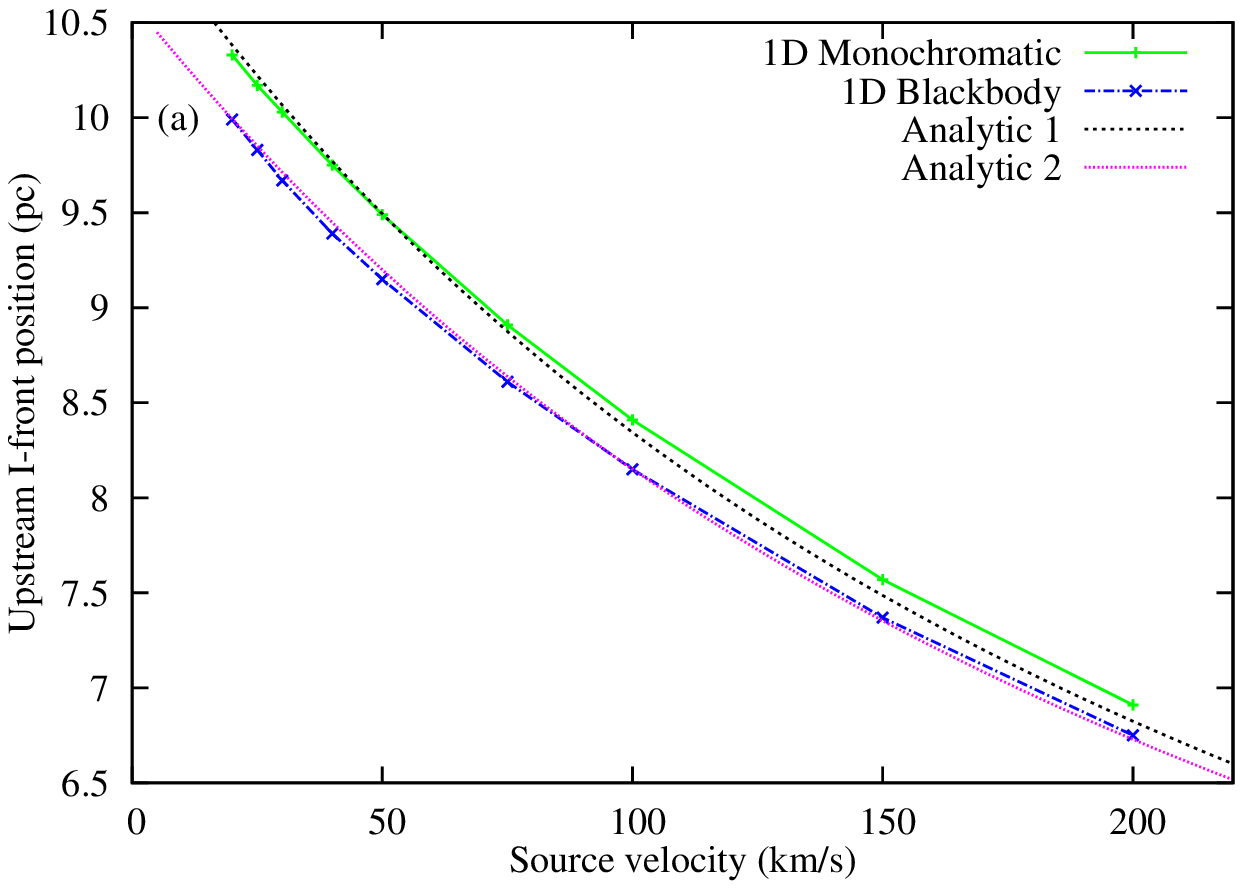}
\includegraphics[width=0.49\textwidth]{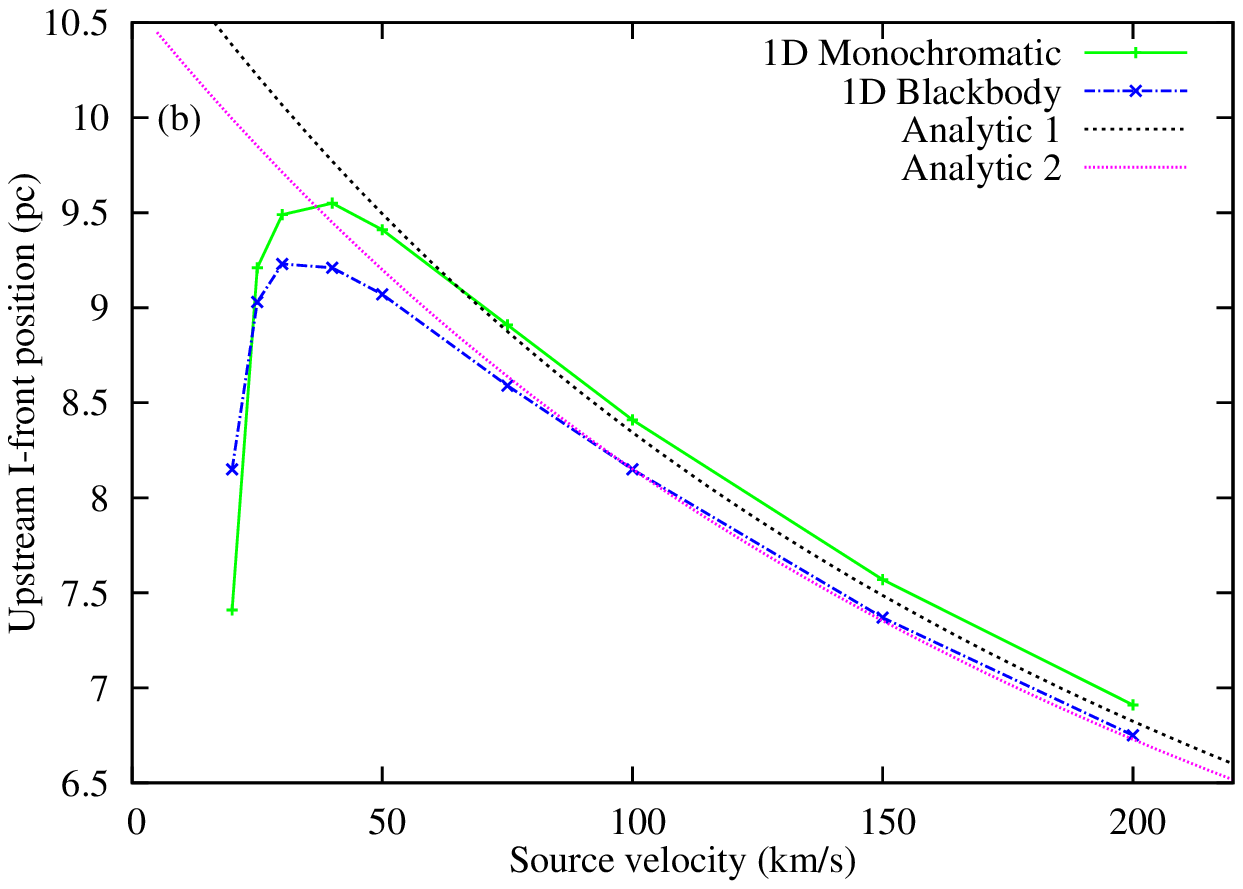}
\caption{
  Distance of the upstream I-front from the star as a function of the source velocity, $v_\star$, at an early time $t=100\,$kyr (a), and a late time corresponding to a stationary state for $v_\star\geq25\,\ensuremath{\mathrm{km}\,\mathrm{s}^{-1}}$ (b).
  The solid (green) curve is for a monochromatic radiation source, and the dot-dashed (blue) curve for a multifrequency radiation source, both normalised to have the same ionizing photon luminosity by number.
  The two dotted lines show the analytic solution (see text) for H\,\textsc{ii} regions with $T=7600\,$K and $6300\,$K, respectively.
  \label{fig:IFpos1D}
  }
\end{figure}

At larger velocities the I-front retreats with increasing $v_\star$.
To quantify this, the position of the I-front (defined as the furthest upstream grid zone with $y_n<0.5$) is plotted as a function of $v_\star$ in Fig.~\ref{fig:IFpos1D} for simulations with multifrequency radiation (the default case) and also for models with monochromatic radiation and the same ionizing photon luminosity (by number of photons).
The model with monochromatic radiation has a hotter interior part of the H\,\textsc{ii} region, hence a smaller recombination rate and a larger H\,\textsc{ii} region.

The results are compared to an analytic solution obtained by equating $v_\star$ with the velocity of the I-front, then solving for the I-front radius that corresponds to this velocity for a static star.
The I-front velocity for a static star with no gas dynamics is \citep{MelIliAlvEA06}
\begin{equation}
v_{\textsc{if}}(t) = \frac{R_s}{3t_r} \frac{\exp(-t/t_r)}{[1-\exp(-t/t_r)]^{2/3}} \;,
\end{equation}
where $t_r\equiv(\alpha_bn_e)^{-1}$ is the recombination time.
The I-front position is
\begin{equation}
R_{\textsc{if}}(t) = R_s[1-\exp(-t/t_r)]^{1/3} \;.
\end{equation}
These two equations can be combined \citep{RagNorCanEA97}, eliminating $t$, to obtain
\begin{equation}
\left(\frac{R_{\textsc{if}}}{R_s}\right)^3 +
 \frac{3t_rv_{\textsc{if}}}{R_s}\left(\frac{R_{\textsc{if}}}{R_s}\right)^2 -1=0 \;.
 \label{eqn:Rif}
\end{equation}
Following \citet{RagNorCanEA97} we equate $v_{\textsc{if}}$ with $v_\star$ to predict the steady-state I-front position in the upstream direction as a function of $v_\star$.
These are the dotted lines in Fig.~\ref{fig:IFpos1D} labelled `Analytic 1/2'.
At early times the actual radius of the I-front agrees very well with the analytic expression, but at later times there are differences for $v_\star<50\,\ensuremath{\mathrm{km}\,\mathrm{s}^{-1}}$, with the difference increasing as the velocity decreases because the dynamical response of the gas grows stronger.

The normalisation of the curve depends on $t_r$, which in turn depends on the temperature-dependent recombination rate $\alpha_b$.
The monochromatic radiation model has a roughly constant temperature in the H\,\textsc{ii} region, whereas the multifrequency radiation model has a cooler interior and a hotter border because of spectral hardening.
This implies a higher recombination rate and therefore a smaller H\,\textsc{ii} region.
The analytic curves were normalised to a H\,\textsc{ii} region temperature of $T=7600\,$K for monochromatic radiation and $6300\,$K for multifrequency radiation.

These 1D results and their comparison with analytic theory provide the basic understanding of the global asymmetry of moving H\,\textsc{ii} regions, which will be confirmed and further explored by multidimensional models in the following sections.
They also allow verification and testing of the algorithms in terms of their convergence properties and timestepping requirements (see Appendix~\ref{sec:app:res}).

\subsection{Comparison to previous work}
An R-type I-front with $v_\star=50\,\ensuremath{\mathrm{km}\,\mathrm{s}^{-1}}$ was modelled by \citet{HenArtWilEA05} as part of a study of steady I-fronts that included advection self-consistently.
They found that the supersonic I-front has a strong temperature peak at the I-front that is absent in the static case, and that the location of the I-front is closer to the radiation source by about 5 per cent.
The temperature peak in our case is perhaps a bit stronger and increases in amplitude with $v_\star$, but is qualitatively similar.
The position of the I-front in our calculations is modified by much more than 5 per cent compared to the static case, although \citet{HenArtWilEA05} use plane-parallel radiation so this cannot be compared directly to our results.

\citet{Ras69} studied H\,\textsc{ii} regions around moving stars assuming isothermal gas and neglecting recombinations, so agreement with our results is not expected.
The best agreement is in the position and properties of the upstream I-front, but the downstream recombination front is of course not captured at all by the \citet{Ras69} calculation.
Subsequent calculations including recombination (but without hydrodynamics) were made by \citet{Thu75} for the H\,\textsc{ii} region around a $v_\star=50\,\ensuremath{\mathrm{km}\,\mathrm{s}^{-1}}$ O star in a diffuse ISM with hydrogen number density $\ensuremath{n_{\textsc{h}}}=0.1\,\ensuremath{\mathrm{cm}^{-3}}$.
The upstream and downstream I-front radii were predicted to be $R_{\mathrm{up}}=140\,$pc and $R_{\mathrm{dn}}=265\,$pc, respectively, compared to $R_s=190\,$pc.
In our case for $v_\star=50\,\ensuremath{\mathrm{km}\,\mathrm{s}^{-1}}$ we find $R_{\mathrm{up}}/R_s=0.85$, significantly larger than the value of 0.74 from \citet{Thu75}; this is simply explained by the different values for $t_r$ and $R_s$ in the two calculations.
\citet{RagNorCanEA97} calculated $R_{\mathrm{up}}/R_s=0.93$ for an O5 star with $v_\star=100\,\ensuremath{\mathrm{km}\,\mathrm{s}^{-1}}$ and $\ensuremath{n_{\textsc{h}}}=1\,\ensuremath{\mathrm{cm}^{-3}}$; this again differs from the values obtained here because of the differing $t_r$ and $R_s$ (we consider an O9.5V star with a much smaller ionizing photon luminosity).
The ratio $R_{\mathrm{up}}/R_s$ decreases as the prefactor in the quadratic term of Equation~(\ref{eqn:Rif}) increases; as long as this is taken into account our results agree well with these previous calculations.
The 1D profiles in \citet{ChiRap96} show much broader I-fronts as a consequence of the much harder radiation spectrum they consider, and so they are not (and are not expected to be) comparable to our results.

\begin{table}
  \centering
  \begin{tabular}{| l | l | c | c |}
    \hline
    ID & $(N_x,N_y,N_z)$ & $\Delta x$ & $\mathitbf{B}$-field ($\mu$G) \\
    \hline
    HD3r1  & $(160,\,160,\,160)$   & 0.32 &  $(0,0,0)$ \\
    HD3r2  & $(320,\,320,\,320)$   & 0.16 &  $(0,0,0)$ \\
    \hline
    BA3r1 & $(160,\,160,\,160)$   & 0.32 &  $(7,0,0)$ \\
    BA3r2 & $(320,\,320,\,320)$   & 0.16 &  $(7,0,0)$ \\
    \hline
    BT3r1 & $(160,\,160,\,160)$   & 0.32 &  $(0,7,0)$ \\
    BT3r2 & $(320,\,320,\,320)$   & 0.16 &  $(0,7,0)$ \\
    \hline
  \end{tabular}
  \caption{
    Simulation properties for three-dimensional calculations.
    Columns show, respectively,
    simulation ID,
    number of grid zones in each direction,
    spatial diameter of a grid zone in parsecs, and
    magnetic field vector in $\mu$G in a reference frame where the motion of the star is along the $x$-axis.
    Models HD3x are hydrodynamic (no magnetic field), BA3x are MHD simulations with a B-field aligned with the direction of motion, BT3x are MHD with a field transverse (perpendicular) to the direction of motion.
    The number following the B-field designation, e.g.~`r2', represents the resolution.
  }
  \label{tab:Sims3D}
\end{table}

%%%%%%%%%%%%%%%%%%%%%%%%%%%%%%%%%%%%%%%%%%%%%%%%%%%%%%%%%%%%%%%%%%%%%
\section{3D simulations}
\label{sec:3D}
%%%%%%%%%%%%%%%%%%%%%%%%%%%%%%%%%%%%%%%%%%%%%%%%%%%%%%%%%%%%%%%%%%%%%
The 3D simulations run are listed in Table~\ref{tab:Sims3D}.
They consist of a 3D domain with $\{x,y,z\}\in[-32.64,18.56]\,$pc with a Cartesian coordinate system, and with the star at the origin.
There are simulations where the ISM magnetic field is aligned with (BA3r1, BA3r2) and perpendicular to (BT3r1, BT3r2) the direction of motion, and simulations with no magnetic field (HD3r1, HD3r2).
All MHD models have a field strength $B=7\,\mu$G, corresponding to a plasma parameter $\beta\equiv8\pi p_g/B^2=0.2$ in the neutral ISM, and $\beta=2.7$ in fully-ionized gas at $T=7000\,$K.
The neutral medium is therefore magnetically dominated, whereas the ionized gas is gas pressure dominated.
Models with a perpendicular field have the field vector in the $x$-$y$ plane.
The simulations have inflow boundary conditions on the upstream boundaries, and only-outflow conditions downstream, and it is assumed that the star `switches on' instantly at $t=0$.
As a result, the initial evolution is not very meaningful because no stars are born in the WNM.
The time for the ISM to advect across the diameter of the H\,\textsc{ii} region ($\approx20\,$pc) is about 0.75 Myr, so we expect the effects of initial conditions to disappear soon after this time.
This is also about the time that gas affected by the star's radiation first leaves the downstream boundary.
A time-stationary state takes $\approx1.5$ Myr to be established.

\begin{figure*}
\centering
\includegraphics[width=0.49\textwidth]{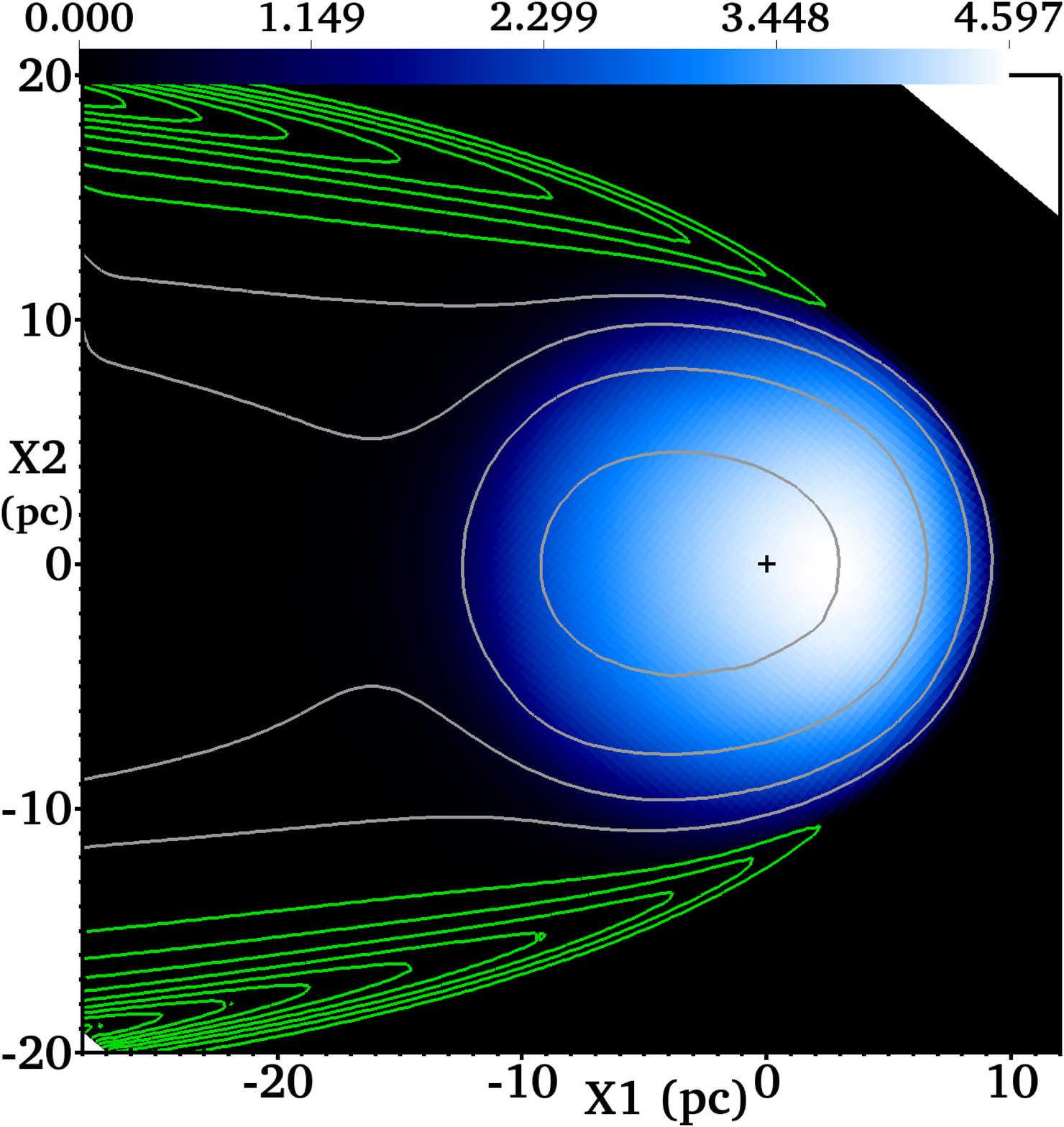}
\includegraphics[width=0.49\textwidth]{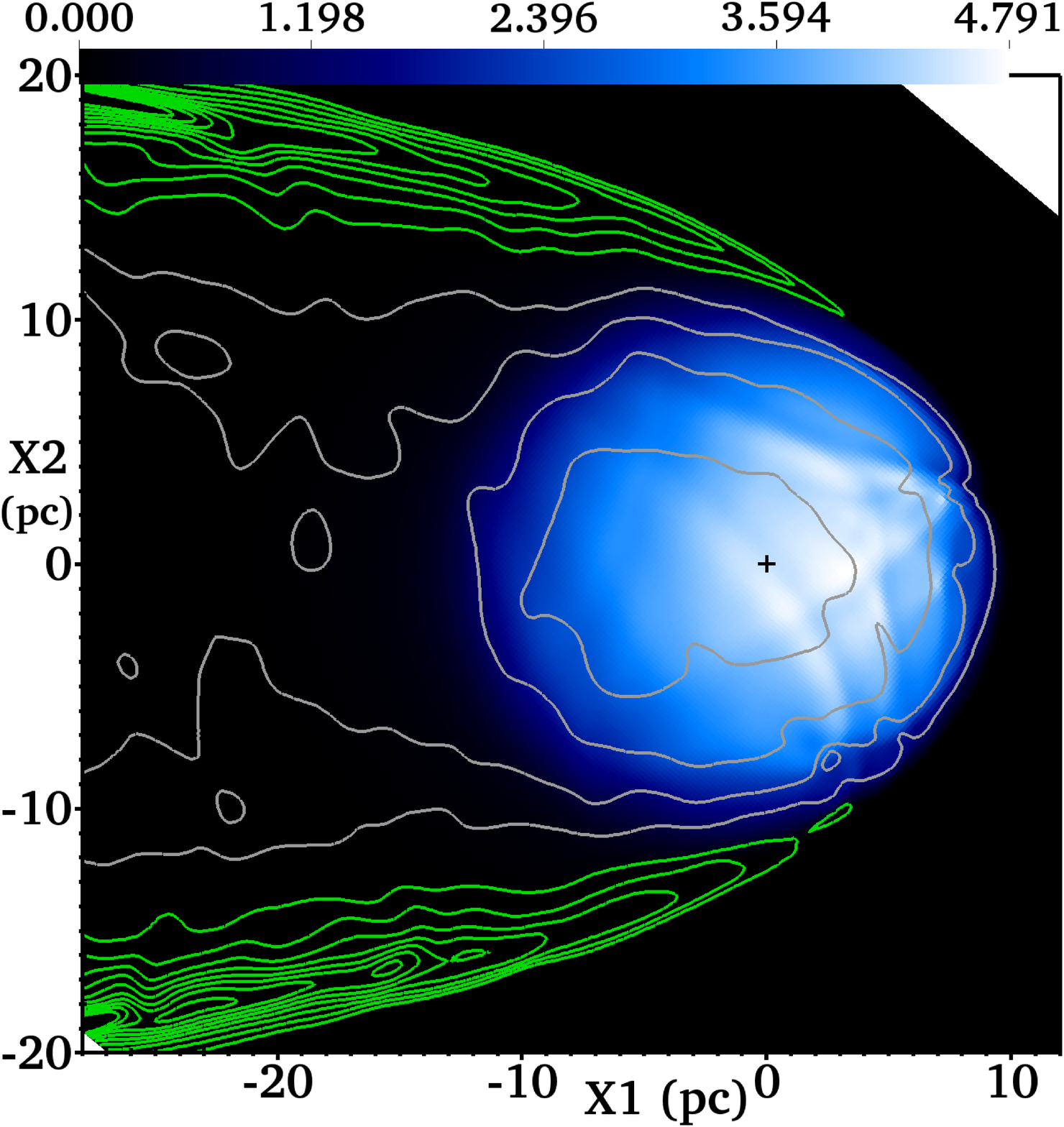}
\caption{
  Projections through the 3D simulations HD3r1 (left) and the higher resolution HD3r2 (right) at $t=4\,$Myr.
  The star is marked with a cross at the origin, and the ISM is flowing past from right to left at $v_\star=26.5\,\ensuremath{\mathrm{km}\,\mathrm{s}^{-1}}$.
  The projection is such that the bulk velocity is fully in the image plane and in the $-\hat{\mathitbf{x}}_1$ direction and perpendicular to $\hat{\mathitbf{x}}_2$.
  Projected H$\alpha$ emitted intensity is plotted on the linear colour scale (in units of $10^{-16}\,\mathrm{erg}\,\mathrm{cm}^{-2}\,\mathrm{s}^{-1}\,\mathrm{arcsec}^{-2}$), and neutral H column density ($N_{\mathrm{HI}}$) as contours.
  The mean column density of the undisturbed grid is subtracted off (to remove grid edge effects), so underdense regions have negative column density and are shown in grey contours with spacing $\Delta N_{\mathrm{HI}}=0.5\times10^{20}\,\ensuremath{\mathrm{cm}^{-2}}$.
  Overdense regions have green contours, again spaced with $\Delta N_{\mathrm{HI}}=0.5\times10^{20}\,\ensuremath{\mathrm{cm}^{-2}}$.
  The zero level contour is omitted for clarity.
  Animations showing time-evolution of these figures are available in the online version of this article.
  \label{fig:simHD3r12}
  }
\end{figure*}

\begin{figure*}
\centering
\includegraphics[width=0.49\textwidth]{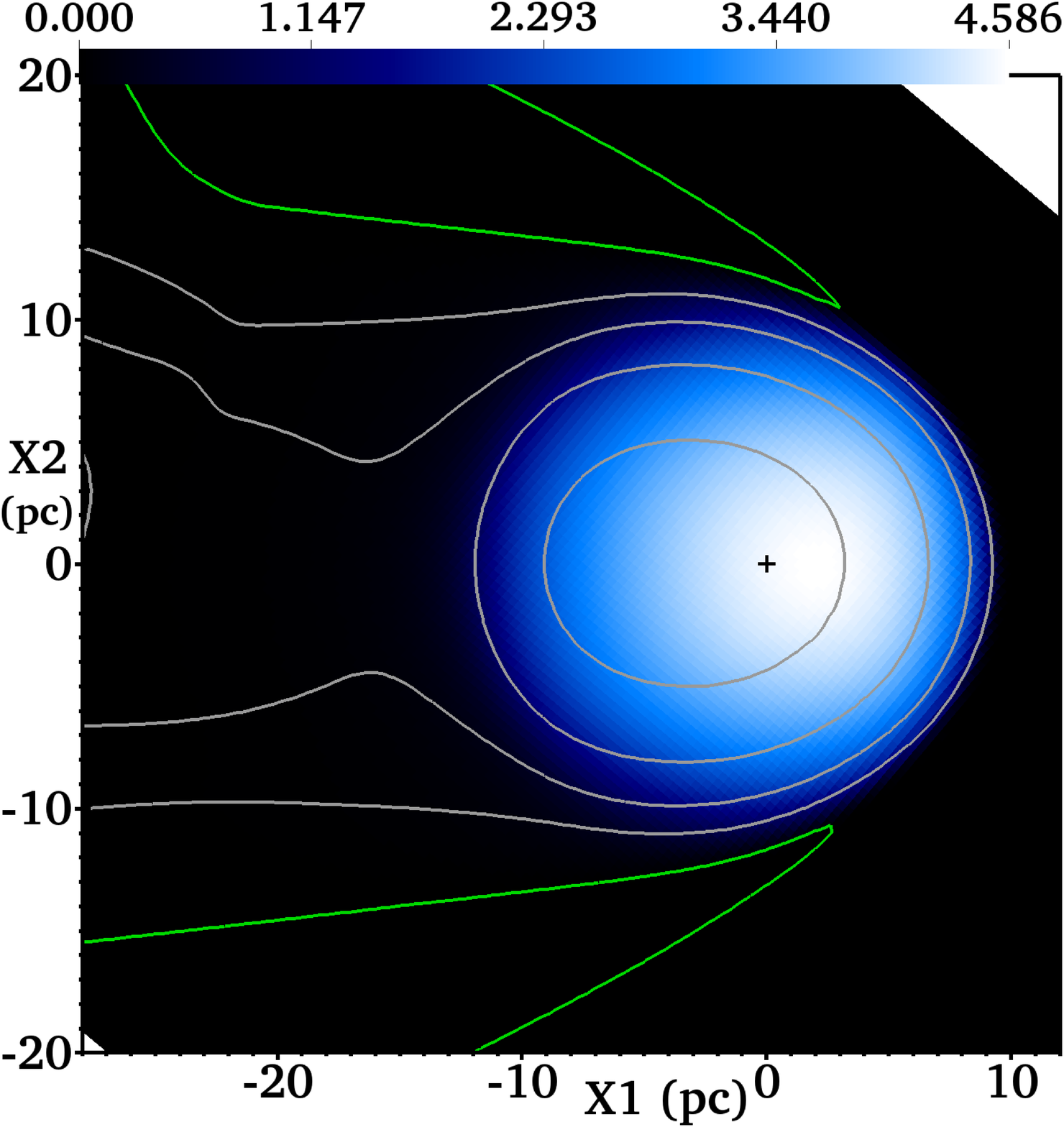}
\includegraphics[width=0.49\textwidth]{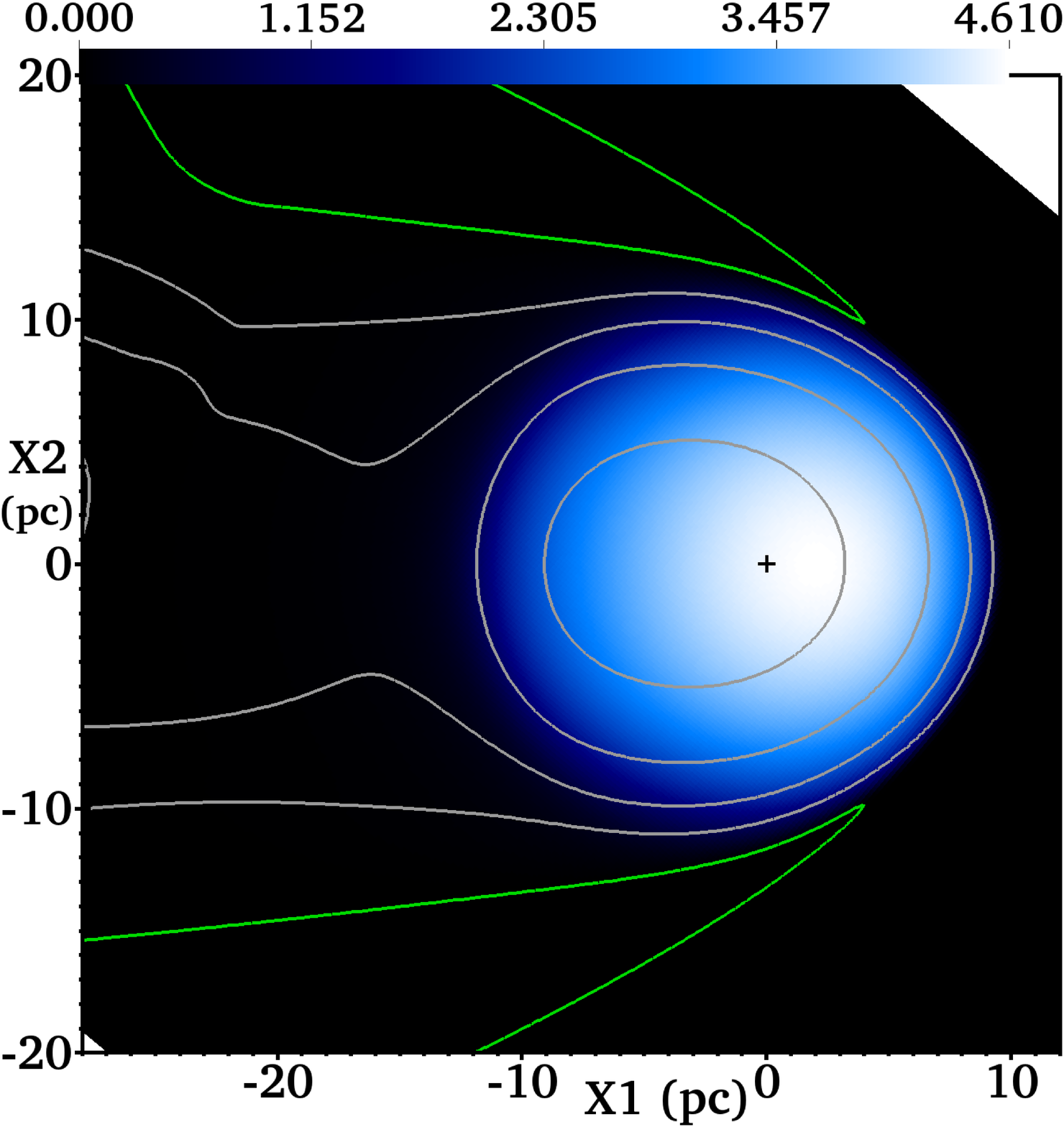}
\caption{
  As Fig.~\ref{fig:simHD3r12}, but for the simulations with a flow-aligned magnetic field, BA3r1 and BA3r2.
  Here the initially uniform magnetic field is fully in the image plane and parallel to the $X_1$ axis.
  Animations showing time-evolution of these figures are available in the online version of this article.
  \label{fig:simBA3r12}
  }
\end{figure*}

\begin{figure*}
\centering
\includegraphics[width=0.49\textwidth]{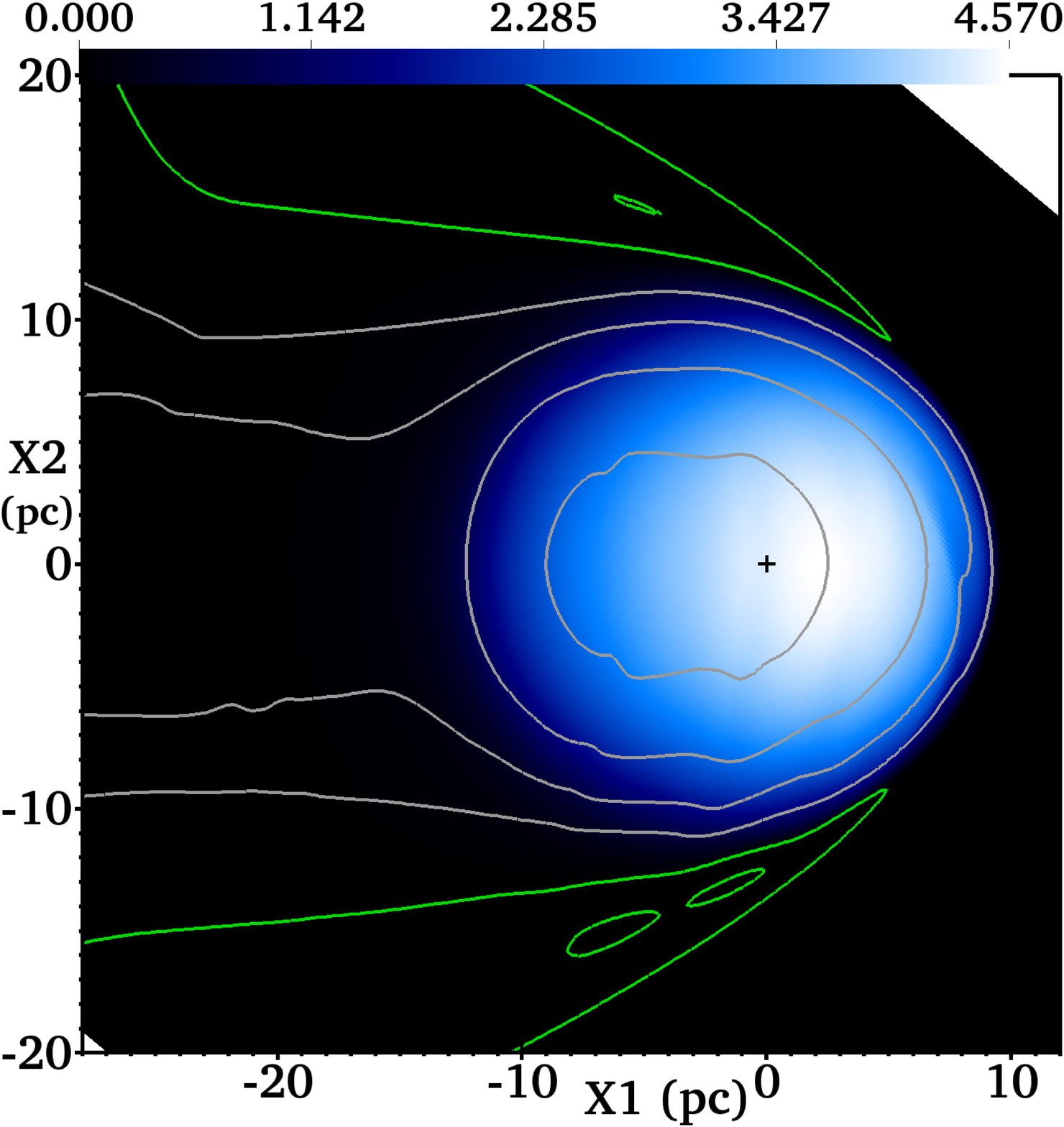}
\includegraphics[width=0.49\textwidth]{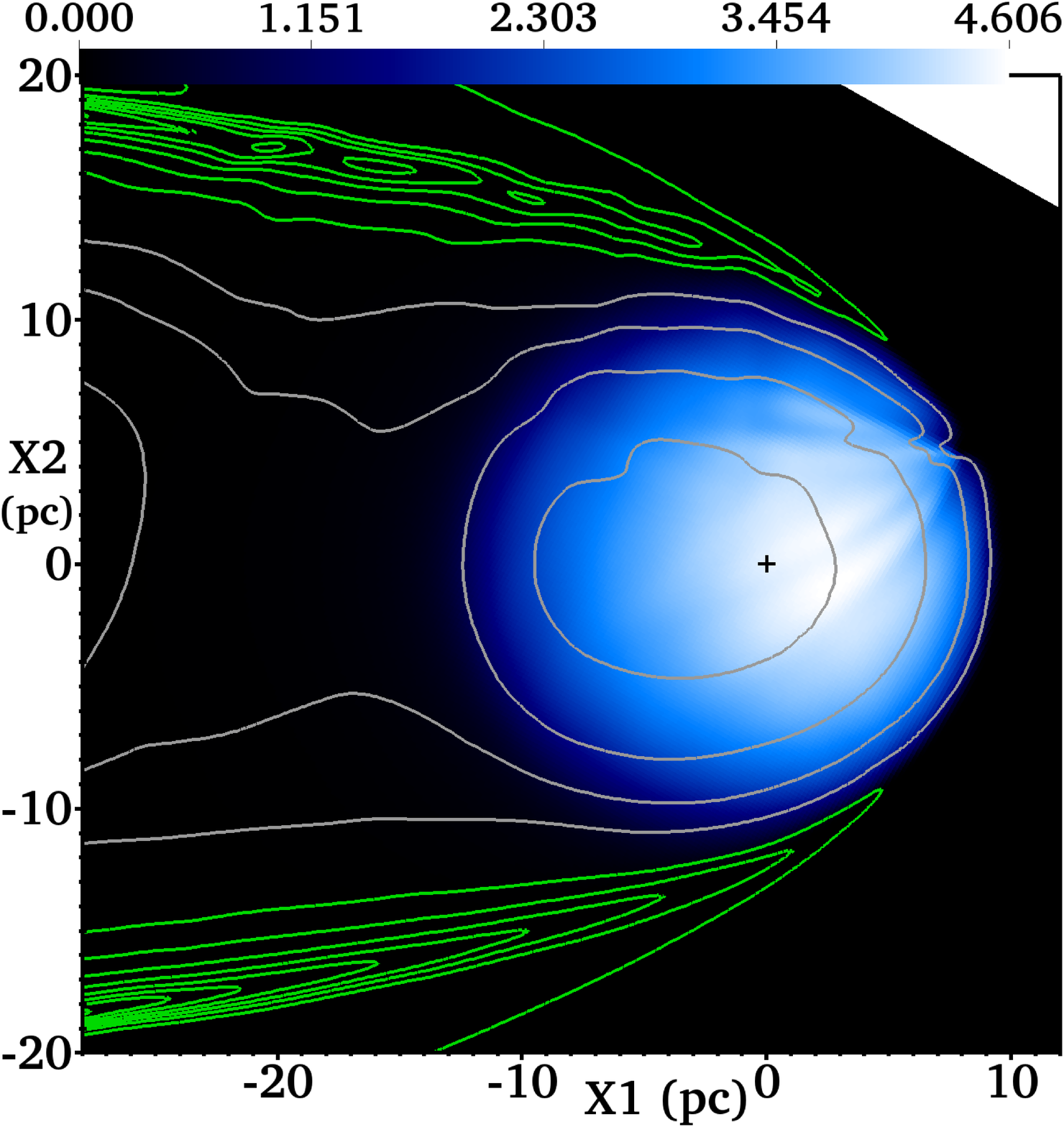}
\caption{
  As Fig.~\ref{fig:simHD3r12}, but for simulation BT3r2 with a magnetic field perpendicular to the gas flow, at $t=4.5\,$Myr, and from two different orientations.
  Both plots have the bulk flow in the image plane as before, but to the left the projection is such that the initial magnetic field is along the line of sight, whereas to the right the field is vertical and in the image plane.
  Animations showing time-evolution of these figures are available in the online version of this article.
  \label{fig:simBT3r2}
  }
\end{figure*}

%%%%%%%%%%%%%%%%%%%%%%%%%%%%%%%%%%%%%%%%%%%%%%%%%%%%%%%%%%%%%%%%%%%%%
\subsection{Observable properties of simulations}
%%%%%%%%%%%%%%%%%%%%%%%%%%%%%%%%%%%%%%%%%%%%%%%%%%%%%%%%%%%%%%%%%%%%%

Projections through snapshots from the hydrodynamic models HD3r1 and HD3r2 are plotted in Fig.~\ref{fig:simHD3r12}, where emitted H$\alpha$ intensity is plotted on a linear colour scale, and H~\textsc{i} column density, $N_{\mathrm{HI}}$, as contours.
The projection is such that the bulk flow is in the image plane, so the line-of-sight (LOS) is not parallel to the grid axes; we have subtracted off the undisturbed background $N_{\mathrm{HI}}$ to remove the effect of this varying LOS depth.
Some edge effects remain for $x<-20\,$pc from the projected edges of the grid domain.
The H$\alpha$ emissivity is taken as
\begin{equation}
j = 2.63\times10^{-33} n_e \ensuremath{n_{\textsc{h}}} (1-y_n) T^{-0.9} \;\;
\mathrm{erg}\,\mathrm{cm}^{-3}\,\mathrm{s}^{-1}\,\mathrm{arcsec}^{-2}
\end{equation}
by interpolation from the values in table 4.4 of \citet{Ost89}.
Absorption is not included because the dominant absorption is from neutral gas near the simulation boundaries and this suffers from edge effects associated with the non-orthogonal projections.  In any case the full simulation box is optically thin to H$\alpha$ radiation, so absorption is a small correction.

The images show some of the same features of previous axisymmetric models of hypersonic stars \citep{RagNorCanEA97}, notably the almost circular H\,\textsc{ii} region and the tail of overdense gas expanding from its lateral edges.
Here, because the Mach number of the flow is much lower, there is noticeable expansion of this overdense shell in a cone shape, and it forms at a smaller angle from the velocity vector of the star.
The $N_{\mathrm{HI}}$ contours show the overdense shell, and also the underdense H\,\textsc{ii} region and the wake behind it.
Underdensity in the H\,\textsc{ii} region is because the gas is ionized (and so does not show up in H\,\textsc{i}) but in the wake where the gas has recombined, the gas is underdense because the H\,\textsc{ii} region has displaced an outward-flowing expanding shell.

The left plot with zone-size $\Delta x=0.32\,$pc (r1) has relaxed to a steady state with a stable I-front, but this is only because of the low numerical resolution.
The right plot with $\Delta x=0.16\,$pc (r2) shows instability in the I-front, resulting in substructure in both H$\alpha$ emission and $N_{\mathrm{HI}}$.
Still higher-resolution 2D simulations in Appendix~\ref{sec:2D} show that the instability is only marginally resolved with resolution r2.
The unstable I-front generates knots of dense gas on small scales; for r2 there are a few of them present in the I-front at any time, and they move slowly away from the symmetry axis around to the sides of the H\,\textsc{ii} region where they join the expanding shell.
This instability was not found by \citet{RagNorCanEA97}, probably because our model has a much lower Mach number and the H\,\textsc{ii} region is in a higher-density environment where shocks are more dissipative.
The evolution of these knots is similar to that depicted in \citet{GarFra96} for the expansion of D-type I-fronts.
In our case the I-front is moving at a constant velocity that is (coincidentally) close to the limit between R-type and D-type (Mach 2 in ionized gas, \citealt{MihMih84}).
In regions where the I-front is expanding there is no upstream shock, but in overdense regions where it is receding, a radiative shock forms that enhances the density and mass of the overdense clumps.
The underdense regions that push neutral gas towards the overdense regions initially resemble the spear-shaped features in previous work \citep{GarFra96,FraGarKurEA07, WhaNor08b} but they soon saturate and develop a bubble shape.
The lifetime of the knots is much longer than the length of time a parcel of gas spends in the knot, so while they are formed from dynamical instability they seem to be quite stable structures.

The dense knots are seeds of strong photoevaporation flows that expand conically into the H\,\textsc{ii} region when they are near the symmetry axis.
The photoevaporation flow would expand spherically for a static ionizing source (cf.\ the Eagle Nebula pillars; \citealt{HesScoSanEA96}) but here the supersonic advection drags the photoevaporation flow downstream, resulting in a cone.
As the dense knots move off-axis to larger angles on the I-front, the photoevaporation flow is no longer symmetric because one side of the flow tries to expand upstream and fails, leading to the H$\alpha$ arcs seen in Fig.~\ref{fig:simHD3r12} that trail downstream from the bottom to the centre of the H\,\textsc{ii} region.
Features like this should be observable in H\,\textsc{ii} regions around moving stars if this I-front instability is present, although ISM inhomogeneities may drive similar photoevaporation flows.

The magnitude of the overdensity and underdensity in neutral H (compared to the mean) is $N_{\mathrm{HI}}\approx3-4\times10^{20}\,\ensuremath{\mathrm{cm}^{-2}}$ and $N_{\mathrm{HI}}\approx-2\times10^{20}\,\ensuremath{\mathrm{cm}^{-2}}$.
The undisturbed background gas density provides $N_{\mathrm{HI}}\approx1.8\times10^{20}\,\ensuremath{\mathrm{cm}^{-2}}$ through the 24 pc diameter of the H\,\textsc{ii} region, so the contrast is only a factor of 2$-$3.
This may be difficult to measure observationally, especially in the Galactic plane where there is strong H\,\textsc{i} background emission.
It should, however, have similar radial velocity to the mean H$\alpha$ emission, so this could help to identify the conical shell.

The upstream I-front is much sharper than the downstream recombination front because of their different character; the recombination length $\ell_r \approx v_\star t_r$ (recall $t_r=(\alpha_b n_e)^{-1}$ is the recombination time) sets the recombination front thickness; the I-front is very thin in comparison.
This is an obvious feature of the H$\alpha$ emissivity plots, and can also be seen in the H$\alpha$ map of $\zeta$ Oph's H\,\textsc{ii} region \citep[][see also Section~\ref{sec:discussion}]{GvaLanMac12}.
In addition the H$\alpha$ emission in our simulations has a clear gradient, being brighter upstream from the star than downstream.
This is because the H\,\textsc{ii} region expands as gas flows downstream, so the recombination rate decreases and the emissivity along with it.
This is just as clear in the HD3r2 plot (see right panel of Fig.~\ref{fig:simHD3r12}) even though the gas is more disturbed.

Equivalent results are shown in Fig.~\ref{fig:simBA3r12} for the MHD simulations with a magnetic field aligned with the stellar velocity, BA3r1 and BA3r2.
In this case the higher resolution model BA3r2 also has a stable I-front and there is no significant difference between the two models.
It has been found for H\,\textsc{ii} regions around static stars \citep{KruStoGar07,ArtHenMelEA11,MacLim11} that D-type I-front expansion along field lines is largely stable, whereas expansion perpendicular to field lines is unstable, producing `beads' of dense gas tied to field lines \citep[see also the 2D studies of][]{Wil02,Wil07}.
Similar results are seen here for R-type I-fronts, although we note that 2D simulations at higher resolution do become unstable even when velocity and magnetic field vectors are aligned (see Appendix~\ref{sec:2D}).
This agrees with the finding \citep{NewAxf67, Wil99} that R-type I-fronts should be generically unstable to the shadowing instability for finite density perturbations, suggesting that the stability of BA3r1 and BA3r2 is at least partly a resolution effect.

Another feature of simulations BA3r1 and BA3r2 is that the expanding neutral shell is much less dense than in the hydrodynamic models, because magnetic pressure resists compression and expansion perpendicular to field lines.
The fast magnetosonic speed is larger than the sound speed so the effective Mach number of the flow is lower, hence the outward-moving shock makes a larger angle with the direction of gas flow.
The fast magnetosonic shock is also less compressive than its hydrodynamical equivalent.
Both of these effects mean that the shell is more diffuse and its column density is lower.

Fig.~\ref{fig:simBT3r2} shows projections through model BT3r2, this time at $t=4.5\,$Myr and from two different projection angles.
Both projections are perpendicular to the bulk gas flow as before, but the left plot is projected such that the background magnetic field is along the LOS whereas in the right plot it is vertical and in the image plane.
The left plot is very similar to the results in Fig.~\ref{fig:simBA3r12} for BA3r2, in that the shell column density is low, the angle it makes with the flow velocity vector is quite large, and the upstream I-front is more stable than the hydrodynamic model HD3r2.
This relates directly to the magnetic field orientation, which in both cases inhibits gas flow in the image plane.
The right-hand plot is much more similar to the hydrodynamic model HD3r2, in that there is instability in the {\changed I-front} and a strongly-compressed shell downstream.

The kink in the upstream H$\alpha$ emission in the left-hand plot seems to cross most of the H\,\textsc{ii} region, and its counterpart in the right-hand image shows the kink end-on as a v-shape notch in the upstream I-front.
It appears that the 3D I-front instability seen in model HD3r2 has reverted to an almost 2D instability in BT3r2, because of the large-scale ordered magnetic field.
It is not clear why the ripple in the I-front would form in this way since its connectivity is across the field lines, not along them.
Gas can only flow freely along field lines, so these unstable clumps must preferentially form in a 2D fashion, but it is not clear why they should be connected across the full length of the I-front.
Higher resolution simulations are required to investigate if this feature is numerical or physical.

\begin{figure*}
\centering
\includegraphics[width=0.49\textwidth]{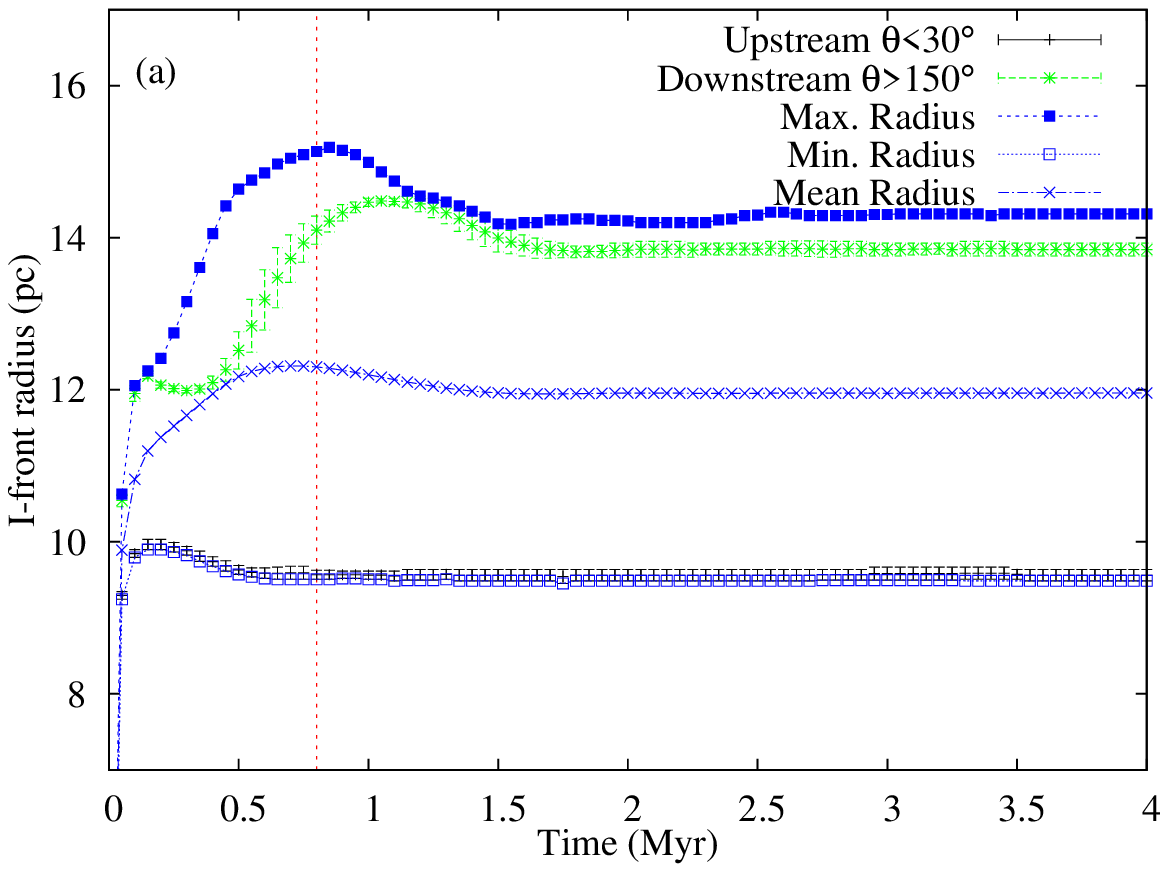}
\includegraphics[width=0.49\textwidth]{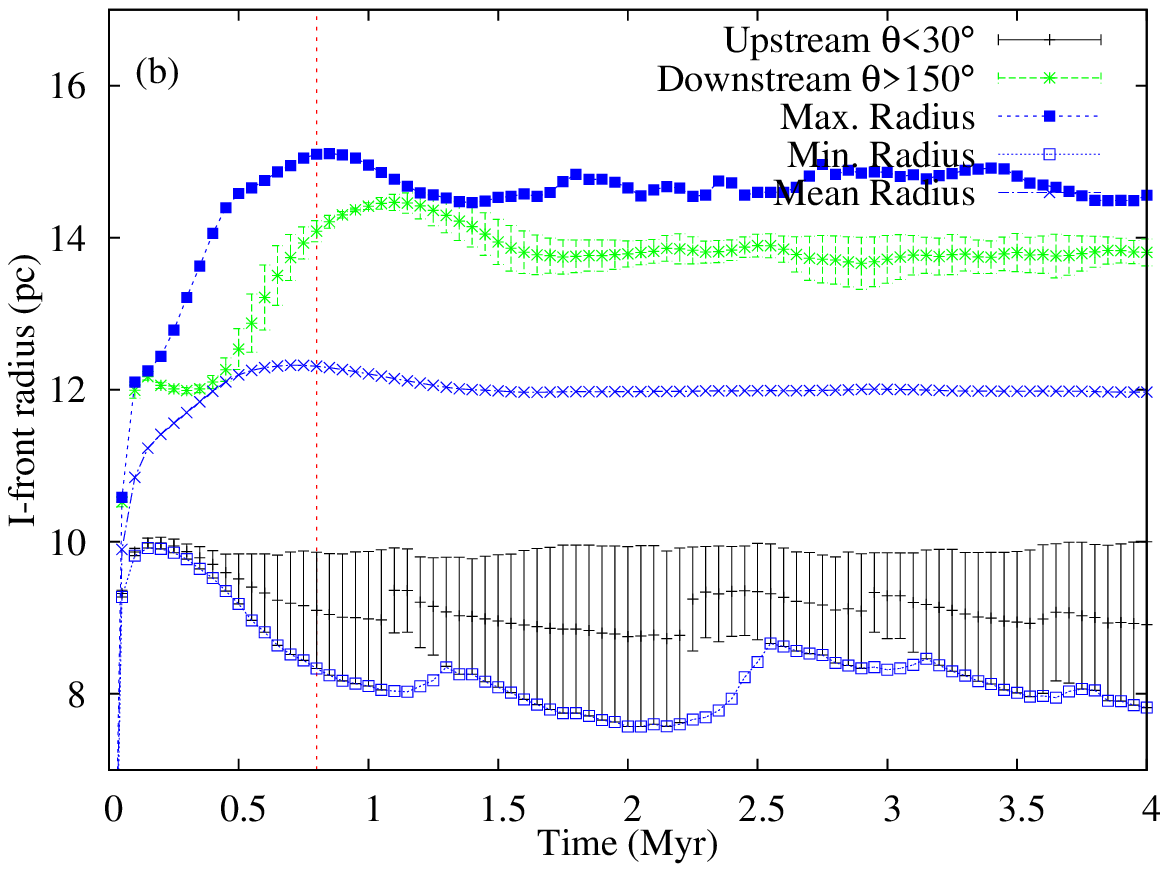} \\
\includegraphics[width=0.49\textwidth]{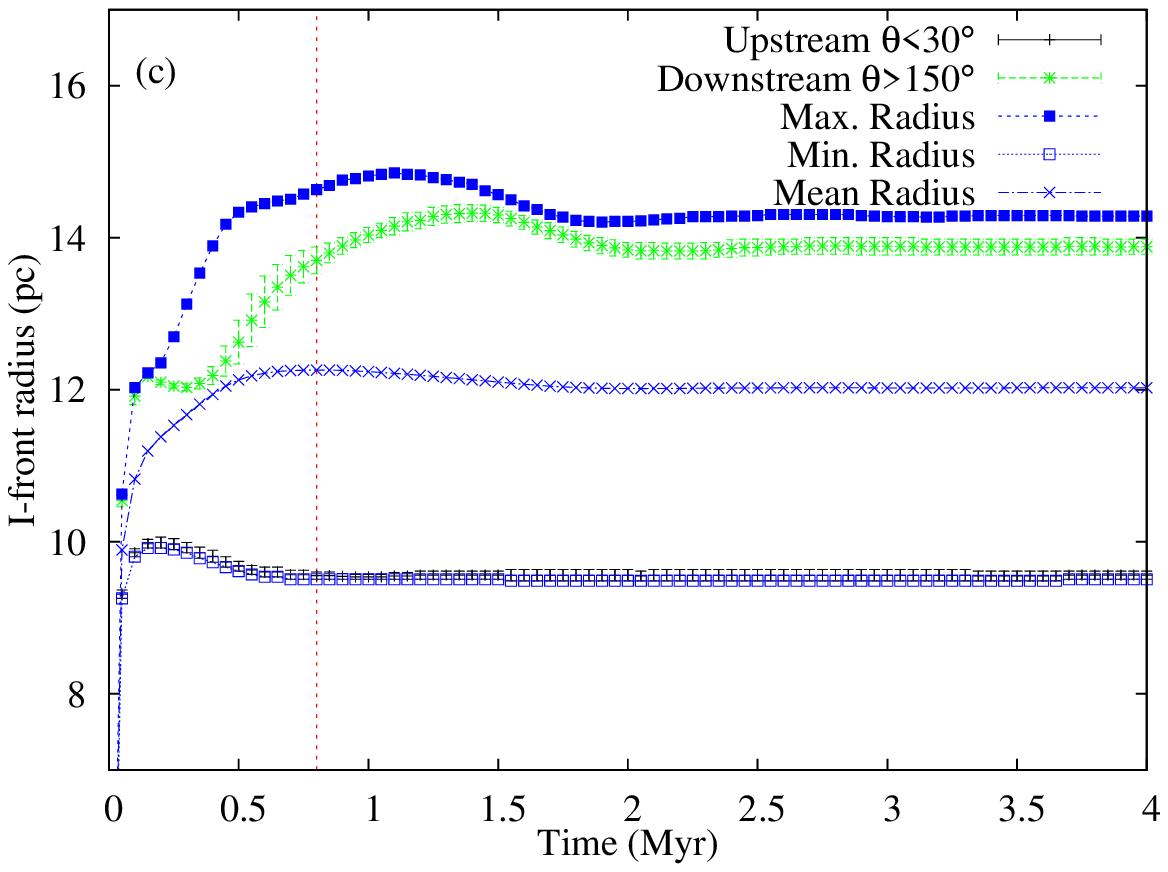}
\includegraphics[width=0.49\textwidth]{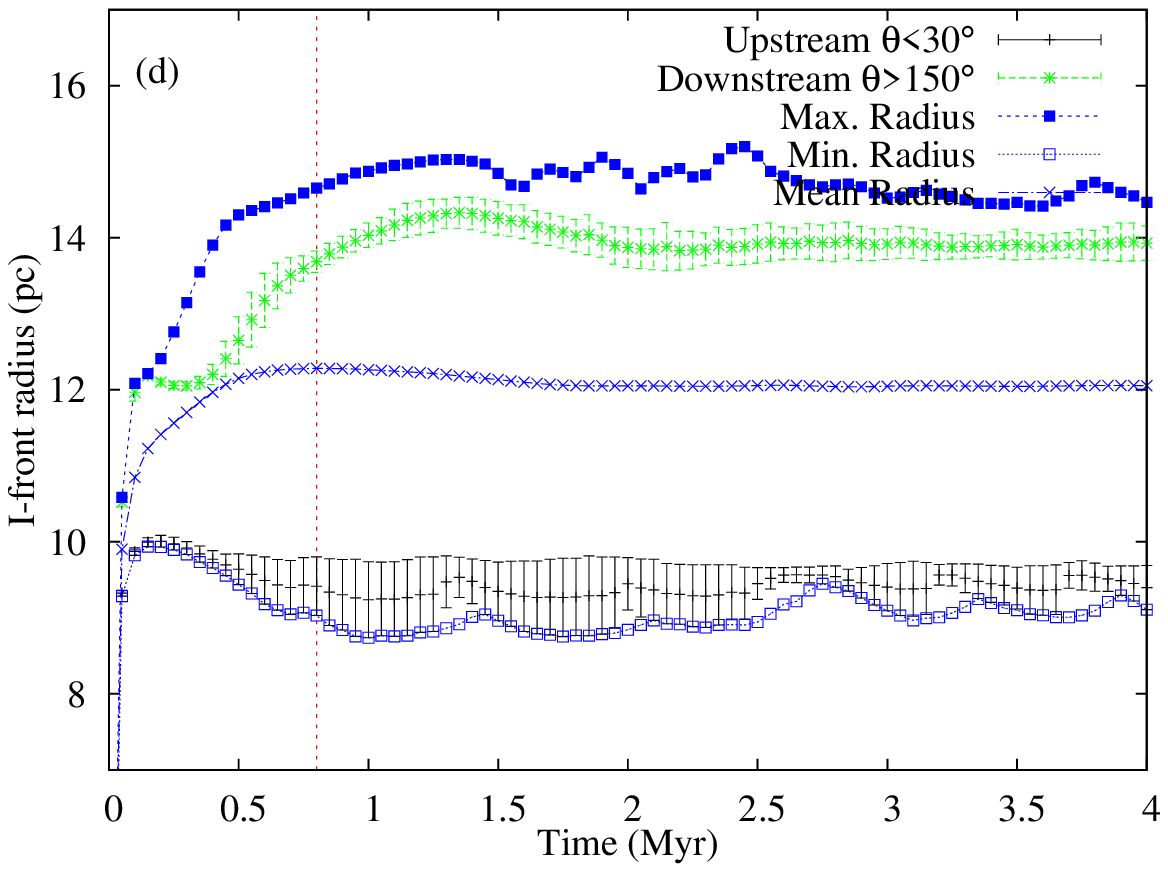} \\
\includegraphics[width=0.49\textwidth]{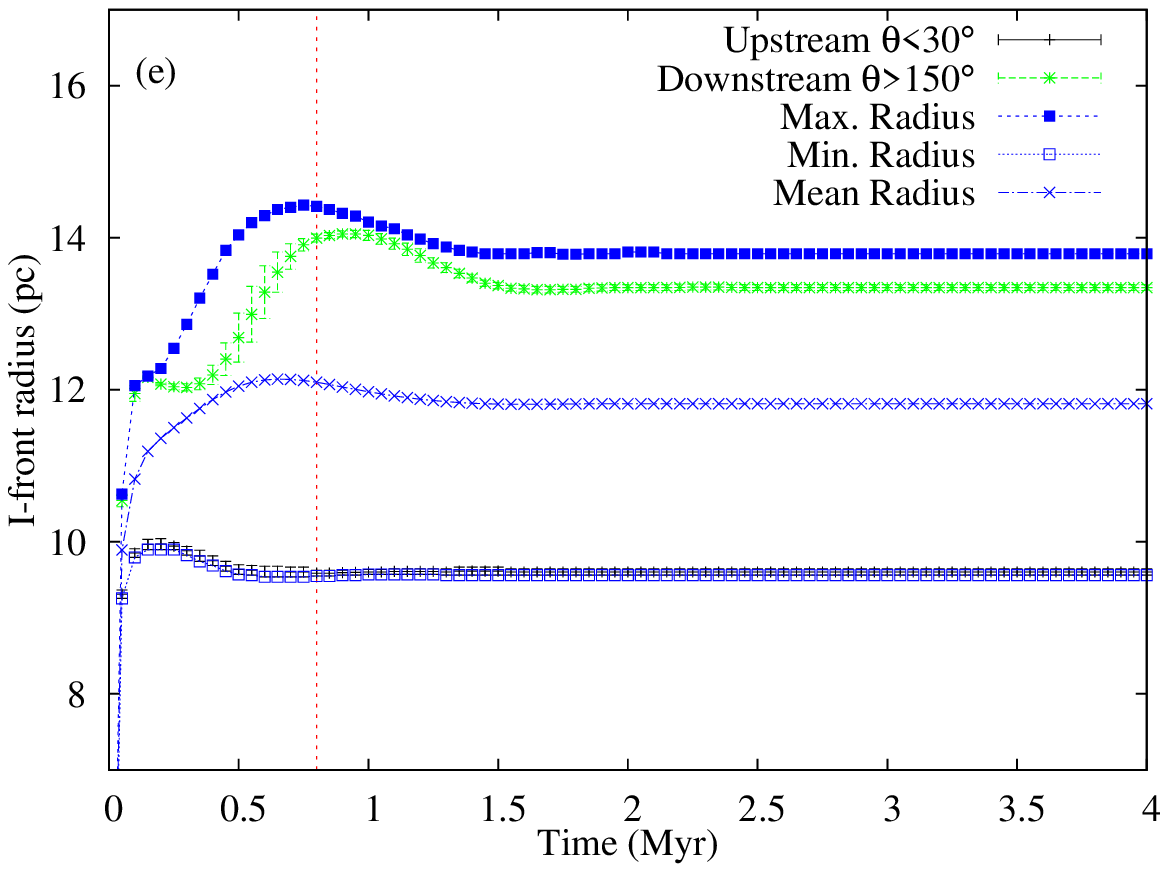}
\includegraphics[width=0.49\textwidth]{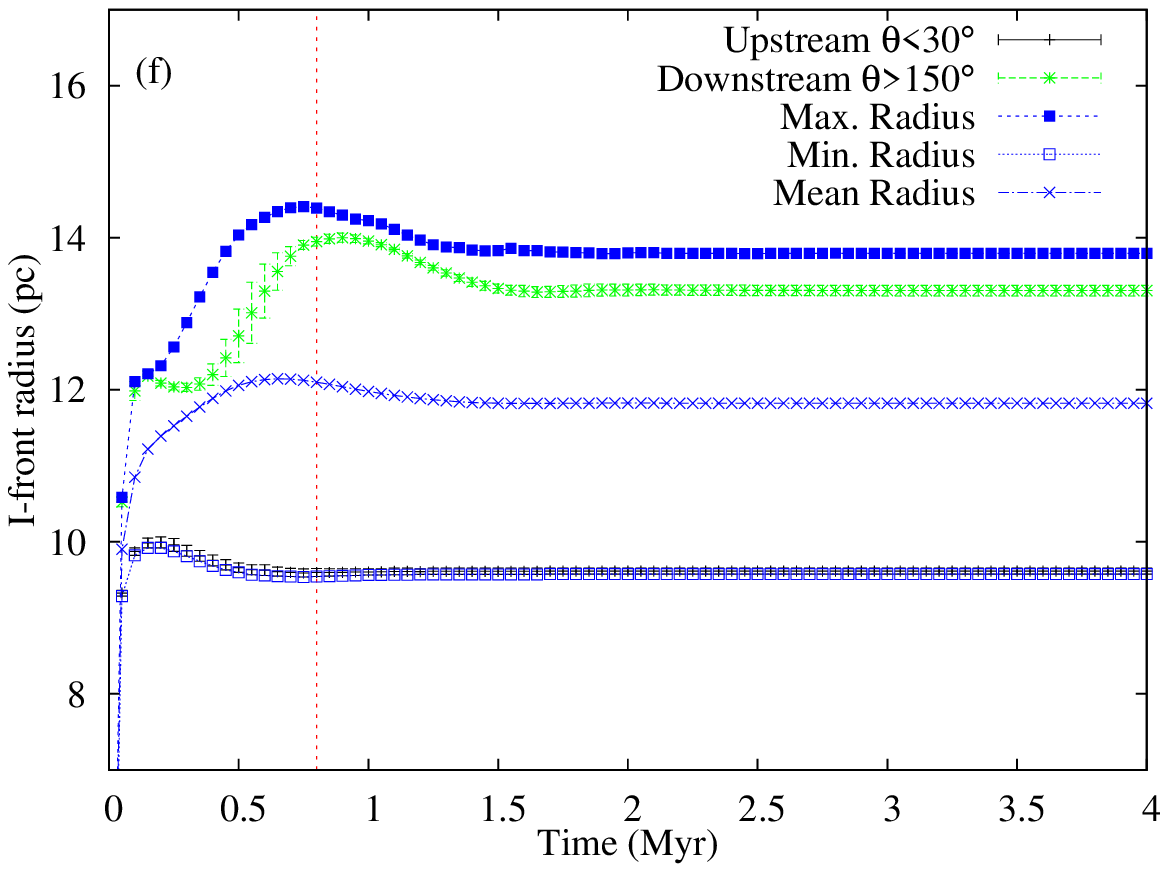} \\
\caption{
  Size of the H\,\textsc{ii} region (measured by H$^+$ fraction $1-y_n>0.5$) for the 3D simulations HD3r1 (a), HD3r2 (b), BT3r1 (c), BT3r2 (d), BA3r1 (e), and BA3r2 (f).
  The black points with error bars show the maximum and minimum radius within 30 degrees of the star's velocity vector in the upstream direction, and the green points show the same in the downstream direction.
  In both cases the midpoint is indicated (rather than the mean).
  The solid blue squares indicate the maximum I-front radius over its full surface, and the empty blue squares the minimum radius, which generally agrees with the minimum upstream radius.
  The blue crosses show the mean H\,\textsc{ii} region radius, obtained by spherically averaging the total volume where $y_n<0.5$.
  \label{fig:HIIregionShape}
  }
\end{figure*}

%%%%%%%%%%%%%%%%%%%%%%%%%%%%%%%%%%%%%%%%%%%%%%%%%%%%%%%%%%%%%%%%%%%%%
\subsection{Shape of H\,\textsc{ii} region}
%%%%%%%%%%%%%%%%%%%%%%%%%%%%%%%%%%%%%%%%%%%%%%%%%%%%%%%%%%%%%%%%%%%%%
Plots of the H\,\textsc{ii} region radius are shown for the different 3D simulations in Fig.~\ref{fig:HIIregionShape}.
The range of values for the upstream and downstream I-front radius are plotted with error bars, the maximum and minimum I-front radius over its full surface is shown with the blue points, and the mean I-front radius is plotted as the blue crosses.
The upstream I-front radius is about 9.5\,pc in all models (compared to 9.83\,pc in the 1D simulations), whereas the Str\"omgren radius is $R_s\approx10.6\,$pc\footnote{A temperature-dependent recombination rate is used, and the H\,\textsc{ii} region is not isothermal, but for a mean temperature $T\approx6300\,$K we find $R_s\approx 10.6\,$pc.}.
The downstream radius is not well-defined because the recombination length is significant, so $y_n=0.5$ is chosen to define the radius.
In this case it is about 13.5$-$14\,pc, so the diameter is 23$-$23.5\,pc, or 8.5$-$11 per cent larger than $2R_s$.
The mean radius calculated from the spherically-averaged ionized volume is about 12 pc in all models ($1.13R_s$).
This larger mean radius is expected because the mean density within the H\,\textsc{ii} region is less than the ISM density.
The maximum radius is always in the range of directions $120^\circ<\theta<150^\circ$ (where $\theta$ measures angles from the star's velocity vector, with the star at the origin), implying that the H\,\textsc{ii} region is flattened in the downstream direction.

Simulations BA3r1 and BA3r2 are basically the same because both are stable and in a steady state at late times.
BT3r2 and HD3r2 have I-front instability, so here the upstream I-front has a range of radii: about $8-10$\,pc for HD3r2 and $8.8-9.8$\,pc for BT3r2.
The maximum and minimum radius also fluctuate because of this instability.

Despite the clear asymmetry in the H\,\textsc{ii} region radius relative to the star, it is much more spherical in H$\alpha$ emission with respect to its own geometric centre.
It it also noteworthy that the star is located downstream of the peak H$\alpha$ intensity, but upstream of the H\,\textsc{ii} region centre.

\begin{figure}
\centering
\includegraphics[width=0.49\textwidth]{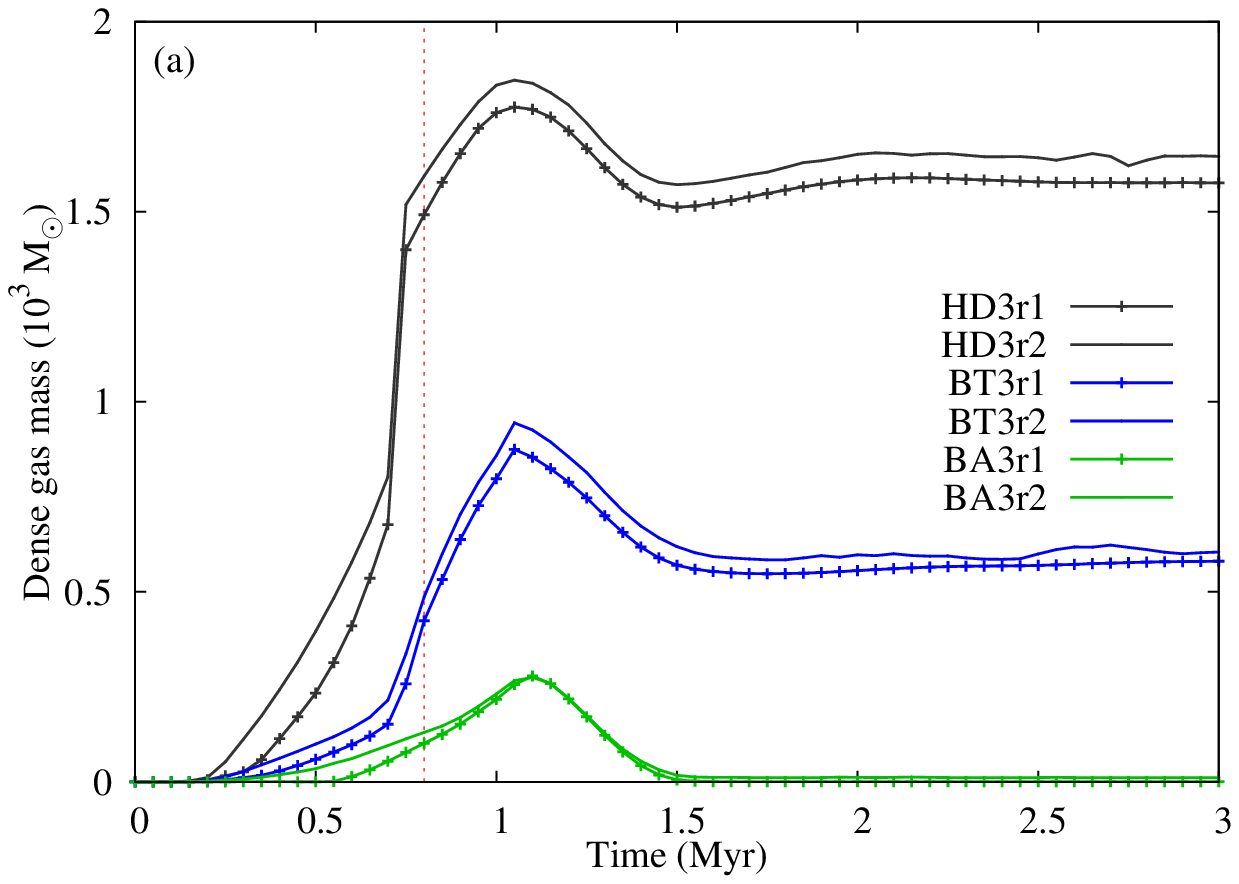}
\includegraphics[width=0.49\textwidth]{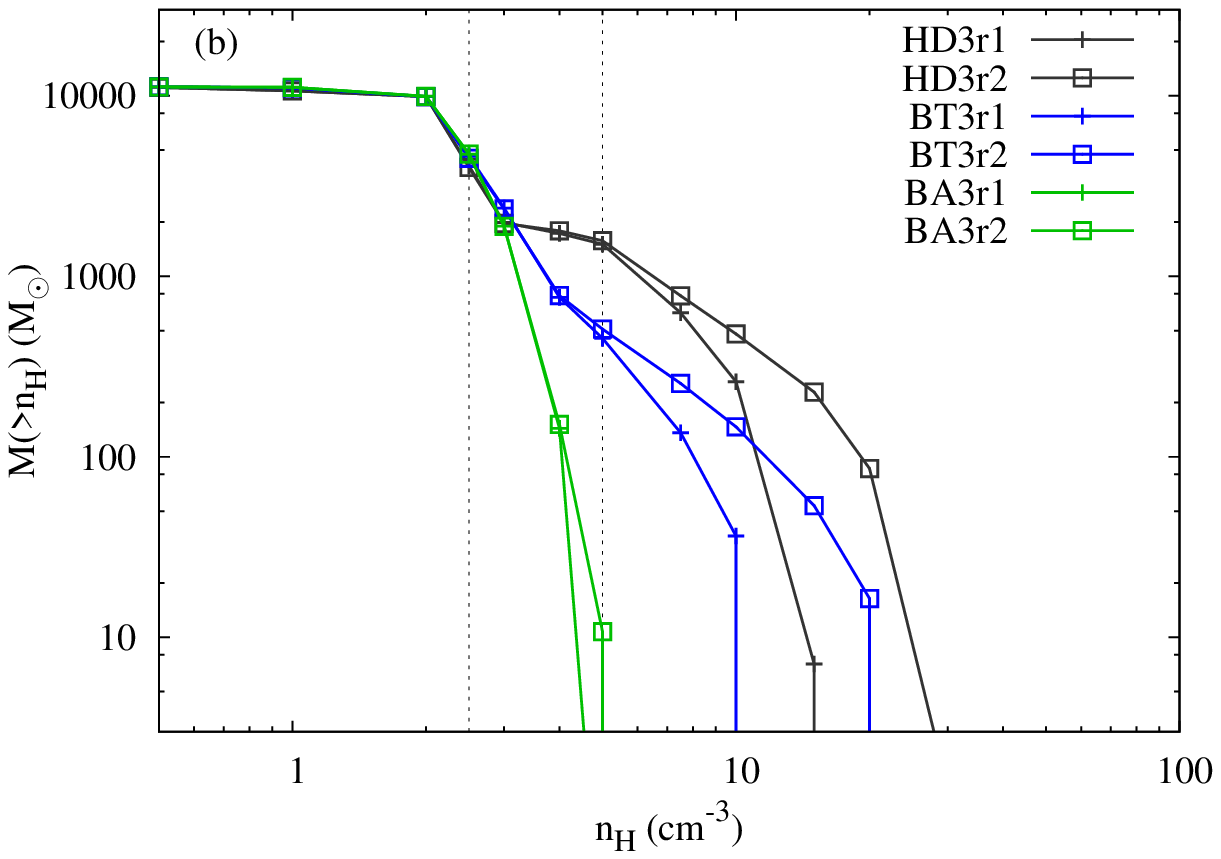}
\caption{
  (a) Time evolution of the mass of gas on the computational domain with $\rho>2\rho_0$ for the 3D simulations, where $\rho_0$ is the background gas density.
  The vertical line at $t=0.8\,$Myr is the point where disturbed gas begins to leave the domain's downstream boundaries.
  (b) Hydrogen number density distribution of the ISM at $t=4\,$Myr in the 3D simulations.
  The vertical lines correspond to $\rho_0$ and $2\rho_0$.
  \label{fig:Density3D}
  }
\end{figure}

%%%%%%%%%%%%%%%%%%%%%%%%%%%%%%%%%%%%%%%%%%%%%%%%%%%%%%%%%%%%%%%%%%%%%
\subsection{H\,\textsc{ii} region feedback on the ISM}
%%%%%%%%%%%%%%%%%%%%%%%%%%%%%%%%%%%%%%%%%%%%%%%%%%%%%%%%%%%%%%%%%%%%%
Figs.~\ref{fig:simHD3r12}-\ref{fig:simBT3r2} show that the H\,\textsc{ii} region has a measurable impact on the ISM that persists once the star has passed.
The quantity of compressed gas as a function of time and the density distribution of gas after 4 Myr are plotted in Fig.~\ref{fig:Density3D} for each of the 3D simulations.
Note that only gas still on the simulation domain is counted, so this measures the mass of dense gas generated in the time from when the gas is compressed to when it leaves the simulation domain ($\approx0.8-1\,$Myr).
The density distribution is shown in terms of gas mass with a density greater than $\ensuremath{n_{\textsc{h}}}$.
After about 1.5\,Myr all models approach a stationary state.

Simulations BA3r1 and BA3r2 have the weakest density enhancement with almost no gas compressed to $\rho>2\rho_0$.
This is because all of the pressure gradient is across field lines, so magnetic pressure and tension resist gas compression.
BT3r1 and BT3r2 have more dense gas because field lines only resist compression in one of the two directions perpendicular to motion, and HD3r1 and HD3r2 have the densest gas because there is no magnetic field to resist gas flows.
In HD3r2 there is $100\,\msun$ of gas that has been compressed by about a factor of 10, and 2D results show that at higher resolution even stronger compression is obtained (Appendix~\ref{sec:2D}).
The lower panel of Fig.~\ref{fig:Density3D} demonstrates that the simulations have converged for weak compression, but that the mass of strongly-compressed gas is very resolution-dependent, as would be expected.
These results suggest that the passage of massive stars through the ISM could compress WNM gas sufficiently for it to make the transition to cold neutral medium (CNM).
They also demonstrate the importance of the ISM magnetic field in determining the compression ratio in these weak radiative shocks.
Fig.~\ref{fig:Density3D}(b) also shows that more than half of the gas in the simulation is underdense compared to the undisturbed upstream gas; the star creates an underdense cylindrical volume in the ISM with cross-section of about 10\,pc radius.

Despite these results, it is not obvious that the H\,\textsc{ii} region feedback effects are significant compared to those expected from stellar winds.
The main stellar feedback mechanism in terms of energetics is photoheating, but this energy is quickly dissipated once the star has passed and the gas has recombined.
For static stars the conversion efficiency of ionizing photon luminosity to kinetic energy was found to be $\lesssim 0.1$ per cent \citep[][]{FreHenYor03} (although they included the ionization energy of H as well as the excess heating energy in this estimate), whereas for stellar winds $\approx 10$ per cent of the input kinetic energy survives dissipative processes and drives motion in the ISM \citep{WeaMcCCasEA77, GarMacLan96, KraFieDieEA13}.

\begin{figure}
\centering
\includegraphics[width=0.49\textwidth]{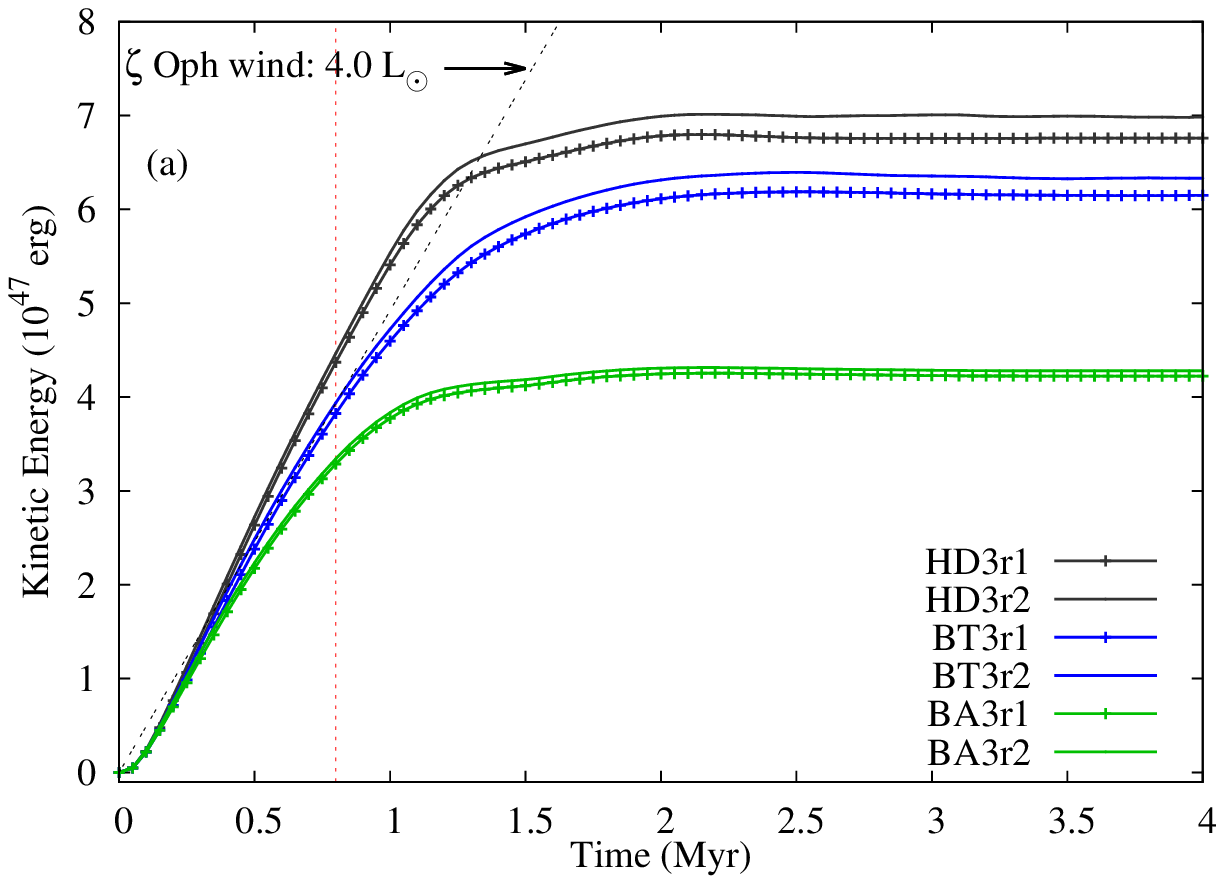}
\includegraphics[width=0.49\textwidth]{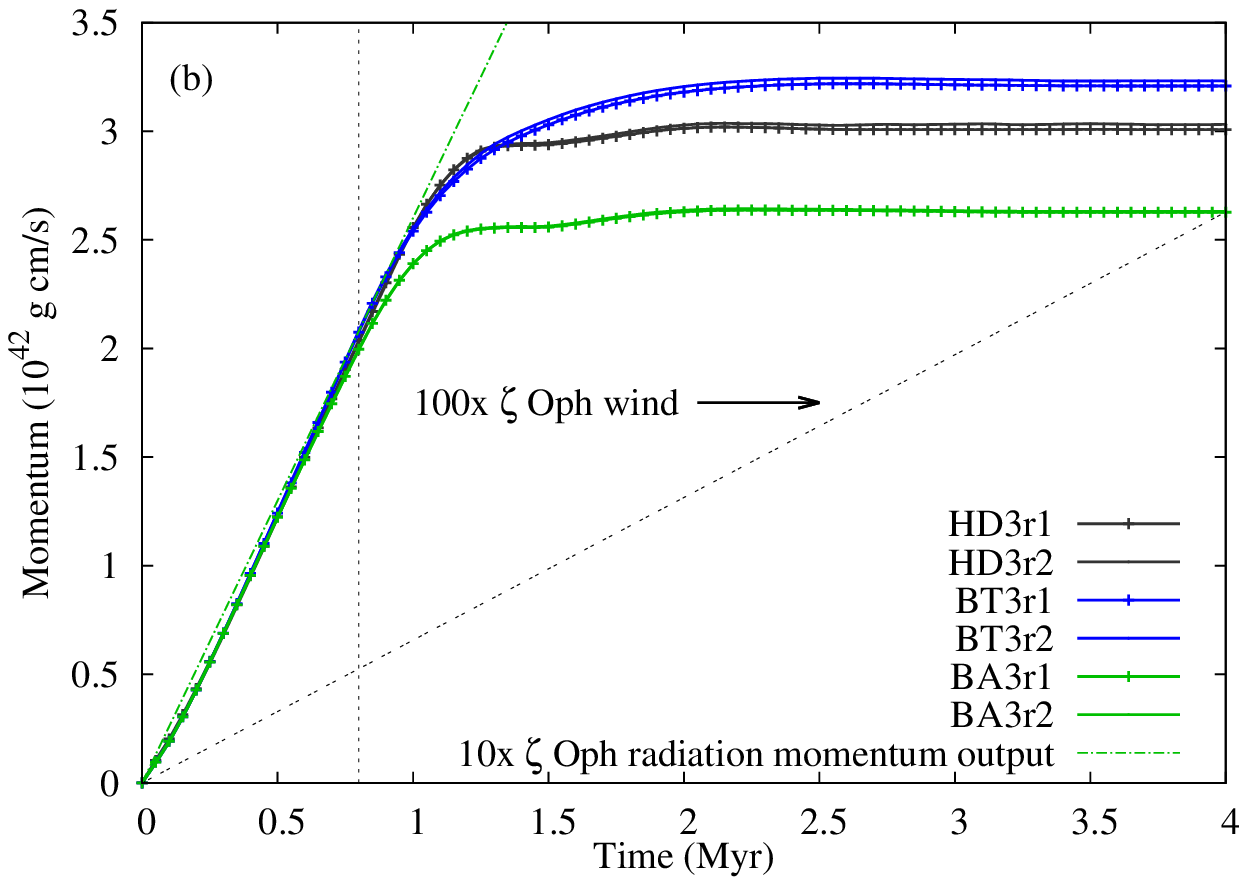}
\caption{
  Kinetic energy (a) and momentum (b) of gas in 3D simulations as a function of time, in the ISM rest frame.
  During the initial phase ($t\lesssim0.8$ Myr, to the left of the vertical dotted line) no disturbed gas leaves the domain and the kinetic energy and momentum increase monotonically.
  At later times a stationary state is reached where the energy input is balanced by gas leaving the domain.
  For comparison, the mechanical luminosity ($4\,\lsun$) of the wind from the nearby massive star $\zeta$ Oph is shown in (a), and its wind momentum input in (b) (multiplied by 100 so it can be distinguished from the $x$-axis) together with the total radiation momentum output from the star's radiation (green dot-dashed line, multiplied by 10 to enable comparison with the ISM momentum).
  \label{fig:KEMom3D}
  }
\end{figure}

%%%%%%%%%%%%%%%%%%%%%%%%%%%%%%%%%%%%%%%%%%%%%%%%%%%%%%%%%%%%%%%%%%%%%
\subsubsection{Kinetic energy feedback}
%%%%%%%%%%%%%%%%%%%%%%%%%%%%%%%%%%%%%%%%%%%%%%%%%%%%%%%%%%%%%%%%%%%%%
The only persistent feedback from the H\,\textsc{ii} region is the part of the photoheating energy that goes into driving kinetic energy in the surroundings, from H\,\textsc{ii} region expansion and I-front instability.
To measure this we calculate the total gas momentum and kinetic energy on the simulation domain as a function of time, with the bulk flow subtracted off, plotted in Fig.~\ref{fig:KEMom3D}.
The slope of this function gives the feedback rate at early times (before disturbed gas leaves the downstream boundary), and the stationary state value at late times divided by the grid-crossing time gives another approximate feedback rate.
The kinetic energy increases somewhat with spatial resolution because of waves/shocks generated by I-front instability, which is more vigorous at higher resolution.
A stronger trend is that the models with aligned magnetic field have lower kinetic energy than those with a transverse field, which in turn lie below the zero field (hydrodynamic) models.
This is because the tension of the aligned magnetic field resists H\,\textsc{ii} region dynamical expansion in all directions perpendicular to the direction of motion.
The perpendicular field models allow expansion along field lines (i.e.~in one of the two perpendicular directions), and hydrodynamic models have no magnetic tension to inhibit expansion.

As a comparison, the kinetic energy plot also shows the mechanical luminosity of the stellar wind driven by $\zeta$ Oph, $L_w = 0.5\dot{M}v_w^2$, using a mass-loss rate of $\dot{M} \approx 2.2\times10^{-8} \,\msun\,\mathrm{yr}^{-1}$ and wind velocity of $v_w\approx1500\,\ensuremath{\mathrm{km}\,\mathrm{s}^{-1}}$ \citep{GvaLanMac12}.
With the stellar luminosity and spectrum used here, the mean energy per ionizing photon is 17.2\,eV, so the mean heating per ionization is 3.6\,eV.
If we consider the kinetic energy of the ISM as a percentage of this heating input power, we obtain an efficiency of conversion from photoheating to kinetic-energy of $\epsilon\approx0.5-1$ per cent.
Even with this low efficiency the kinetic energy gained from the H\,\textsc{ii} region is comparable to that from the stellar wind, and in reality is probably larger because much of the wind kinetic energy gets lost through dissipative processes.

%%%%%%%%%%%%%%%%%%%%%%%%%%%%%%%%%%%%%%%%%%%%%%%%%%%%%%%%%%%%%%%%%%%%%
\subsubsection{Momentum feedback}
%%%%%%%%%%%%%%%%%%%%%%%%%%%%%%%%%%%%%%%%%%%%%%%%%%%%%%%%%%%%%%%%%%%%%
Perhaps a truer measure of the feedback is the rate of momentum input to the ISM, because momentum is conserved regardless of the dissipative properties of the gas.
Here the resolution dependence is not apparent and the differences between the models are also less pronounced, but a similar dependence on magnetic field orientation is seen.
For comparison, we also show the momentum input we would obtain from the wind of $\zeta$ Oph, with parameters given above.
Its momentum input rate is multiplied by 100 to show it on the same scale, so evidently it is insignificant compared to the H\,\textsc{ii} region's dynamical expansion.
The rate of momentum input from H\,\textsc{ii} region expansion (measured as the slope of the line for $t<0.8\,$Myr) is also about 10 times larger than the total momentum output rate of the radiation from $\zeta$ Oph, $(L/c)t$, plotted as the green dot-dashed line in Fig.~\ref{fig:KEMom3D} for a luminosity of $L=6.4\times10^4\,\lsun$ \citep[][scaled to a distance of 112\,pc]{HowSmi01}.
Each photon would therefore need to have $\geq10$ scatterings before escaping the H\,\textsc{ii} region in order to significantly affect the dynamics presented here.
This is unlikely given the low ISM density.

%%%%%%%%%%%%%%%%%%%%%%%%%%%%%%%%%%%%%%%%%%%%%%%%%%%%%%%%%%%%%%%%%%%%%
\subsubsection{Comparison to supernova feedback}
%%%%%%%%%%%%%%%%%%%%%%%%%%%%%%%%%%%%%%%%%%%%%%%%%%%%%%%%%%%%%%%%%%%%%
By both measures (kinetic energy and momentum) the H\,\textsc{ii} region expansion has stronger feedback than the stellar wind for an O9.5\,V star, even though it is moving supersonically through the ISM.
This feedback is not very violent, and may not be as easily observed as a stellar wind bow shock, but it is more important in terms of energetics.
Compared to the kinetic energy of the eventual supernova explosion of $\zeta$ Oph, $E_{\mathrm{sn}}\approx10^{51}\,$erg, both feedback effects are not very important.
If we take the main sequence lifetime of $\zeta$ Oph to be $\tau_{\mathrm{ms}}\approx7\,$Myr then the total kinetic energy imparted to the ISM from the H\,\textsc{ii} region is $\approx3.8\times10^{-3} E_{\mathrm{sn}}$.
The momentum input to the ISM is, however, similar to that of a supernova, at $8.8\times10^4\,\msun\,\ensuremath{\mathrm{km}\,\mathrm{s}^{-1}}$.
If we take the supernova to have $4\,\msun$ of ejecta with velocity $5000\,\ensuremath{\mathrm{km}\,\mathrm{s}^{-1}}$ (giving a kinetic energy of $10^{51}\,$erg), then we see that the H\,\textsc{ii} region imparts about $4\times$ as much momentum to the ISM during the star's main sequence lifetime.

The large momentum generation of the H\,\textsc{ii} region can be understood simply with the following argument.
The mass flux through the upstream I-front is given by
\begin{equation}
F_{\mathrm{m}} = \pi R_{\textsc{hii}}^2 \rho_0 v_\star = 1060 \,\msun\,\mathrm{Myr}^{-1} \;,
\end{equation}
where $R_{\textsc{hii}}=12$\,pc has been used as the mean radius of the H\,\textsc{ii} region.
The maximum gas velocity (in the ISM rest frame) for the simulations is typically $v=10-13\,\ensuremath{\mathrm{km}\,\mathrm{s}^{-1}}$, so over the star's main sequence lifetime this gives a total gas momentum of about $F_{\mathrm{m}}v\tau_{\mathrm{ms}}=7.4-9.6\times10^{4}\msun\,\ensuremath{\mathrm{km}\,\mathrm{s}^{-1}}$.

The calculations here have only considered a single stellar mass, ISM density, and stellar space velocity.
It is not trivial to estimate how the relative importance of the different feedback processes scale with all of these parameters, particularly stellar mass, but it should be explored in future work.

%%%%%%%%%%%%%%%%%%%%%%%%%%%%%%%%%%%%%%%%%%%%%%%%%%%%%%%%%%%%%%%%%%%%%
\subsection{ISM near the star}
%%%%%%%%%%%%%%%%%%%%%%%%%%%%%%%%%%%%%%%%%%%%%%%%%%%%%%%%%%%%%%%%%%%%%

\begin{figure*}
\centering
\includegraphics[width=0.49\textwidth]{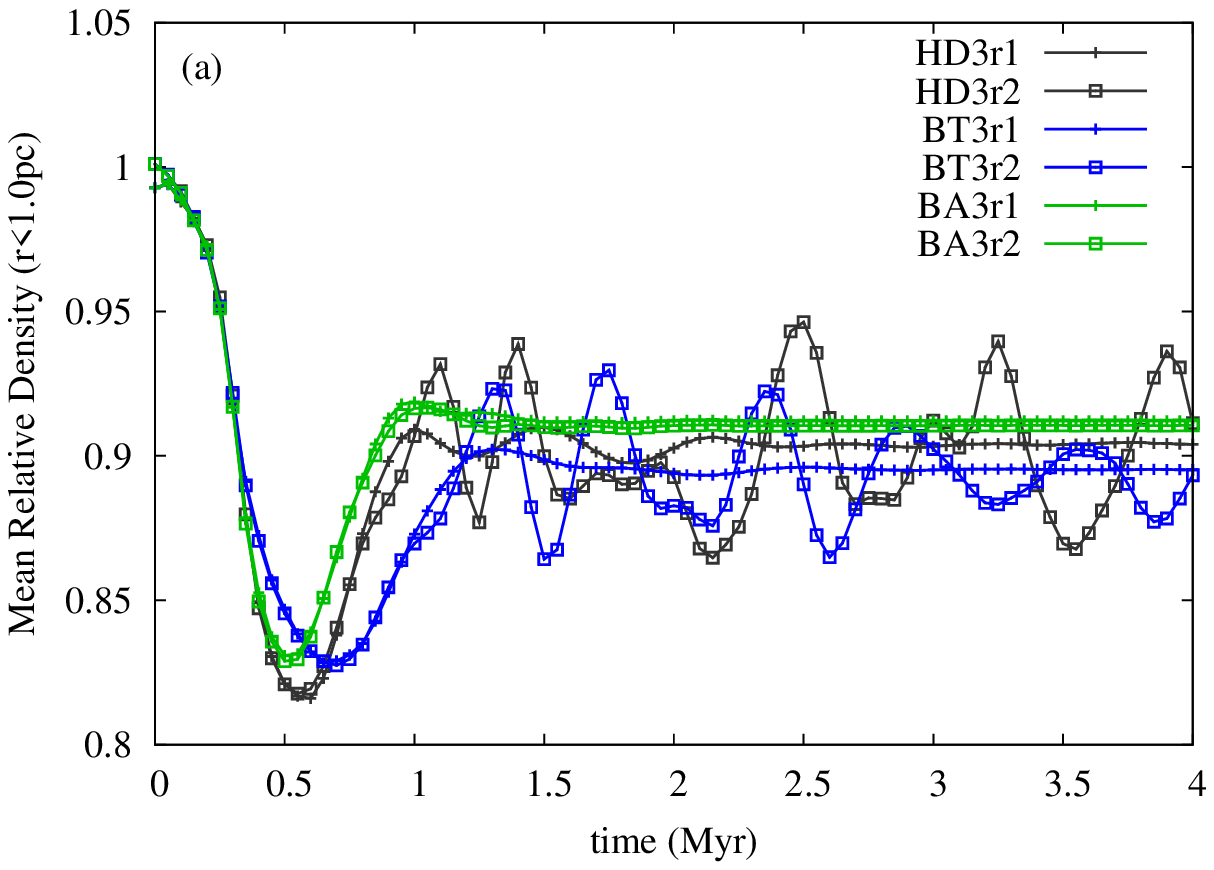}
\includegraphics[width=0.49\textwidth]{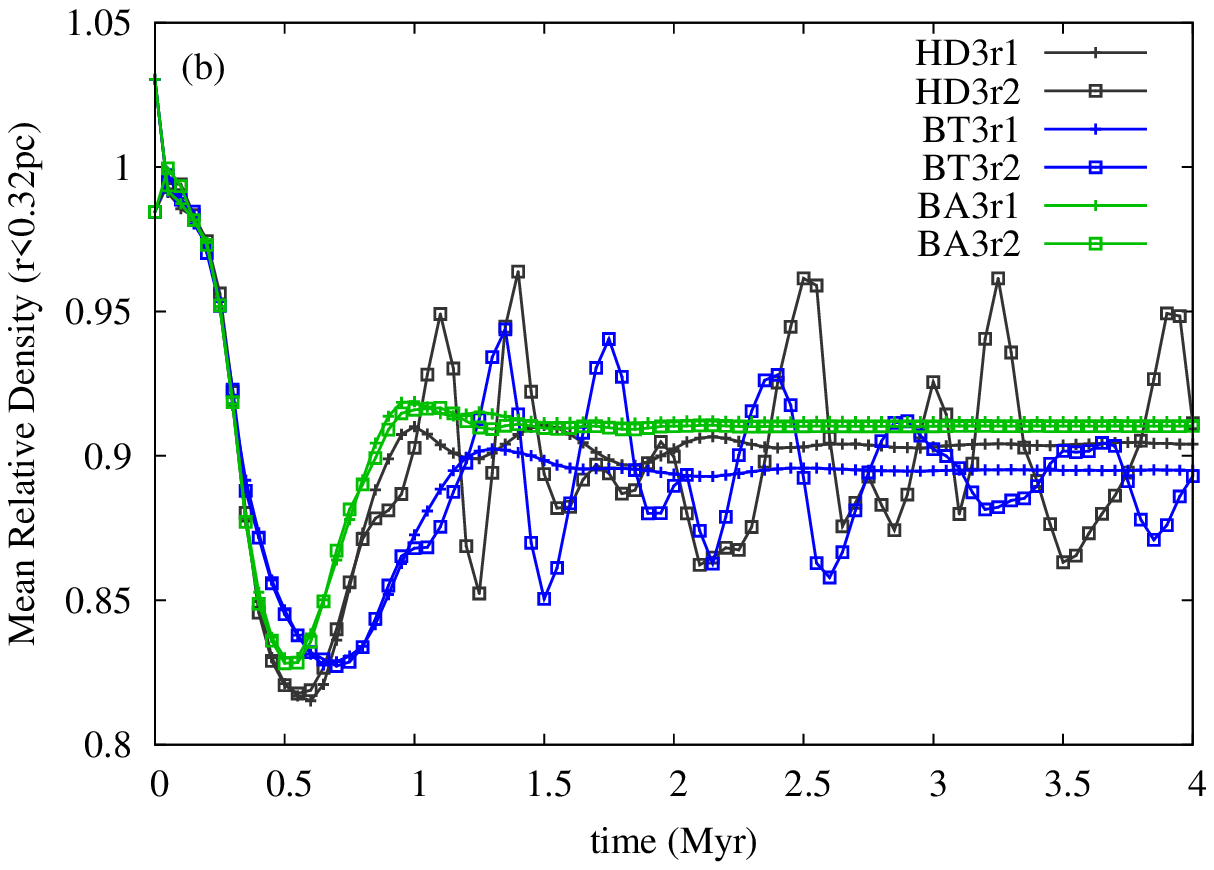} \\
\includegraphics[width=0.49\textwidth]{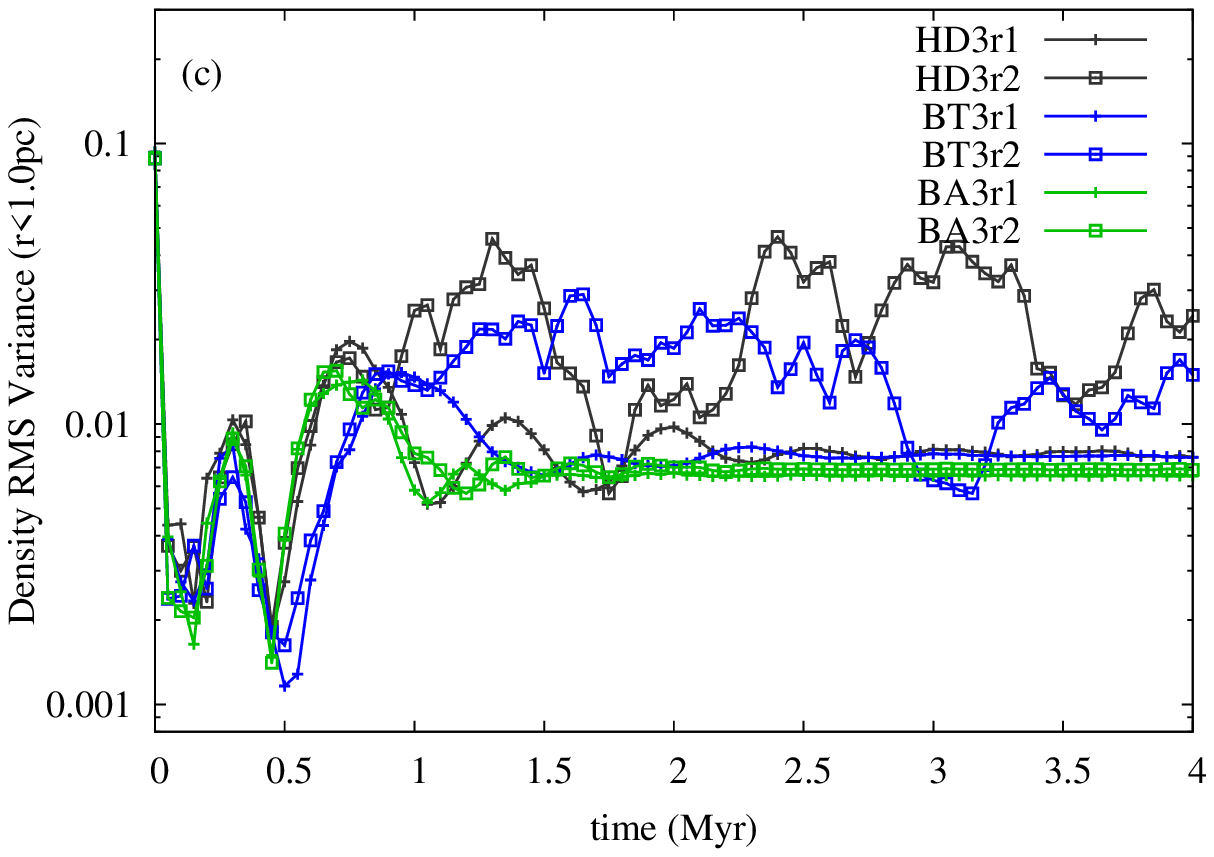}
\includegraphics[width=0.49\textwidth]{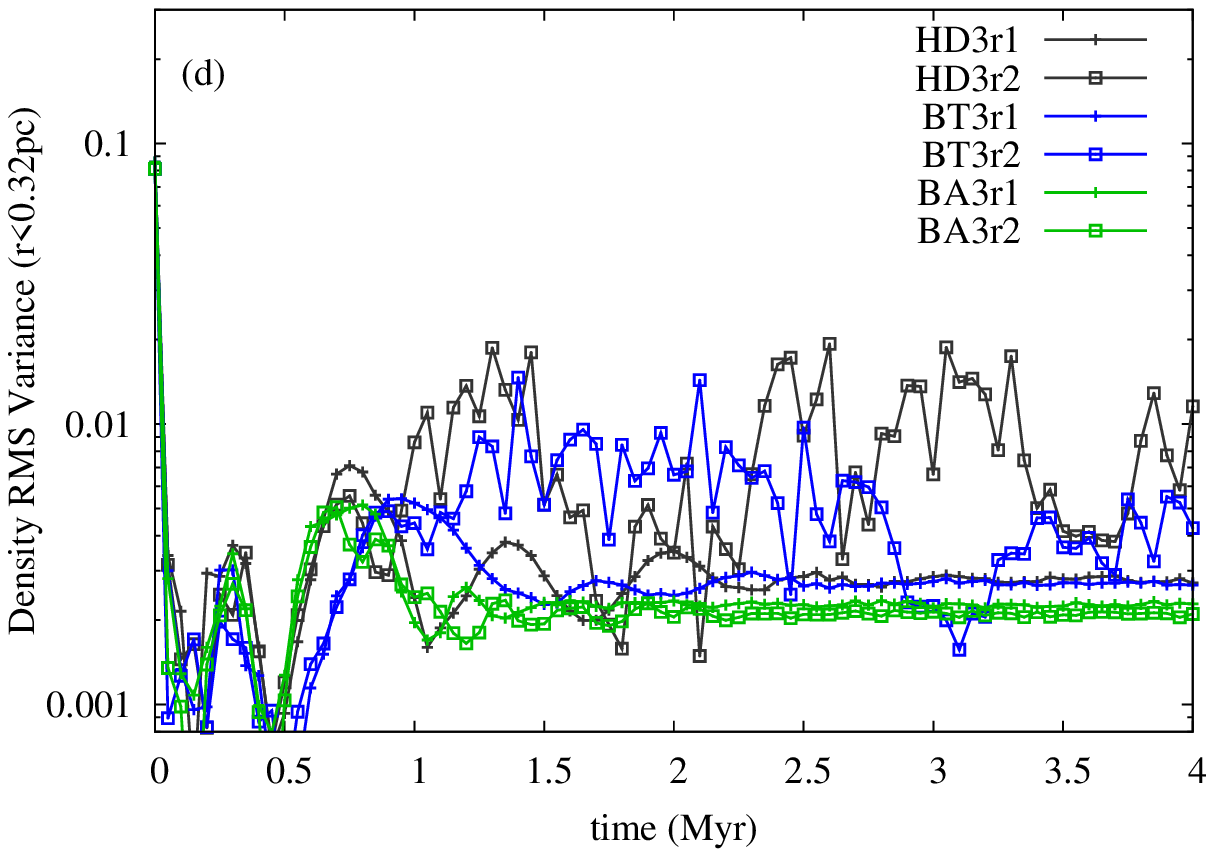} \\
\includegraphics[width=0.49\textwidth]{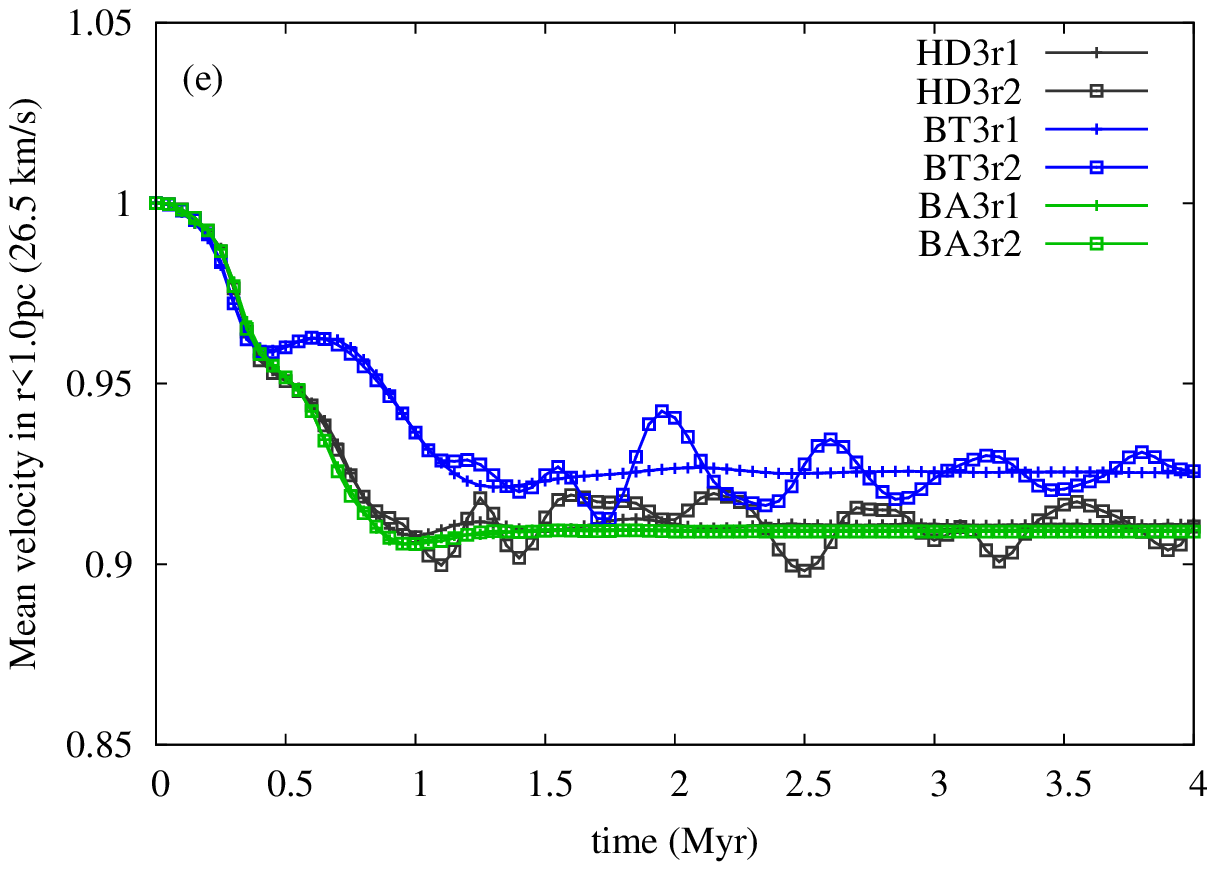}
\includegraphics[width=0.49\textwidth]{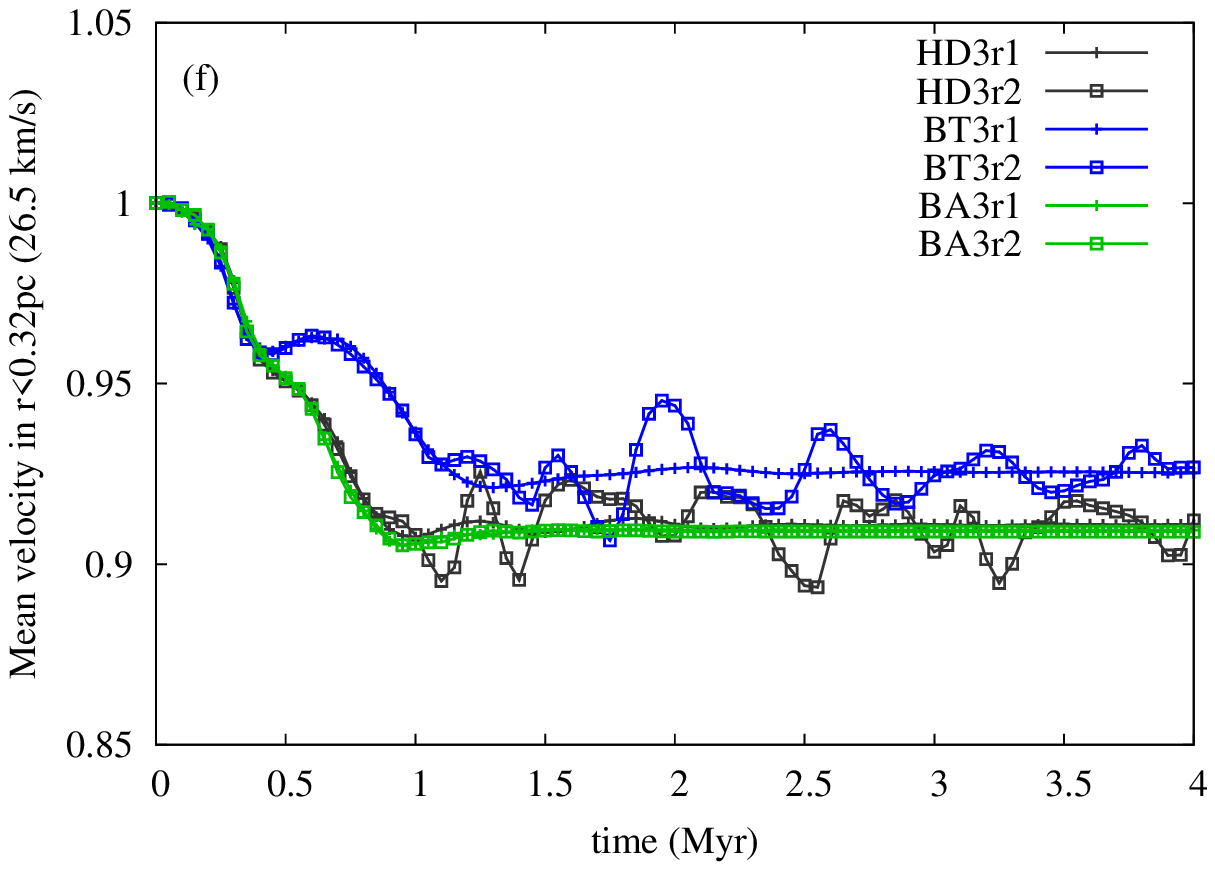}
\caption{
  The mean density (a and b), RMS density variance (c and d), and mean gas velocity (e and f) as a function of time for a volume within 1.0 pc (left [a,c,e]) or 0.32 pc (right [b,d,f]) of the star.
  The density is normalised to the mean ISM input density of $\ensuremath{n_{\textsc{h}}}=2.5\,\ensuremath{\mathrm{cm}^{-3}}$,
  and the velocity to the undisturbed bulk velocity of $v_\star=26.5\,\ensuremath{\mathrm{km}\,\mathrm{s}^{-1}}$.
  Different lines refer to the 3D simulations in Table~\ref{tab:Sims3D}.
  \label{fig:NearStar3D}
  }
\end{figure*}

Models of bow shocks around hot stars usually assume (for simplicity) that the H\,\textsc{ii} region does not affect the ISM density or dynamics significantly \citep[e.g.][]{ComKap98}.
This is clearly true in the hypersonic limit, and false in the subsonic limit, but the effects for intermediate $v_\star$ have not so far been explored.
We analysed the mean gas properties near the star for all simulation snapshots to investigate their spatial and temporal variations.
Two averaging volumes have been used: $r<1.0\,$pc %, $r<0.64\,$pc,
and $r<0.32\,$pc, corresponding to typical bow shock sizes.
Within these volumes the 
mean density $\langle\rho\rangle = (1/N)\sum_{i=1}^{N} \rho_i$, 
RMS density variance $\sigma = \sqrt{(\langle\rho^2\rangle-\langle\rho\rangle^2)}/\langle\rho\rangle$,
and mean velocity (in the star's rest frame) $\langle v\rangle = (1/N)\sum_{i=1}^{N} |v_i|$
of the ISM were calculated.
Here $N$ is the number of grid zones in the averaging volume.

Plots of these three quantities as a function of time are shown in Fig.~\ref{fig:NearStar3D}.
At early times ($t<1\,$Myr) the H\,\textsc{ii} region undergoes large-scale changes in shape leading to a decrease in density and subsequent increase.
This is a strongly damped oscillation driven by the initial out-of-equilibrium state.
The low-resolution simulations, HD3r1, BA3r1, and BT3r1, show no later time variation because they have reached a steady state.
The higher resolution aligned magnetic field model BA3r2 also reaches a steady state, with values very similar to BA3r1.
Models BT3r2 and HD3r2, however, show persistent fluctuations in gas properties near the star, because of the I-front instability and the waves it creates in the H\,\textsc{ii} region.
The mean gas density $\langle\rho\rangle$ is $\approx10$ per cent below the ISM mean density because the H\,\textsc{ii} region has expanded slightly.
The mean value is independent of the averaging volume for low resolution models, indicating that a linear density gradient through the volume is a good approximation.
At higher resolution, the temporal fluctuations in $\langle\rho\rangle$ are stronger on smaller scales, as expected if there are isolated planar shocks passing through.

The RMS density variance is larger for a larger averaging volume because there are more cells and so more independent volume elements.
For the small averaging volume the density values are strongly correlated because of numerical diffusion.
The RMS fluctuations in density are about 1$-$5 per cent for this spatial resolution, although this is expected to rise with higher resolution simulations that have stronger I-front instability (see Appendix~\ref{sec:2D}).

Interestingly, the velocity flowing past the star is $7.5-10$ per cent smaller than the ISM bulk velocity.
This is caused by the I-front jump conditions for an R-type I-front, which predict a small density increase and velocity decrease according to Equation~(\ref{eqn:WeakR}).
The H\,\textsc{ii} region temperature is $T_1\approx 6300\,$K, so the isothermal sound speed is about $c_1\approx9\,\ensuremath{\mathrm{km}\,\mathrm{s}^{-1}}$ giving $\mathcal{M}_1\approx3$, and we expect $\rho_1/\rho_0\approx1.1$ and $v_1/v_0\approx0.9$.
The velocity decrease is seen in Fig.~\ref{fig:NearStar3D} but the density increase is not.
Inspection of slices through the simulations shows that at the upstream I-front this density increase is found, but that the H\,\textsc{ii} region expansion has reduced the density at the position of the star.
This multi-dimensional effect was of course not seen in the 1D simulations (see Fig.~\ref{fig:1Dprofiles}).

These changes to the ISM properties generated by the H\,\textsc{ii} region dynamics are relatively small perturbations to the bulk flow properties, and are not expected to disrupt or greatly perturb the stellar wind bow shock that should be present.
It remains to be seen, however, what effect an inhomogeneous ISM will have on this picture.
In the cloud-zapping regime \citep{Ber89} where clumps are photoionized by R-type I-fronts, the H\,\textsc{ii} region should homogenise the ISM \citep[e.g.][]{McKvBurLaz84}, but if clumps are large enough to trap the I-front then a photoevaporation flow can be set up with velocity up to $v\lesssim30\,\ensuremath{\mathrm{km}\,\mathrm{s}^{-1}}$, which could strongly affect the bow shock.

%%%%%%%%%%%%%%%%%%%%%%%%%%%%%%%%%%%%%%%%%%%%%%%%%%%%%%%%%%%%%%%%%%%%%
\section{Discussion} \label{sec:discussion}
%%%%%%%%%%%%%%%%%%%%%%%%%%%%%%%%%%%%%%%%%%%%%%%%%%%%%%%%%%%%%%%%%%%%%
\begin{figure*}
\centering
\includegraphics[width=1.0\textwidth]{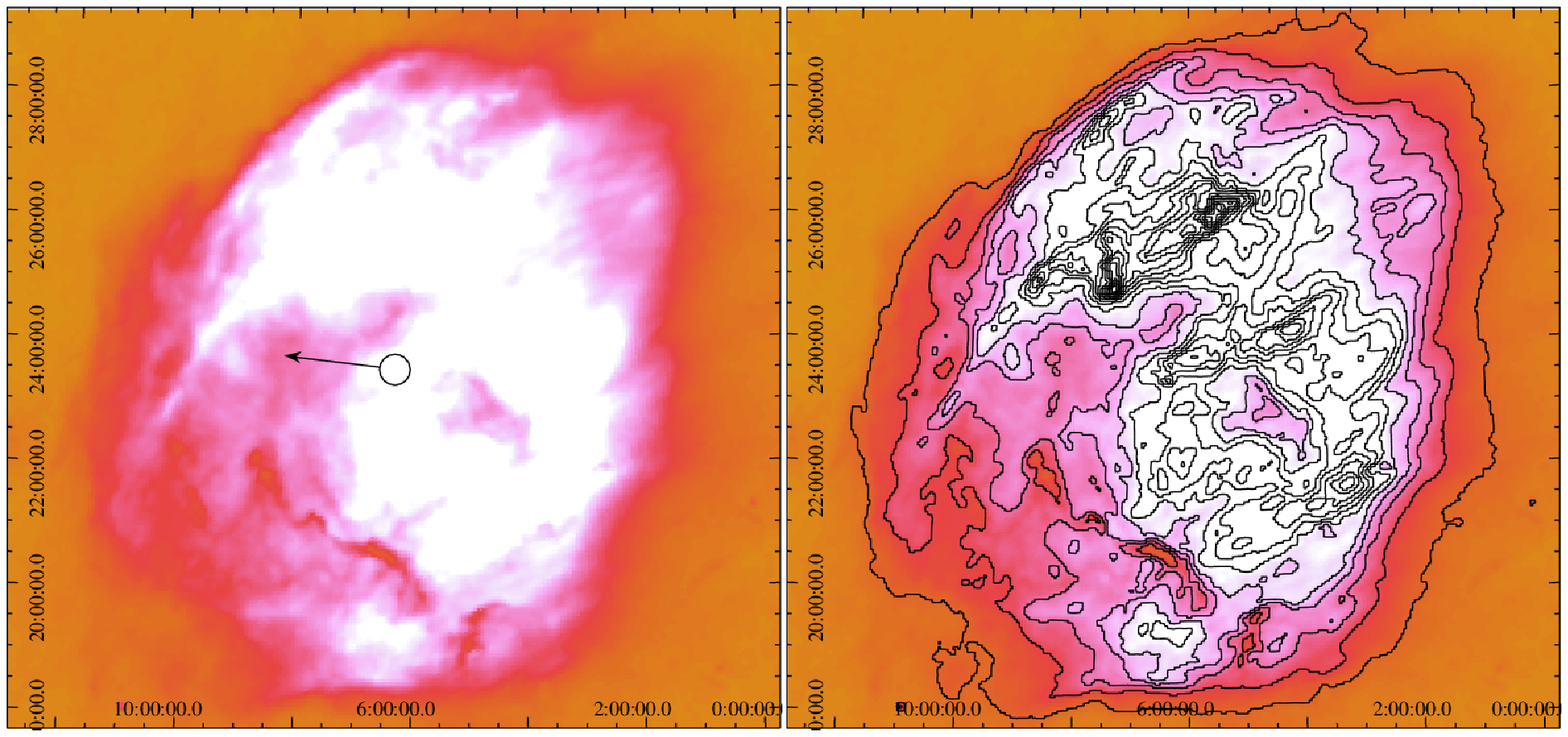}
\caption{
  Left: SHASSA H$\alpha$ image of the H\,{\sc ii} region around $\zeta$\,Oph.
  The star is at the centre of the circle, and the arrow indicates its direction of motion relative to the ISM.
  Right: The same image with linearly spaced brightness contours (in arbitrary units) highlighting the brightness asymmetry in the upstream and downstream directions (see text for details).
  The image is oriented with Galactic longitude (in units of degrees) increasing to the left and Galactic latitude increasing upwards.
  At a distance of 112 pc, 1 degree corresponds to $\approx1.93$ pc. 
  \label{fig:ZetaOph}
  }
\end{figure*}

%%%%%%%%%%%%%%%%%%%%%%%%%%%%%%%%%%%%%%%%%%%%%%%%%%%%%%%%%%%%%%%%%%%%%
\subsection{Comparison to the H\,\textsc{ii} region of $\zeta$ Oph}
%%%%%%%%%%%%%%%%%%%%%%%%%%%%%%%%%%%%%%%%%%%%%%%%%%%%%%%%%%%%%%%%%%%%%
The nearest example of an H\,\textsc{ii} region around an exiled O star is that of $\zeta$ Oph,
{\changed Sh\,2-27}.
In Fig.~\ref{fig:ZetaOph} we show the H$\alpha$ image of this H\,\textsc{ii} region originating from the Southern H$\alpha$ Sky Survey Atlas \citep[SHASSA;][]{GauMcCRosEA01}\footnote{The Southern H-Alpha Sky Survey Atlas (SHASSA) is supported by the National Science Foundation.}.
A detailed comparison to our simulations is beyond the scope of this paper, and will be pursued in future work; here we note some of the morphological similarities to (and differences from) the simulated H\,\textsc{ii} regions.
The most striking difference is the clumpy nature of the H$\alpha$ emission in the observational image, most likely related to underlying ISM inhomogeneity and/or foreground patchy extinction.
The ridge of strong absorption towards the bottom of the H\,\textsc{ii} region is correlated with a molecular cloud \citep*[Complex 4 in][]{dGeBroTha90}, indicating significant extinction due to foreground clouds that may or may not be physically associated with the H\,\textsc{ii} region.
Complex 1(E) from \citet{dGeBroTha90} also provides some extinction around $(l,b)=(5^\circ,23^\circ)$, downstream from and just below the star.
Even allowing for patchy extinction, it still seems clear that underlying ISM inhomogeneity is a significant component of the $\zeta$ Oph H\,\textsc{ii} region \citep[cf.][]{WooHafReyEA05}.

Nevertheless, if we consider only the upper half of the H\,\textsc{ii} region where there is less foreground extinction, there are definite similarities to our simulated H$\alpha$ maps.
The arc of the upstream I-front is sharp, whereas the downstream recombination front has a much more gradual decrease in intensity with distance from the star.
In addition, the downstream quadrant is generally lower intensity than the upstream quadrant.
The distance to the upstream I-front is smaller than that to the downstream recombination front ($R_{\mathrm{up}}/R_{\mathrm{dn}}\approx3/4$), and the H\,\textsc{ii} region has its largest radius at about 120$-$150$^\circ$ from the star's velocity vector.
All of these qualitative features are also found in the simulation results.

{\changed
The line-of-sight magnetic field through Sh\,2-27 has been measured using Faraday rotation by \citet*{HarMadGae11}, and found to be reasonably strong at 6.1\,$\mu$G (with some uncertainty arising from assumptions about gas clumping and distance).
If the field is this strong then the compression factor of the neutral gas shell found in our simulations would be significantly reduced by the magnetic pressure, and an H\,\textsc{i} column density map of Sh\,2-27 would resemble the left panel of Fig.~\ref{fig:simBT3r2} more than the right panel.
If the plane-of-sky magnetic field is much weaker than 6.1\,$\mu$G, then strongest shell compression is along the line-of-sight, in which case the shell may be observable in position-velocity H\,\textsc{i} data.
}

In \citet{GvaLanMac12}, where the mass-loss rate of $\zeta$ Oph was estimated based on the sizes of its H\,\textsc{ii} region and bow shock, it was assumed that the mean H\,\textsc{ii} region gas density is the same as the mean density at the bow shock, and furthermore that the ISM upstream from the bow shock remains at rest.
We have shown in Fig.~\ref{fig:NearStar3D} that these assumptions are reasonably well justified, in that the ISM density near the star ($\rho_i$) is only about 10 per cent lower than that of the undisturbed ISM ($\rho_0$), and the ISM has been accelerated only slightly in passing through the R-type I-front (to $v_i=2-2.4\,\ensuremath{\mathrm{km}\,\mathrm{s}^{-1}}$ in the upstream direction).
These are both small modifications of the upstream ram pressure in the star's rest frame, reducing $\rho_i v_i^2$ by about 25 per cent.
The bow shock standoff distance, $R_{\mathrm{\textsc{so}}}\equiv \sqrt{\dot{M}v_w/4\pi\rho_iv_i^2}$, is then increased by about 12 per cent.
The analytic model in \citet{GvaLanMac12} can therefore be applied with only minor corrections even for the relatively slowly-moving exile $\zeta$ Oph.

%%%%%%%%%%%%%%%%%%%%%%%%%%%%%%%%%%%%%%%%%%%%%%%%%%%%%%%%%%%%%%%%%%%%%
\subsection{ISM inhomogeneity}
%%%%%%%%%%%%%%%%%%%%%%%%%%%%%%%%%%%%%%%%%%%%%%%%%%%%%%%%%%%%%%%%%%%%%
We have not addressed ISM clumping in this work, assuming as a first step that the ISM is smooth.
A large clumping factor could affect the \citet{GvaLanMac12} analysis because $R_s$ is determined by recombination which scales with $\rho^2$, whereas  $R_{\mathrm{\textsc{so}}}$ is determined by ram pressure which is linear in $\rho$.
A very clumpy medium would therefore have a smaller ratio $R_s/R_{\mathrm{\textsc{so}}}$ than the maximal value obtained for a smooth medium.
Inspection of the observed H$\alpha$ maps in Fig.~\ref{fig:ZetaOph} strongly suggests some degree of underlying inhomogeneity in the ISM.
This remains a caveat to the \citet{GvaLanMac12} analysis, and will be studied further in future work.

This question has also been approached from another angle by \citet{WooHafReyEA05}, who used H$\alpha$, [N\,\textsc{ii}] and [S\,\textsc{ii}] emission maps from the H\,\textsc{ii} region around $\zeta$ Oph to constrain the porosity (or equivalently clumpiness) of the ISM through which it is passing.
They used a radiative transfer code to predict the emission properties of a static density field consisting of clumps of various sizes and filling factors embedded in a low-density medium.
They compared the circularly-averaged predictions to observations, for H$\alpha$ brightness and the [N\,\textsc{ii}]/H$\alpha$ and [S\,\textsc{ii}]/H$\alpha$ line ratios, as a function of distance from the star.
It was found that some clumping of the ISM was required to explain the spatial distribution of the H\,\textsc{ii} region surface brightness in these lines, given the modelling assumptions.

Figs.~\ref{fig:simHD3r12}-\ref{fig:simBT3r2} show, however, that circular symmetry is a bad approximation for H$\alpha$ emission because the upstream I-front has very different properties to the downstream recombination front.
This should also hold for other emission lines \citep[cf.][]{HenArtWilEA05}.
It would be very interesting to repeat the \citet{WooHafReyEA05} analysis, but with a 2D spatially-resolved comparison using a method such as that proposed by \citet{WooBarErcEA13}.
Even with this refinement, however, it is not clear that the supersonic relative motion between the star and the ISM can be neglected in the radiative transfer modelling because ionization equilibrium is never attained (although a stationary state is).
Supersonic advection compresses the upstream {\changed I-front} significantly \citep{RagNorCanEA97,HenArtWilEA05} and significantly changes the temperature and ionization structure of the downstream H\,\textsc{ii} region border from an I-front to a recombination front \citep{NewAxf68, WilDys96}.
No static model will be able to capture the downstream recombination front correctly, or the spatial compression of the upstream I-front.
The degree to which this is important should be investigated in future work where we will include more ions in the microphysics network, allowing us to predict line ratios directly.

\begin{figure}
\centering
\includegraphics[width=0.49\textwidth]{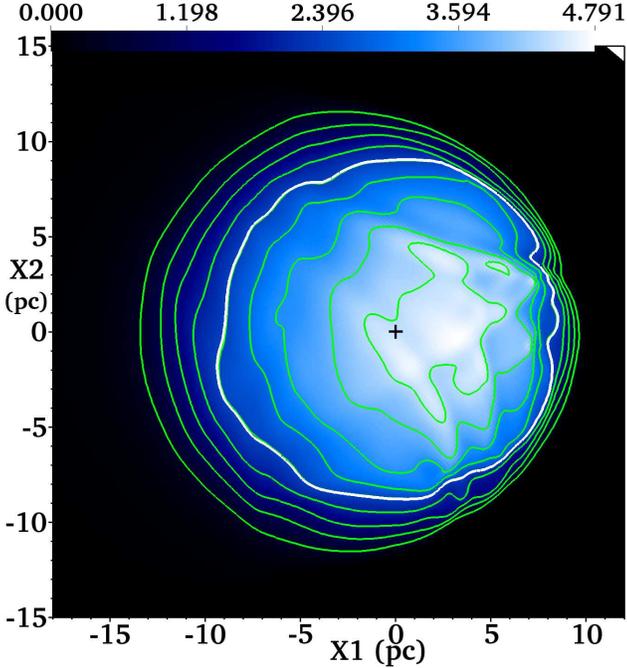}
\caption{
  Projected H$\alpha$ surface brightness of simulation HD3r2 at $t=4\,$Myr (using the same linear intensity scale as Fig.~\ref{fig:simHD3r12}), with 9 contours of equal intensity from 0.1 to 0.9 of the maximum in steps of 0.1.
  The 50 per cent contour is plotted in white for clarity.
  As before, the star is moving from left to right, and is indicated by the cross at the origin.
  \label{fig:Ha_contours}
  }
\end{figure}

%%%%%%%%%%%%%%%%%%%%%%%%%%%%%%%%%%%%%%%%%%%%%%%%%%%%%%%%%%%%%%%%%%%%%
\subsection{Isolated O and B stars in the Small Magellanic Cloud}
%%%%%%%%%%%%%%%%%%%%%%%%%%%%%%%%%%%%%%%%%%%%%%%%%%%%%%%%%%%%%%%%%%%%%
Recently \citet{OeyLamKusEA13} presented observations of 14 single-star H\,\textsc{ii} regions in the Small Magellanic Cloud (SMC), for each of which the ionizing source is a late O or early B star located at least 28\,pc from its nearest O or B type neighbour.
These were interpreted as candidate static stars that formed in isolation, largely on the basis of a lack of stellar wind bow shock, the circular shape of the H\,\textsc{ii} region, and the approximately central location of the star within the H\,\textsc{ii} region.

Regarding the shape of the H\,\textsc{ii} region, we have demonstrated that, at least for $v_\star=26.5\,\ensuremath{\mathrm{km}\,\mathrm{s}^{-1}}$, the H\,\textsc{ii} region remains approximately circular in projected H$\alpha$ emission.
The main difference for an exile is that the upstream edge is sharper and has higher surface brightness than the downstream edge (see Fig.~\ref{fig:Ha_contours} which shows an H$\alpha$ emission map of simulation HD3r2 overlaid with linearly-spaced contours from 10 to 90 per cent of the peak emission).
A number of the H\,\textsc{ii} regions in the \citet{OeyLamKusEA13} sample have similar asymmetries in their H$\alpha$ emission, although we have no way to constrain whether the asymmetries are correlated with their unknown proper motion. 
Similar results to ours were also obtained by \citet{RagNorCanEA97} for a much higher-velocity exile ($v_\star=100\,\ensuremath{\mathrm{km}\,\mathrm{s}^{-1}}$) in a somewhat lower-density medium ($n=1\,\ensuremath{\mathrm{cm}^{-3}}$).
Both of these results show that the shape of the H\,\textsc{ii} region in H$\alpha$ emission is not a strong indicator of stellar motion unless we are discussing properties at the 10$-$25 per cent level \citep[but see][for results from much harder-spectrum X-ray sources in a denser medium]{ChiRap96}.
Furthermore there is a well-known degeneracy between H\,\textsc{ii} region asymmetry produced by ISM density gradients and by stellar motion \citep[see][for a detailed investigation]{ArtHoa06}.
When the likely presence of ISM inhomogeneity is added to this, it may be difficult to draw strong conclusions regarding stellar motion from H\,\textsc{ii} region shapes in H$\alpha$ emission maps.
A better probe of stellar motion is probably the neutral shell that forms around the H\,\textsc{ii} region.
For $v_\star\gtrsim20-30\,\ensuremath{\mathrm{km}\,\mathrm{s}^{-1}}$ there will be no upstream or downstream dense shell, whereas a static star should have a shell of swept-up ISM in nearly all directions.

The offset of the star from the centre of the H\,\textsc{ii} region depends on how the centre is defined.
Identifying the H\,\textsc{ii} region border by eye, many of the stars in the \citet{OeyLamKusEA13} sample are offset from the H\,\textsc{ii} region centre by a similar degree to the star in our simulations.
This is a weak statement, however, and should be made more quantitative.
From Fig.~\ref{fig:Ha_contours} the star is significantly offset downstream from the 90 per cent intensity contour, and significantly offset upstream from the 10 per cent contour, but is very close to the centre of the (almost circular) 50 per cent contour.

The \citet{OeyLamKusEA13} H\,\textsc{ii} region sample is clearly an important dataset for probing the formation process and environment of isolated massive stars, but we argue that further quantitative predictions for the properties of H\,\textsc{ii} regions around moving stars (with a range of spatial velocities) are required before conclusions can be drawn about whether these stars are exiles or formed in situ.

%%%%%%%%%%%%%%%%%%%%%%%%%%%%%%%%%%%%%%%%%%%%%%%%%%%%%%%%%%%%%%%%%%%%%
\subsection{Implications of our results}
%%%%%%%%%%%%%%%%%%%%%%%%%%%%%%%%%%%%%%%%%%%%%%%%%%%%%%%%%%%%%%%%%%%%%
We have seen that the dynamical response of the ISM to a passing exiled star is that an overdense expanding conical shell is swept up, leaving an underdense wake behind it.
Stellar winds are not included in our simulations because the range of scales is too large, but hot shocked-wind material will also gather in the underdense wake behind the star.
It is therefore plausible that exiled massive stars create underdense `tunnels' through the ISM which can then be maintained by energy from supernova explosions \citep[cf.][]{CoxSmi74}, since the tunnels always originate at star clusters.
The momentum feedback of the H\,\textsc{ii} region expansion is at least as strong as that of a supernova for the parameters considered here, and will be even stronger for earlier O stars.
This suggests that exiled massive stars can make a significant contribution to the dynamics of gas in galaxy disks.
Further work quantifying the dependence of this feedback mechanism on stellar mass, ISM density, and metallicity are warranted to deduce its importance at a global level.

An important point, first made by \citet{RagNorCanEA97} and emphasised here, is that the H\,\textsc{ii} regions around exiled stars remain roughly circular, and the star stays near the centre of the H\,\textsc{ii} region unless it is moving with a really extreme velocity \citep[also for H\,\textsc{ii} regions around X-ray sources;][]{ChiRap96}.
The H\,\textsc{ii} region produced by a moving star remains quite spherical because it is always dynamically young, in contrast to the often-complex shapes of old H\,\textsc{ii} regions around static stars.
This is only true for H\,\textsc{ii} regions in which the stellar wind bow shock does not absorb a significant fraction of the star's ionizing photon output.
In a dense medium the H\,\textsc{ii} region can be trapped by the bow shock \citep{MacvBurWooEA91,ArtHoa06} for two reasons: the bow shock standoff distance decreases with increasing density less rapidly ($R_{\mathrm{\textsc{so}}}\propto \rho^{-1/2}$) than the H\,\textsc{ii} region radius ($R_s\propto\rho^{-2/3}$), and gas compression in the bow shock increases with density because of the shorter cooling time.

\citet{ConKra12} suggested that runaway massive stars could have contributed significantly to the reionization of the universe because they move out of their small protogalaxies within their main sequence lifetime, and so all of their ionizing radiation escapes their parent galaxy.
Dynamical feedback effects from H\,\textsc{ii} regions should become much stronger at low metallicity because the H\,\textsc{ii} region is hotter and less thermal energy is dissipated through radiative cooling.
We suggest that, in addition to the process considered by \citet{ConKra12}, the exiled stars also created lower-density tunnels from the centres of protogalaxies to their outskirts, increasing the escape fraction of ionizing photons and potentially providing channels through which supernova-enriched gas can more easily escape to the intergalactic medium (IGM).
Of course the exiled stars themselves also explode in the IGM, leading to in situ enrichment and the generation of intergalactic shockwaves.
\citet{GvaGuaPor09} suggested that high-velocity ($v_\star>200\,\ensuremath{\mathrm{km}\,\mathrm{s}^{-1}}$) O and B stars ejected from galactic disks could be the ionizing sources of extraplanar H\,\textsc{ii} regions observed $\approx0.5-1.5\,$kpc above some galactic disks \citep[e.g.][]{TueRosElwEA03}.
Such stars could travel outside the virial radius of a $10^8\,\msun$ galaxy at redshift $z\sim10$ \citep[see][]{ConKra12}.
The small size and lower escape velocities of high redshift galaxies make exiled stars much more important at high redshift than in the local universe.

%%%%%%%%%%%%%%%%%%%%%%%%%%%%%%%%%%%%%%%%%%%%%%%%%%%%%%%%%%%%%%%%%%%%%
\section{Conclusions} \label{sec:conclusions}
%%%%%%%%%%%%%%%%%%%%%%%%%%%%%%%%%%%%%%%%%%%%%%%%%%%%%%%%%%%%%%%%%%%%%
We have studied the dynamics of H\,\textsc{ii} regions around supersonically-moving hot stars with multidimensional MHD simulations including non-equilibrium photoionization.
While the response of the ISM to the H\,\textsc{ii} region is not violent, it is significant and leaves a long-lasting imprint on the ISM once the star has passed.
Only one stellar velocity has been considered, $v_\star=26.5\,\ensuremath{\mathrm{km}\,\mathrm{s}^{-1}}$, matching the peculiar velocity of the nearby O9.5\,V star $\zeta$ Oph.
We have also considered only one ionizing photon luminosity, again matching observations of this star.

The H\,\textsc{ii} region expansion leaves a cone-shaped shell in the wake behind the star, truncated on the H\,\textsc{ii} region edge at the angle where the normal velocity of gas through the I-front reaches Mach 2 (with respect to ionized gas).
This shell should have column density of a few $\times10^{20}\,\ensuremath{\mathrm{cm}^{-3}}$ in neutral H and may be observable with kinematic H\,\textsc{i} data \citep[or dust emission;][]{RagNorCanEA97}.
Directly downstream from the star is an underdense wake that should be filled by hot gas from the stellar wind (not modelled here).
The gas expansion driven by the H\,\textsc{ii} region should make the stellar wind's wake longer-lived than would be the case without this expansion, because this creates a lower density and pressure environment.

The shocked shell is affected by the presence and orientation of a large-scale ISM magnetic field.
Both H\,\textsc{ii} region expansion and shock compression are strongly inhibited perpendicular to magnetic field lines, leading to a less overdense shell with a much lower H\,\textsc{i} column density.
The larger wavespeeds in a magnetised plasma also give the shell a wider opening angle.
The underdense wake, by contrast, is not significantly affected by the ISM magnetic field.

Kinetic energy is generated in the ISM from the H\,\textsc{ii} region's expansion at a rate comparable to the mechanical luminosity of the stellar wind from $\zeta$ Oph.
When we take into account that (at least for static stars) about 90 per cent of the wind energy is lost to dissipation \citep[e.g.][]{GarMacLan96}, the H\,\textsc{ii} region expansion has stronger feedback energetically than the stellar wind.
This is also true of the momentum feedback, where the H\,\textsc{ii} region expansion generates more than $100\times$ the momentum in the ISM than the stellar wind.
The momentum input rate is also about $10\times$ the total radiation momentum from the stellar luminosity.
The kinetic energy feedback rate is affected somewhat ($\sim50$ per cent level) by the ISM magnetic field, but momentum feedback is very little affected.
Compared to the star's eventual supernova explosion, the total kinetic energy feedback from the H\,\textsc{ii} region over the star's main sequence lifetime is $>100\times$ less, but the momentum feedback is up to $4\times$ larger.

The unstable upstream I-front perturbs gas within the H\,\textsc{ii} region, but the perturbations have an amplitude of only $\approx 5-10$ per cent near the star and are unlikely to affect the bow shock to any extent (although further simulations are needed to ensure the perturbations do not grow to non-linear amplitudes with higher resolution).

A density gradient is created by the H\,\textsc{ii} region expansion, such that the ionized gas is densest near the upstream I-front and decreases downstream.
This results in H$\alpha$ emission maps that are brighter upstream from the star than downstream.
In addition the upstream I-front is very thin whereas the downstream recombination front is broad, and this is also be reflected in H$\alpha$ emission maps.
Similar features can be seen in H$\alpha$ emission from the H\,\textsc{ii} region around $\zeta$ Oph.
From this we conclude that gas dynamics and non-equilibrium ionization are important ingredients in the observational properties of H\,\textsc{ii} regions from supersonically-moving exiles.
Radiative transfer models that assume ionization equilibrium or ignore the relative motion between the star and the ISM may have only limited applicability to stars such as $\zeta$ Oph.

We have not studied the effects of ISM inhomogeneity and turbulent motions in this work.
Rather we have provided a baseline study to see what the dynamics generated purely by the photoionization process are.
In future work we will include clumpy and turbulent ISM structure \citep[cf.][]{MelArtHenEA06,ArtHenMelEA11} to investigate the degree to which this substructure is destroyed or enhanced by the passage of a hot star.

It does not seem to have been appreciated until now that H\,\textsc{ii} regions around exiled O stars can be very good laboratories for studying the physics of {\changed I-fronts}, in particular their stability.
The early R-type expansion of H\,\textsc{ii} regions around newly formed stars is deeply embedded in molecular clouds and so is difficult to observe, and both the I-front velocity and the stellar radiation spectrum are time-varying.
Runaway stars, by contrast, are typically in low-extinction environments and have a constant velocity R-type I-front that (at least in regions of constant ISM density) should relax to a steady state.
Some of them, for example $\zeta$ Oph, are also much closer to us than the nearest massive star-forming regions, so we can observe their I-fronts in much more detail and strongly test theoretical predictions and numerical simulations.

%%%%%%%%%%%%%%%%%%%%%%%%%%%%%%%%%%%%%%%%%%%%%%%%%%%%%%%%%%%%%%%%%%%%%
\section*{Acknowledgments}
%%%%%%%%%%%%%%%%%%%%%%%%%%%%%%%%%%%%%%%%%%%%%%%%%%%%%%%%%%%%%%%%%%%%%
JM is funded by a fellowship from the Alexander von Humboldt Foundation.
This work was supported by the Deutsche Forschungsgemeinschaft priority program 1573, `Physics of the Interstellar Medium'.
The authors acknowledge the John von Neumann Institute for Computing for a grant of computing time on the JUROPA supercomputer at J\"ulich Supercomputing Centre.
We thank M.S.~Oey for useful comments on a draft version of this paper.
JM acknowledges the Nordita program on Photo-Evaporation in Astrophysical
Systems (June 2013), which enabled useful discussions about the results obtained in this work.
{\changed We thank the referee, Alex Raga, for useful suggestions which improved the presentation of the paper.}

%%%%%%%%%%%%%%%%%%%%%%%%%%%%%%%%%%%%%%%%%%%%%%%%%%%%%%%%%%%%%%%%%%%%%
\bibliography{../../../../../documentation_misc/bibtex/refs}
%%%%%%%%%%%%%%%%%%%%%%%%%%%%%%%%%%%%%%%%%%%%%%%%%%%%%%%%%%%%%%%%%%%%%

\appendix

%%%%%%%%%%%%%%%%%%%%%%%%%%%%%%%%%%%%%%%%%%%%%%%%%%%%%%%%%%%%%%%%%%%%%
\section{Resolution effects in 1D simulations}
\label{sec:app:res}
%%%%%%%%%%%%%%%%%%%%%%%%%%%%%%%%%%%%%%%%%%%%%%%%%%%%%%%%%%%%%%%%%%%%%
\begin{figure}
\centering
\includegraphics[width=0.42\textwidth]{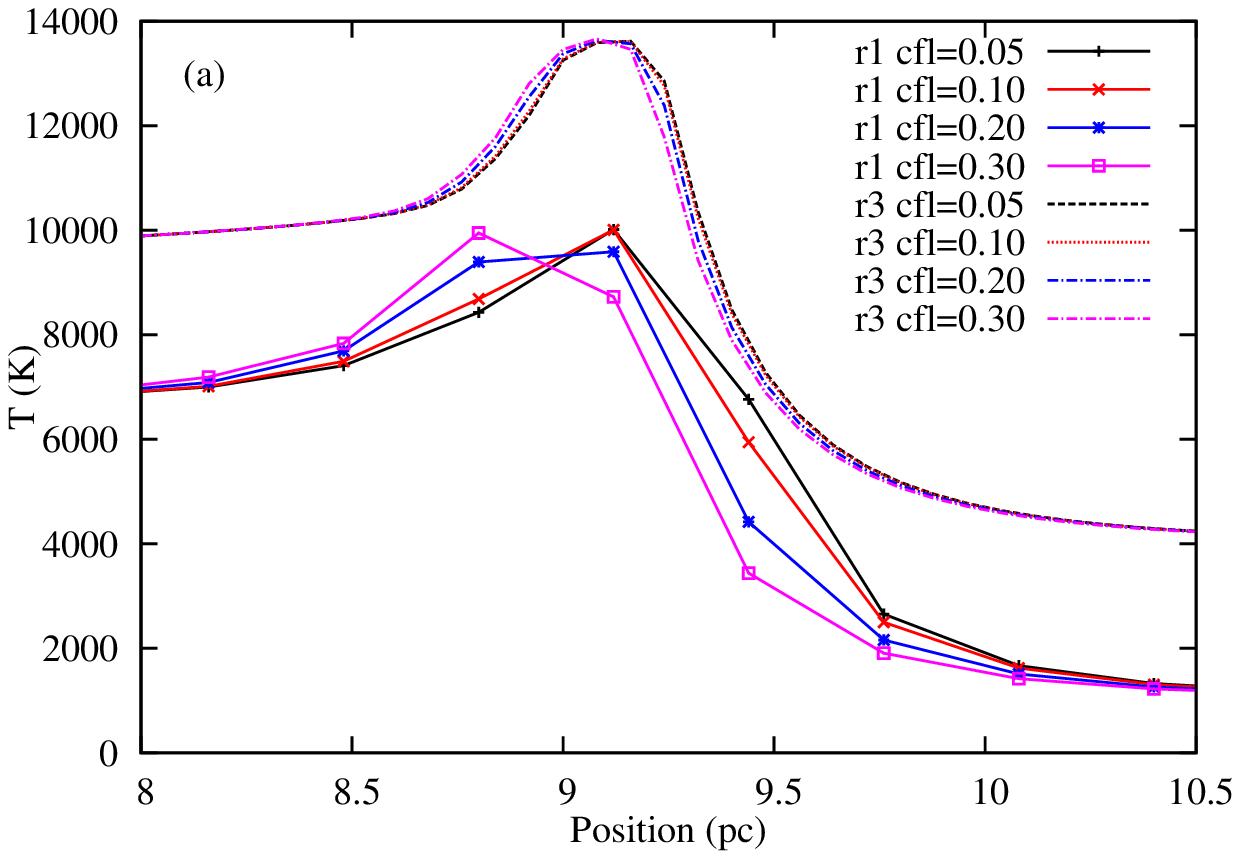}
\includegraphics[width=0.42\textwidth]{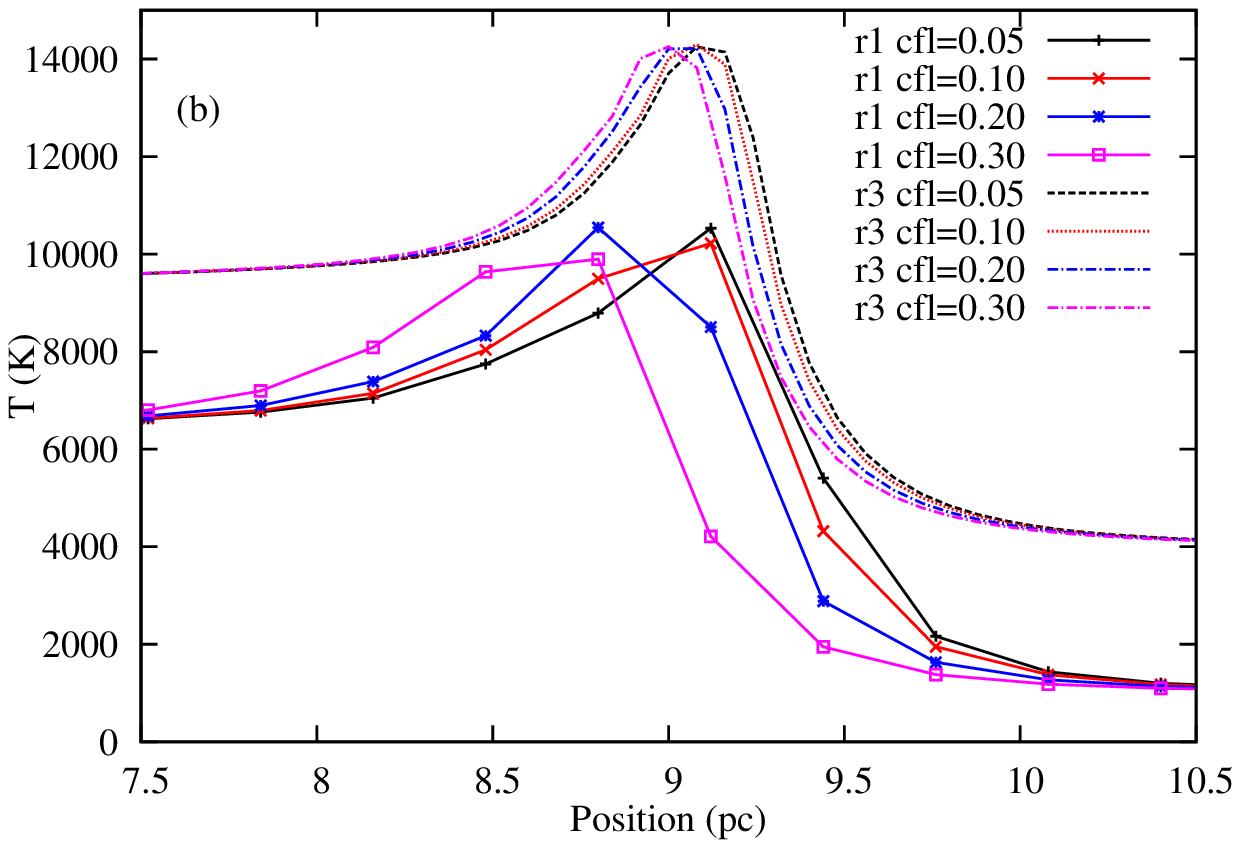}
\includegraphics[width=0.42\textwidth]{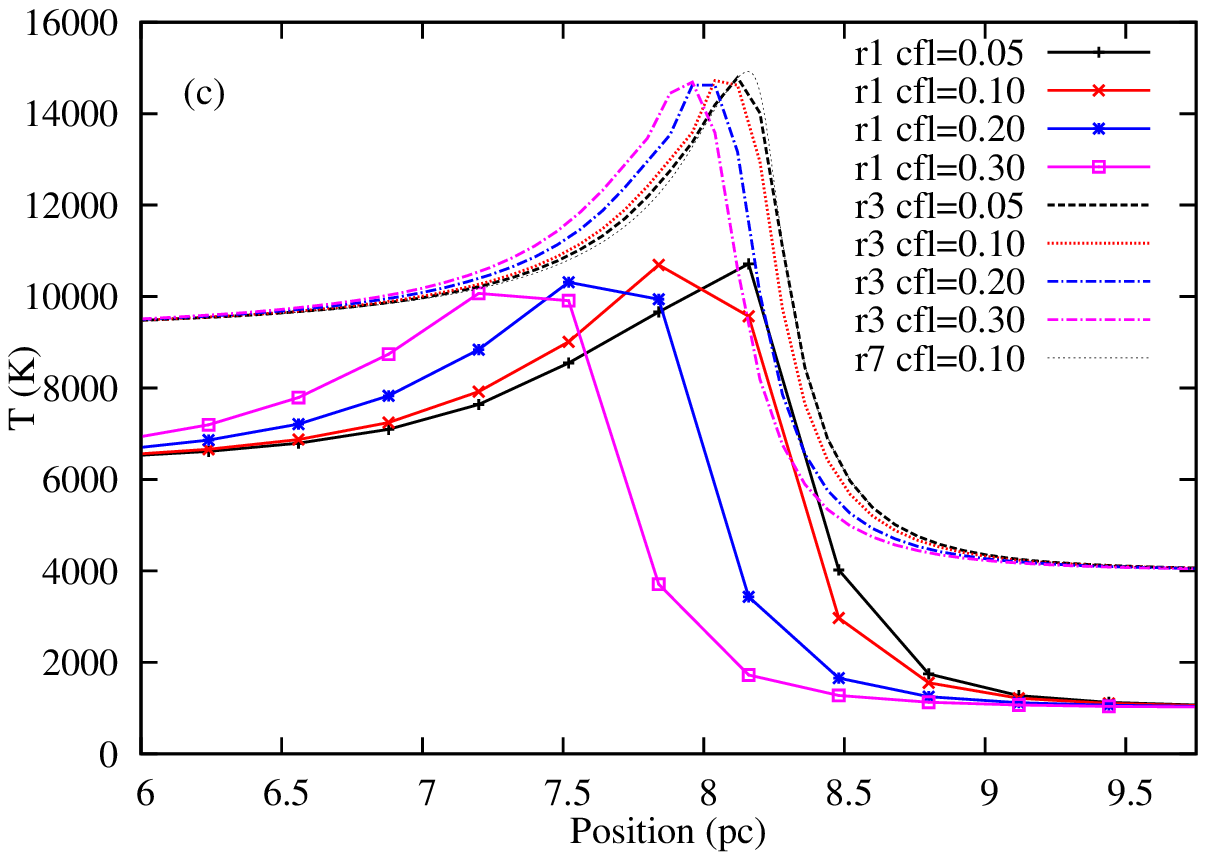}
\includegraphics[width=0.42\textwidth]{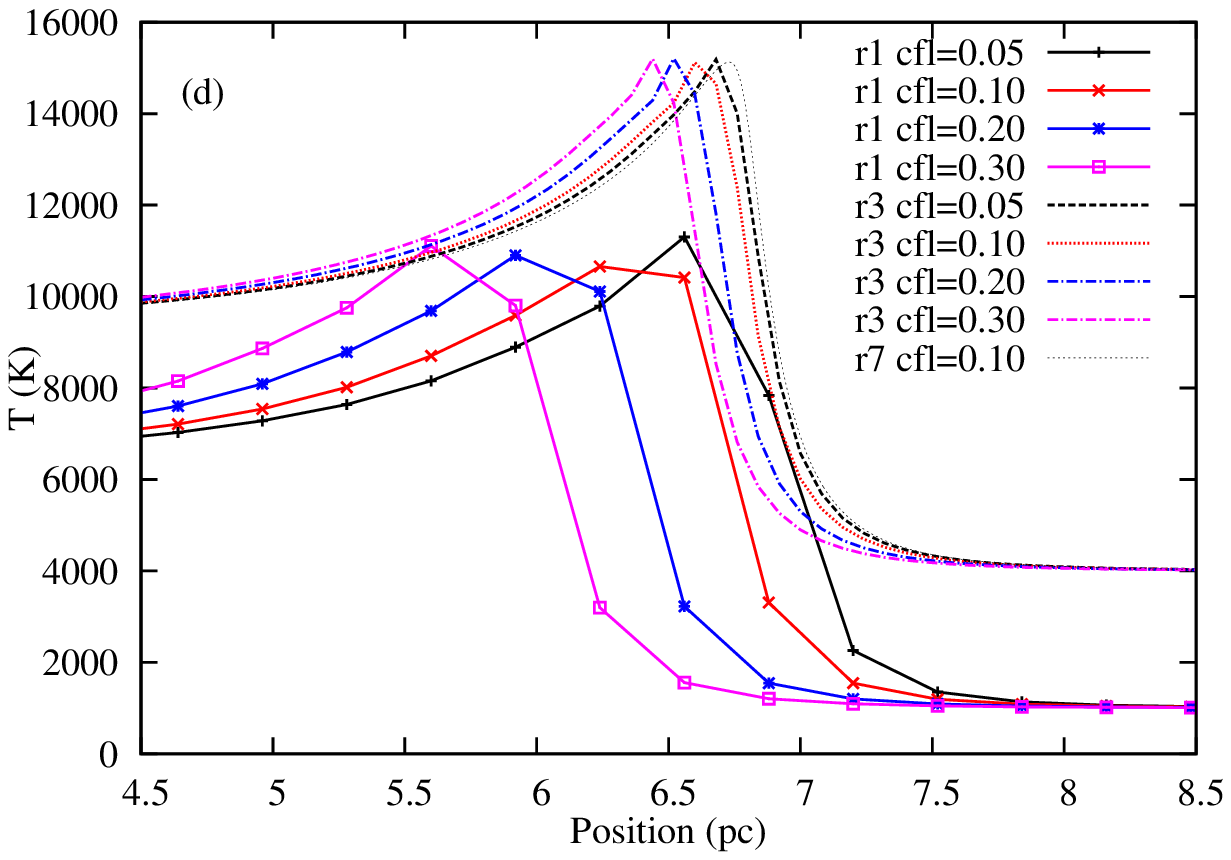}
\caption{
  Gas temperature at the upstream I-front for 1D simulations at different spatial and temporal resolutions as indicated, for $v_\star=25\,\ensuremath{\mathrm{km}\,\mathrm{s}^{-1}}$ (a), $50\,\ensuremath{\mathrm{km}\,\mathrm{s}^{-1}}$ (b), $100\,\ensuremath{\mathrm{km}\,\mathrm{s}^{-1}}$ (c), and $200\,\ensuremath{\mathrm{km}\,\mathrm{s}^{-1}}$ (d).
  The higher resolution models (r3) have been offset vertically by 3000 K for clarity.
  The dotted line in the lower two curves is the highest resolution simulation (r7) with $C_{\mathrm{cfl}}=0.1$, showing the converged solution.
  \label{fig:1Dcfl}
  }
\end{figure}

\begin{figure}
\centering
\includegraphics[width=0.42\textwidth]{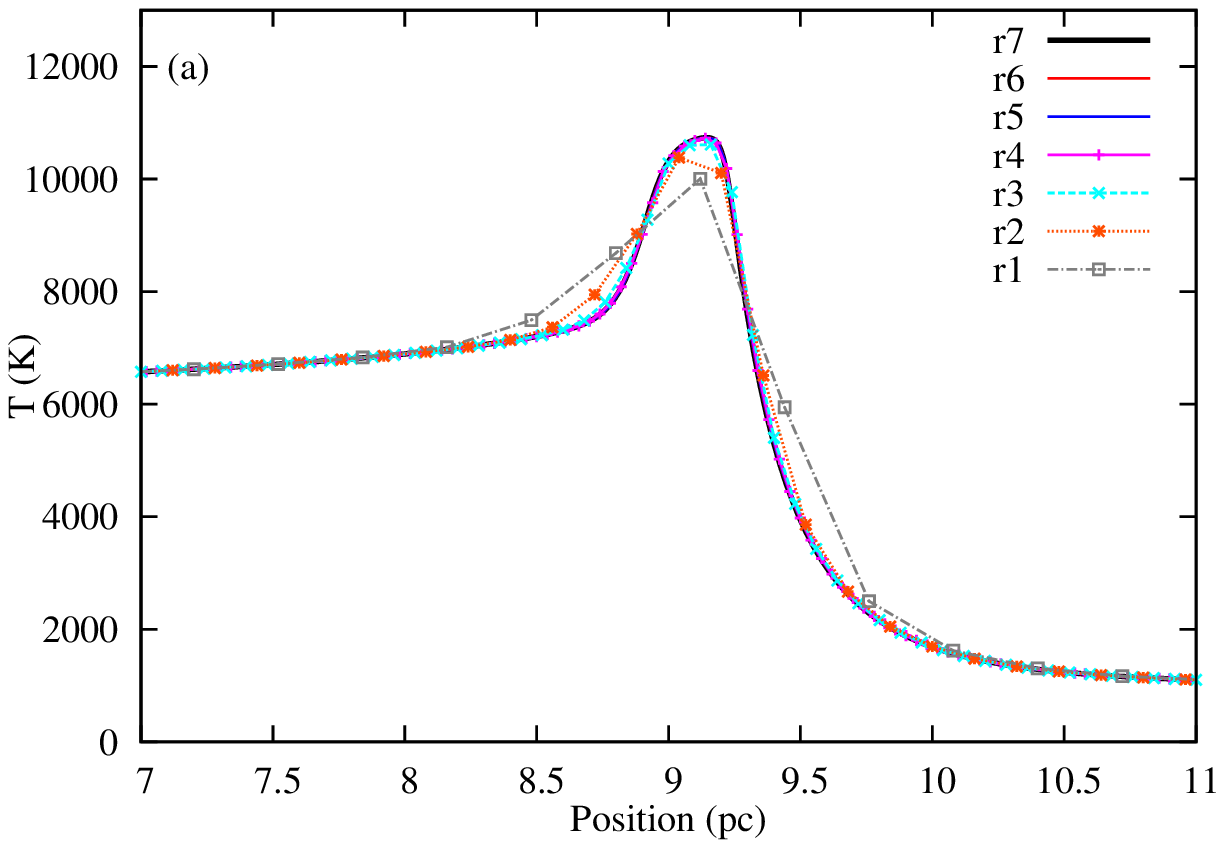}
\includegraphics[width=0.42\textwidth]{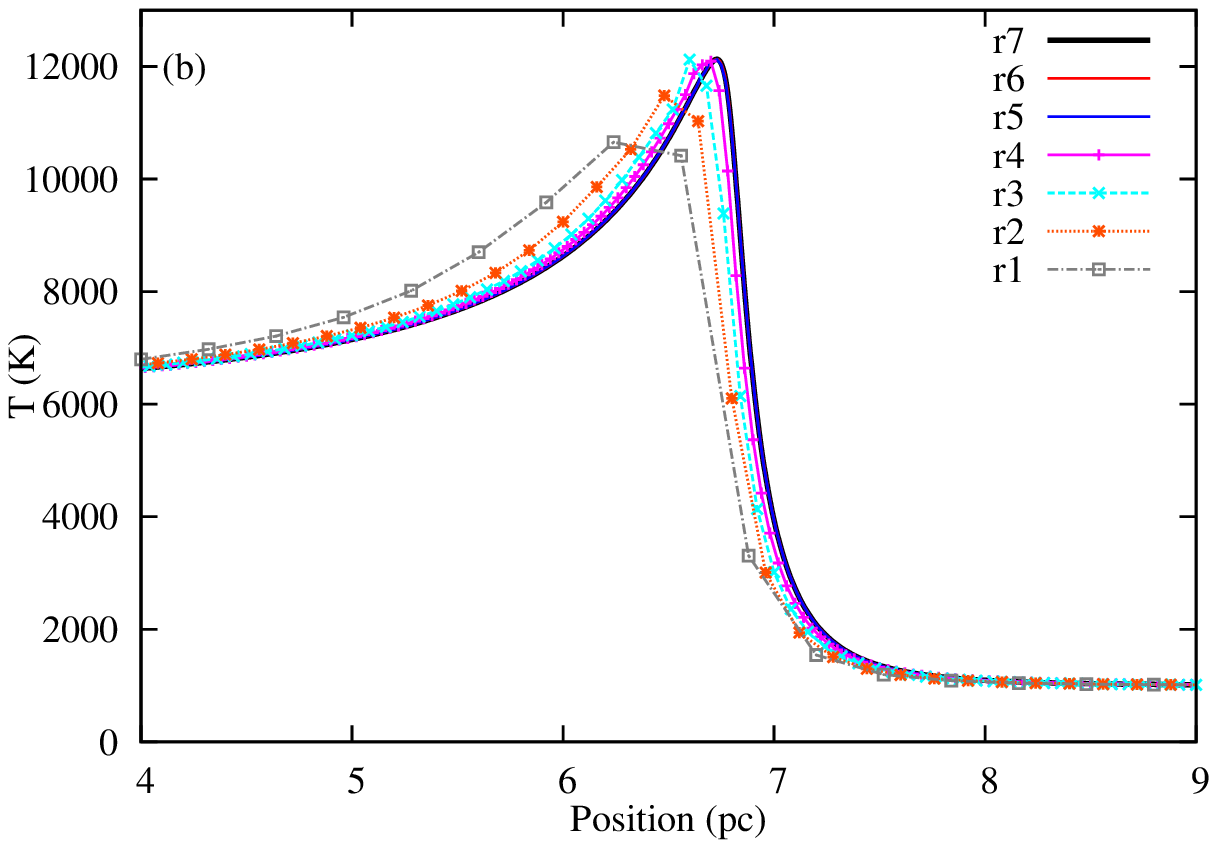}
\caption{
  Gas temperature at the upstream I-front for 1D simulations at different spatial resolutions as indicated, for $C_{\mathrm{cfl}}=0.1$ with $v_\star=25\,\ensuremath{\mathrm{km}\,\mathrm{s}^{-1}}$ (a) and $200\,\ensuremath{\mathrm{km}\,\mathrm{s}^{-1}}$ (b).
  The higher resolution models (r5-r7) spatially resolve the I-front and are indistinguishable on this plot.
  \label{fig:1Dres}
  }
\end{figure}

\begin{figure}
\centering
\includegraphics[width=0.42\textwidth]{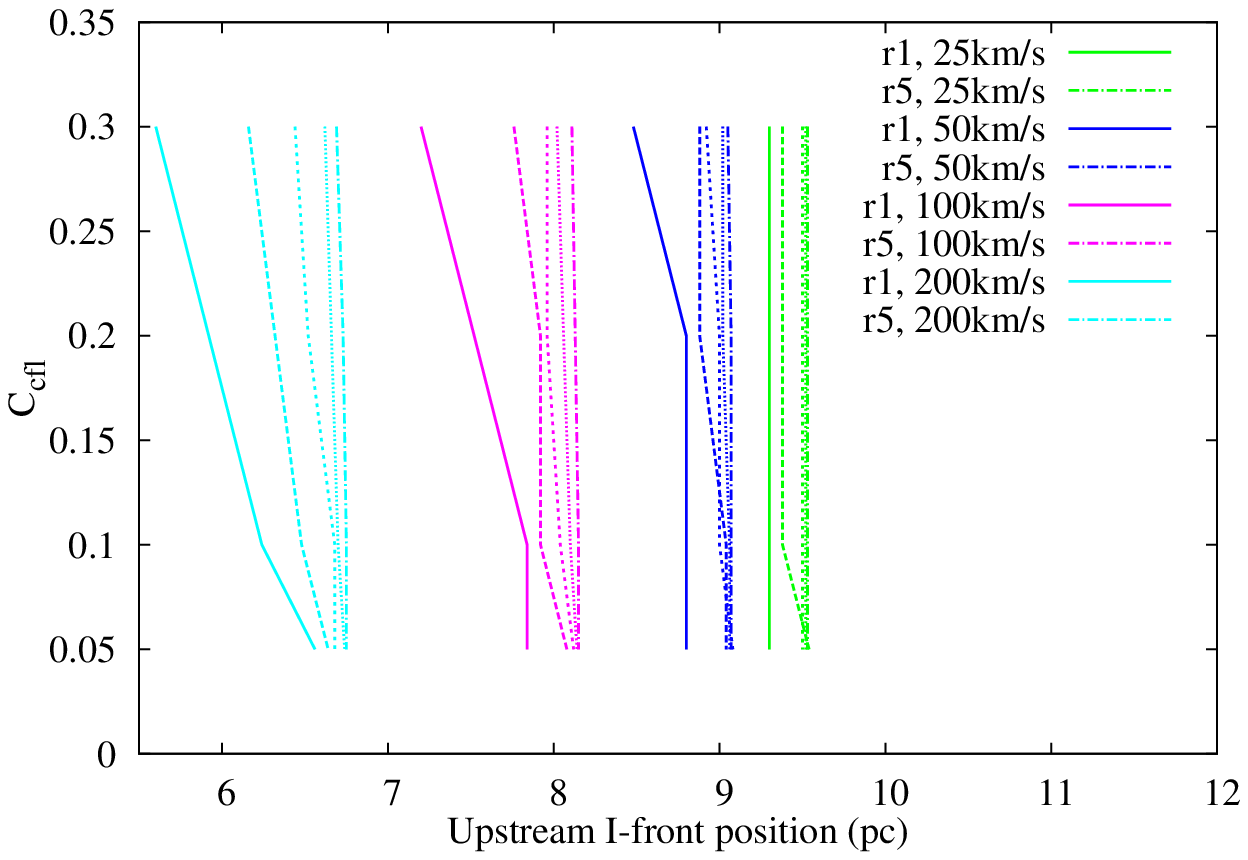}
\caption{
  The position of the I-front (traced by the largest radius such that $y_n>0.5$) for different $v_\star$ and $C_{\mathrm{cfl}}$.
  For each velocity the two labelled lines (solid and dot-dashed) are for resolutions r1 and r5, and they bracket the unlabelled lines for r2, r3 and r4.
  Ideally all lines would be vertical, and values for each velocity would lie on top of each other.
  The finite zone-size means that lines can move horizontally from discreteness error, and a change in slope is from time-integration errors.
  For large velocities, the different spatial resolutions are separated by more than one grid zone ($\Delta x=0.32\,$pc for r1) for large $C_{\mathrm{cfl}}$.
  The curves for $v_\star=25\,\ensuremath{\mathrm{km}\,\mathrm{s}^{-1}}$ have been offset by $+0.5\,$pc for clarity.
  \label{fig:1D_Vel_CFL}
  }
\end{figure}

As discussed in Section~\ref{sec:1D}, a CFL number of $C_{\mathrm{cfl}}=0.1$ was required for the simulations presented in the text.
This choice is justified here, by varying the temporal and spatial resolution of the 1D simulations.
The problem is that for $\mathcal{M}>2$, the H\,\textsc{ii} region structure relaxes to a stationary state with respect to the computational grid (even though gas is continuously recombining and being ionized), so the microphysics timescales go to infinity.
In this case the CFL condition is the only active timestep restriction on the simulation.
The effects of varying $C_{\mathrm{cfl}}$ on the upstream I-front position are shown in temperature plots for two different spatial resolutions (r1 and r3) in Fig.~\ref{fig:1Dcfl}, for four different values of $v_\star$.
The errors increase with $v_\star$ for a given $C_{\mathrm{cfl}}$ and $\Delta x$.
The thin dotted line in the higher velocity figures shows the highest resolution model (r7) with $C_{\mathrm{cfl}}=0.1$; in this case the solution has converged, but we have had to spatially resolve the I-front (cf.\ \citealt{CanPor11}).
Even for $v_\star=25\,\ensuremath{\mathrm{km}\,\mathrm{s}^{-1}}$ the lowest resolution simulations have an incorrect I-front position with $C_{\mathrm{cfl}}=0.3$.

Fig.~\ref{fig:1Dres} shows the gas temperature through the upstream I-front at different spatial resolutions for $C_{\mathrm{cfl}}=0.1$ and $v_\star=25\,\ensuremath{\mathrm{km}\,\mathrm{s}^{-1}}$ (above) and $200\,\ensuremath{\mathrm{km}\,\mathrm{s}^{-1}}$ (below).
It shows that the temperature peak is well modelled for $C_{\mathrm{cfl}}=0.1$ for the low velocity model, but for $v_\star=200\,\ensuremath{\mathrm{km}\,\mathrm{s}^{-1}}$ the solution lags behind the true solution at low spatial resolution.
Once the I-front is spatially resolved (with $\Delta \tau \leq 1$ per zone) the solution has converged.
The positions of the I-front are plotted for different $C_{\mathrm{cfl}}$ in Fig.~\ref{fig:1D_Vel_CFL}, showing that even for $v_\star=100\,\ensuremath{\mathrm{km}\,\mathrm{s}^{-1}}$ the I-front position is modelled accurately with $C_{\mathrm{cfl}}=0.1$, but larger velocities are not.
It is concluded that $C_{\mathrm{cfl}}=0.1$ provides nearly resolution-independent I-front positions and temperatures, at least for $v_\star<100\,\ensuremath{\mathrm{km}\,\mathrm{s}^{-1}}$.

Such constraints were not necessary for the test calculations in \citet{Mac12} that did not include any hydrodynamics.
In that case the accuracy of the solution was independent of the I-front velocity, but for a static ISM and for an I-front that was moving rapidly with respect to the grid.
It seems that the strong advection requires tight coupling to the microphysics integration to obtain an accurate solution (at least for this algorithm).

%%%%%%%%%%%%%%%%%%%%%%%%%%%%%%%%%%%%%%%%%%%%%%%%%%%%%%%%%%%%%%%%%%%%%
\section{Resolution effects in 2D simulations}
\label{sec:2D}
%%%%%%%%%%%%%%%%%%%%%%%%%%%%%%%%%%%%%%%%%%%%%%%%%%%%%%%%%%%%%%%%%%%%%

\begin{figure*}
\centering
  \includegraphics[width=0.49\textwidth]{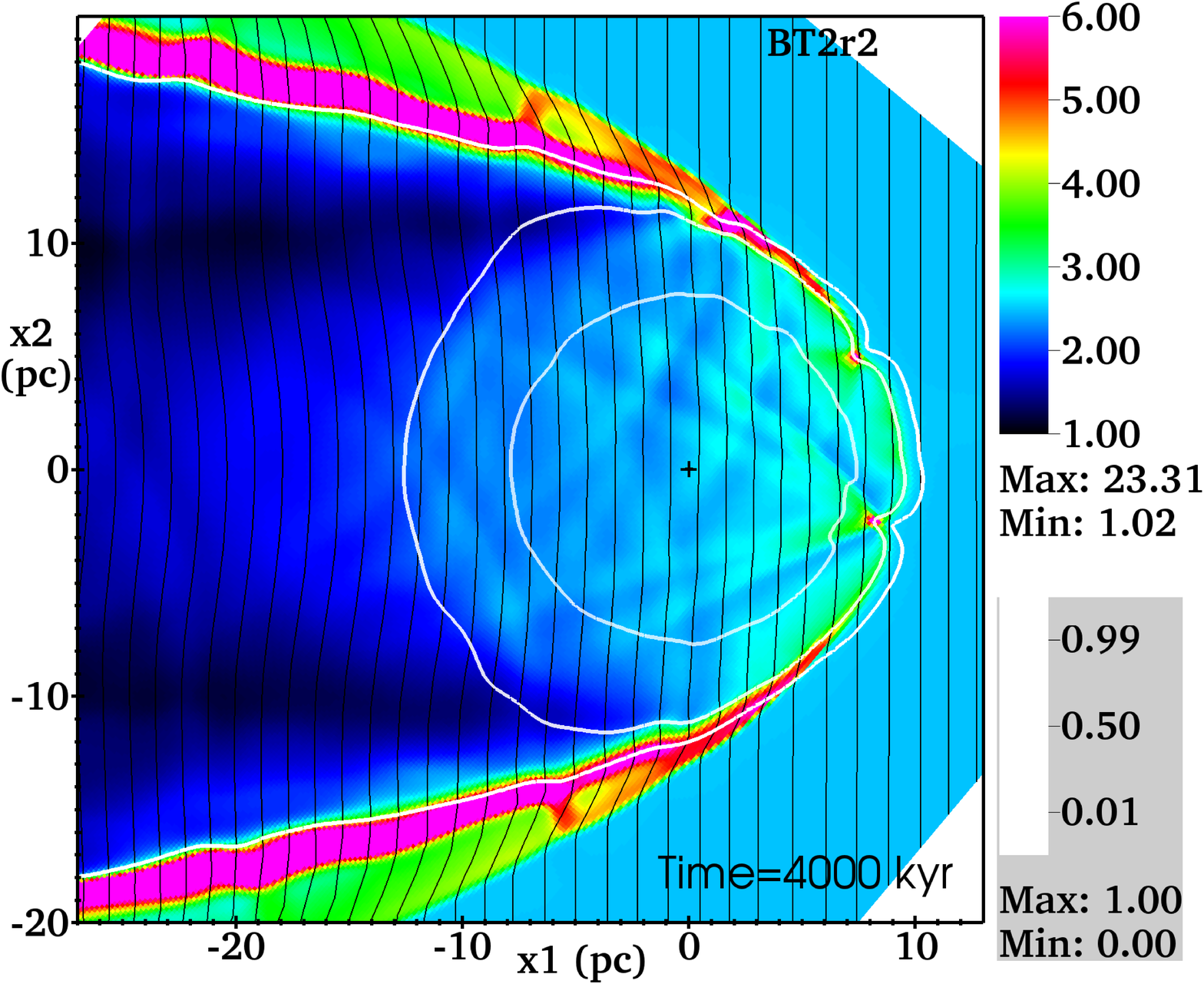}
  \includegraphics[width=0.49\textwidth]{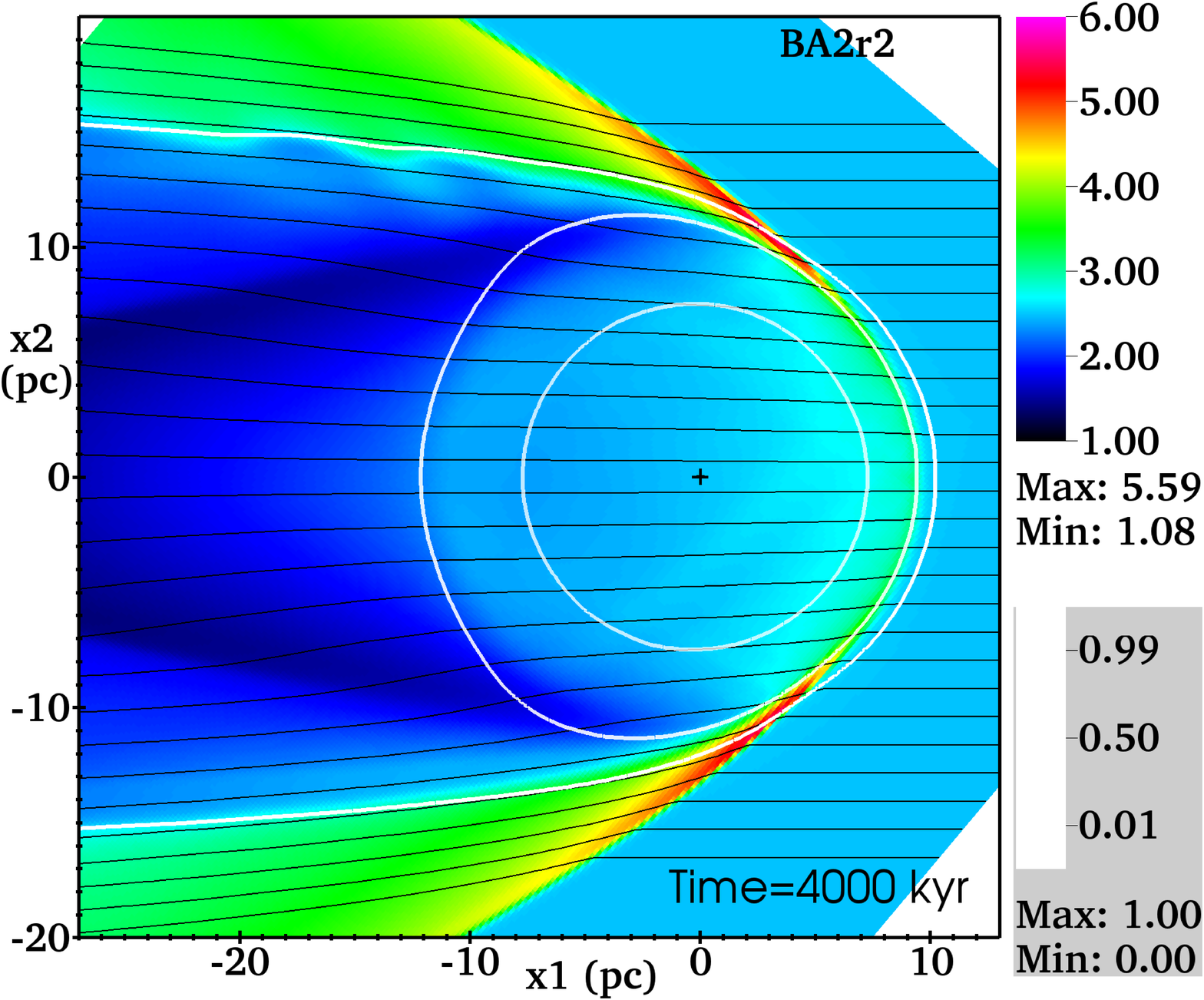}
\caption{
  Snapshots from the 2D simulations BT2r2 (left) and BA2r2 (right) at $t=4\,$Myr.
  The star is at the origin, and the ISM is flowing past from right to left at 26.5 \ensuremath{\mathrm{km}\,\mathrm{s}^{-1}}.
  The grid is rotated so that the bulk flow is along the $-\hat{\mathitbf{x}}_1$ direction and perpendicular to the $\hat{\mathitbf{x}}_2$ direction.
  The colour image shows H number density, \ensuremath{n_{\textsc{h}}} (in \ensuremath{\mathrm{cm}^{-3}}), with the linear colour scale to the right and the maximum and minimum values on the grid indicated just below.
  Three contours of ion fraction $(1-y_n)=\{0.99,\,0.5,\,0.01\}$ are overlaid in white from the origin outwards, and streamlines with the magnetic field orientation are shown in black.
  The streamline spacing is uniform at $\mathitbf{x}_2=0$ pc for the left plot and so does not indicate field strength, whereas the starting point is at $\mathitbf{x}_1=10$ pc for the right plot (where the field is uniform) and so the downstream field spacing does indicate strength.
  \label{fig:simBX2r2}
  }
\end{figure*}
\begin{figure*}
\centering
\includegraphics[width=0.49\textwidth]{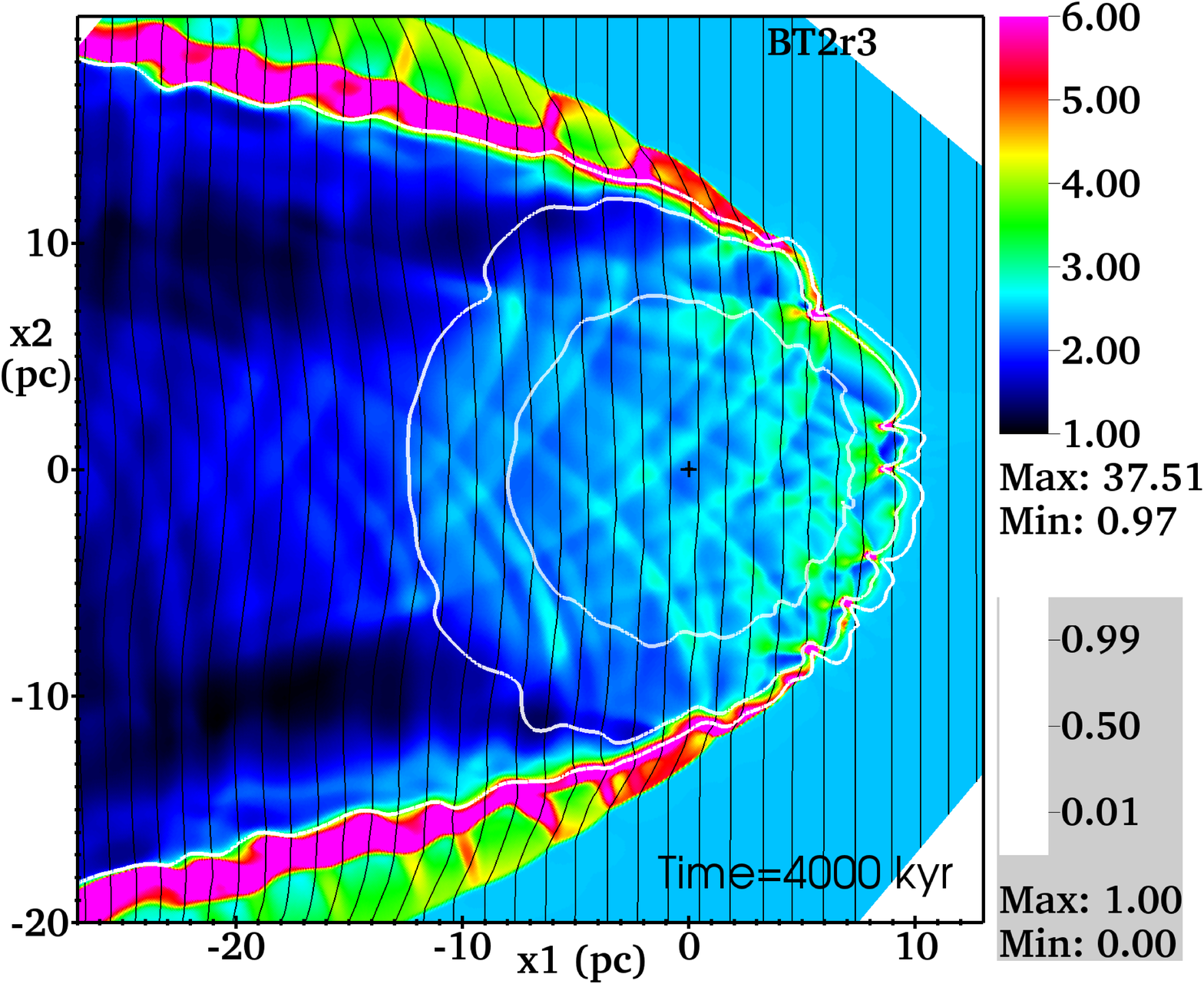}
\includegraphics[width=0.49\textwidth]{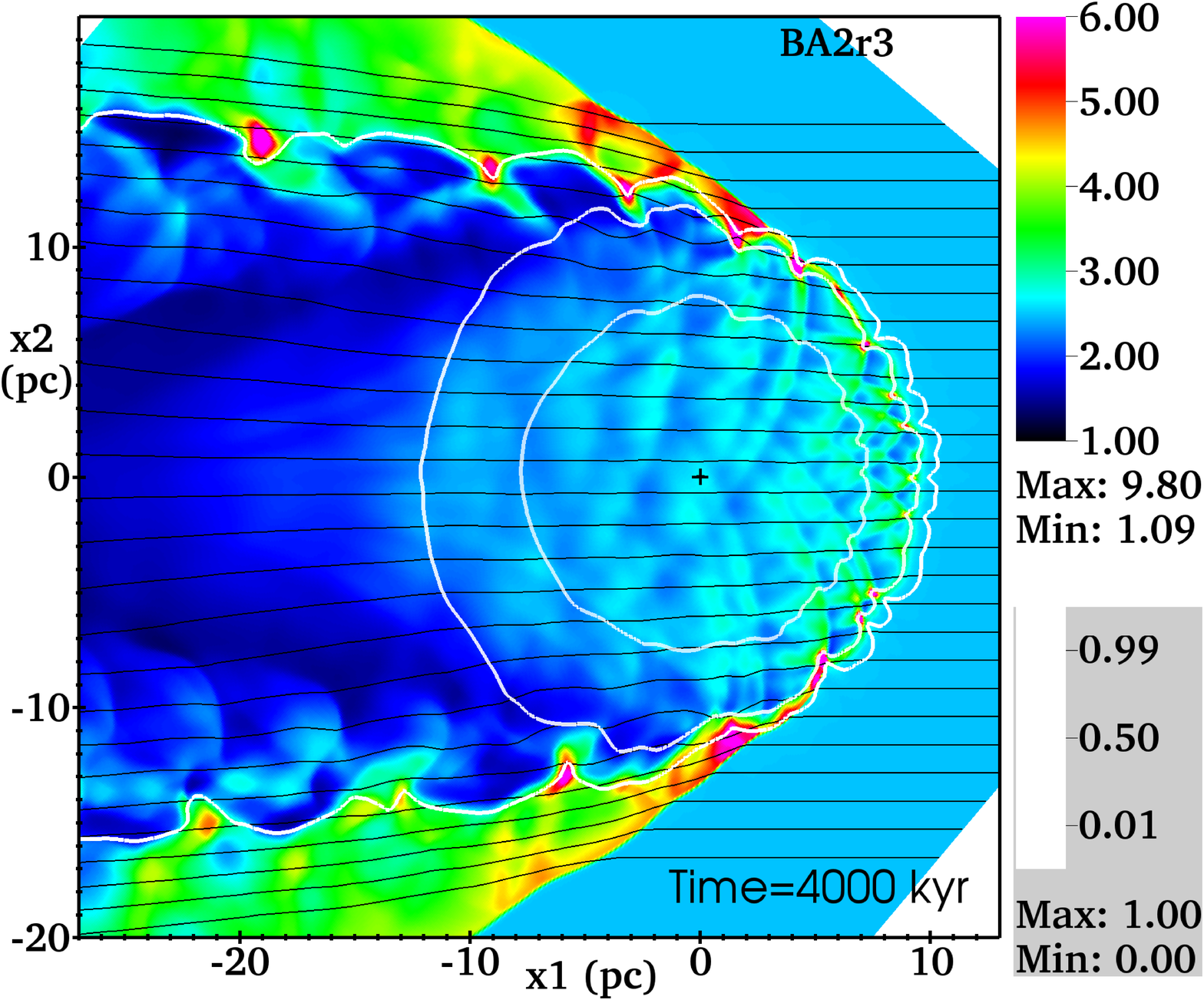}
\caption{
  As Fig.~\ref{fig:simBX2r2} but this time for higher resolution models BT2r3 (left) and BA2r3 (right).
  \label{fig:simBX2r3}
  }
\end{figure*}
\begin{figure*}
\centering
\includegraphics[width=0.49\textwidth]{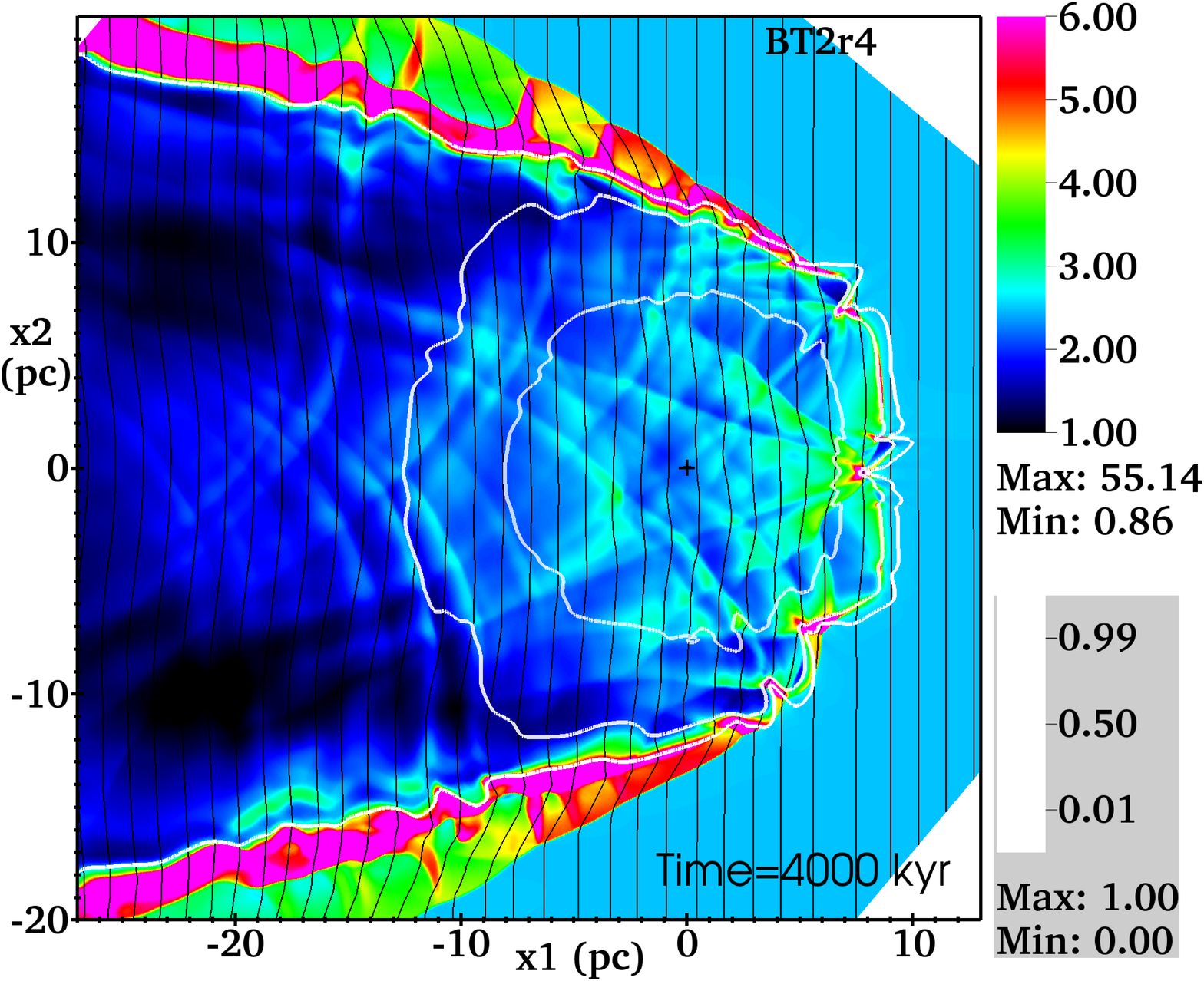}
\includegraphics[width=0.49\textwidth]{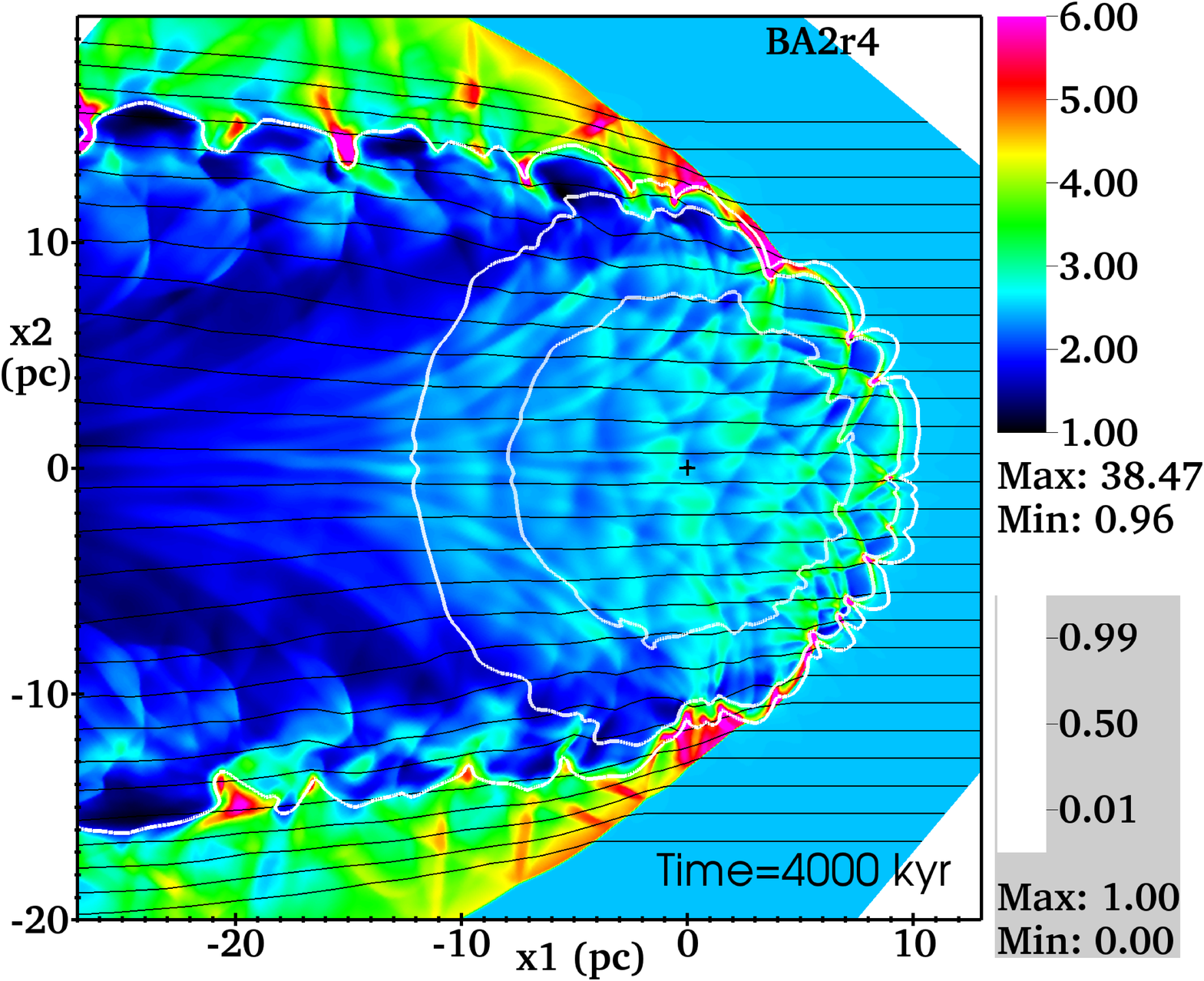}
\caption{
  As Fig.~\ref{fig:simBX2r3} but this time for higher resolution models BT2r4 (left) and BA2r4 (right).
  \label{fig:simBX2r4}
  }
\end{figure*}

\begin{figure}
\centering
\includegraphics[width=0.49\textwidth]{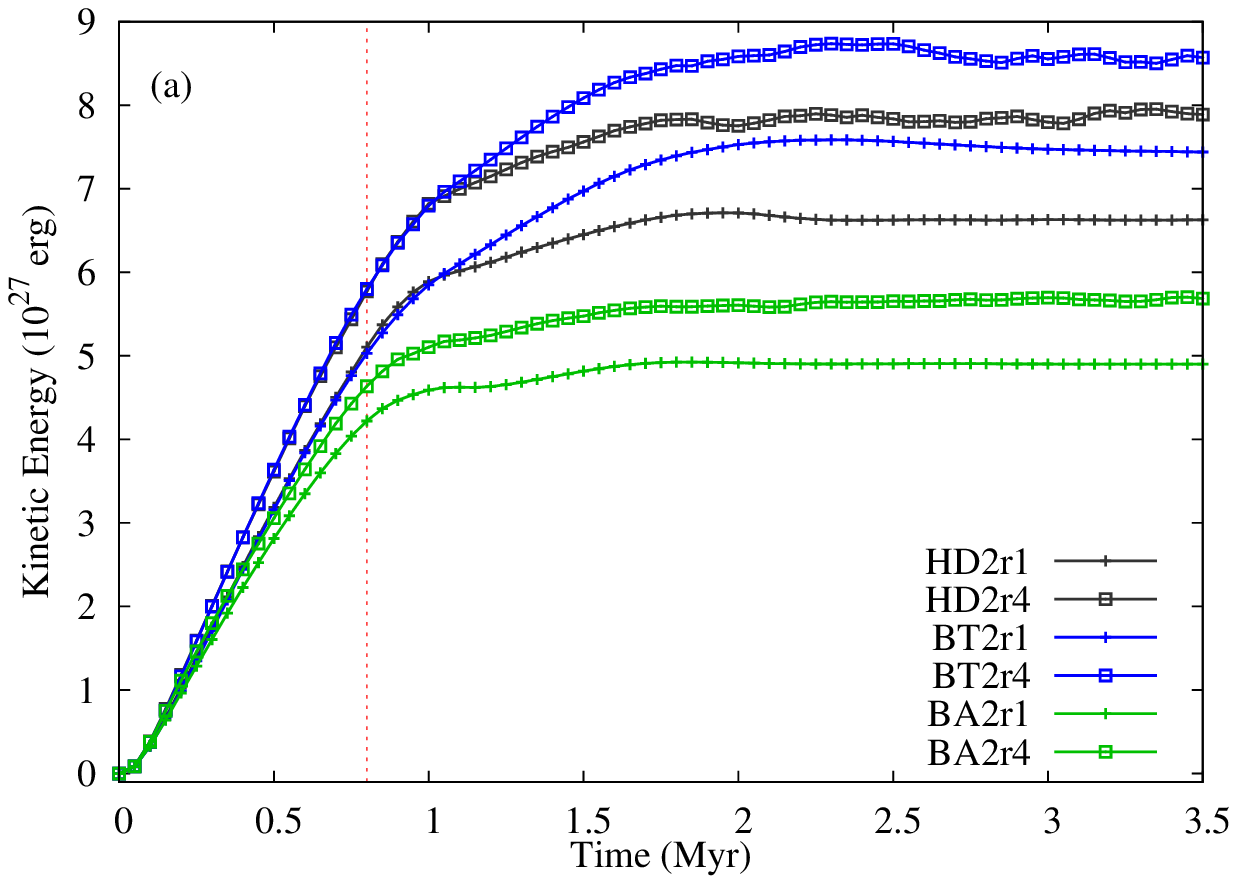}
\includegraphics[width=0.49\textwidth]{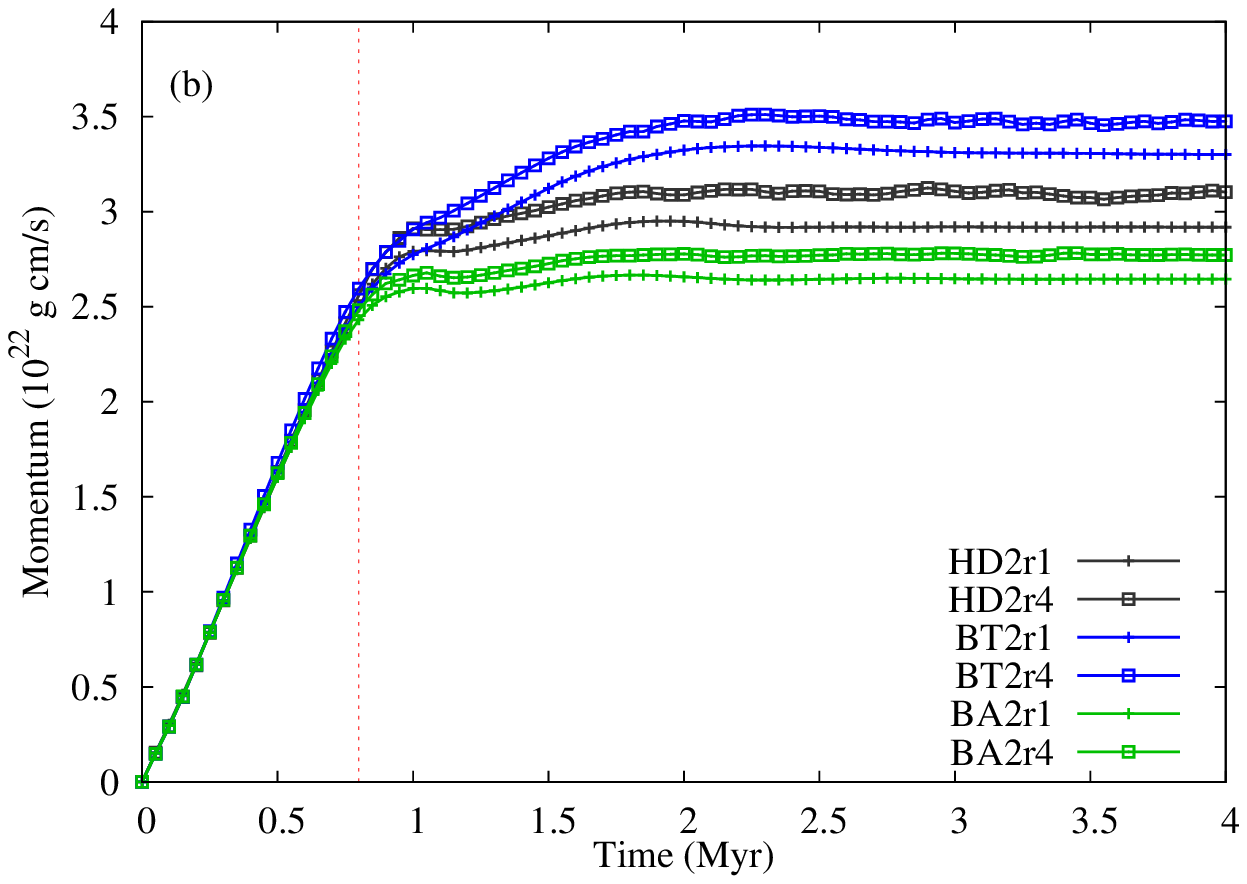}
\caption{
  Kinetic energy (a) and momentum (b) of gas in 2D simulations as a function of time, in the ISM rest frame.
  During the initial phase ($t\lesssim0.8$ Myr) no disturbed gas leaves the domain and the momentum increases monotonically.
  At later times a stationary state is reached where the momentum generated by the H\,\textsc{ii} region is balanced by gas leaving the domain.
  \label{fig:KEMom2D}
  }
\end{figure}

\begin{figure}
\centering
\includegraphics[width=0.49\textwidth]{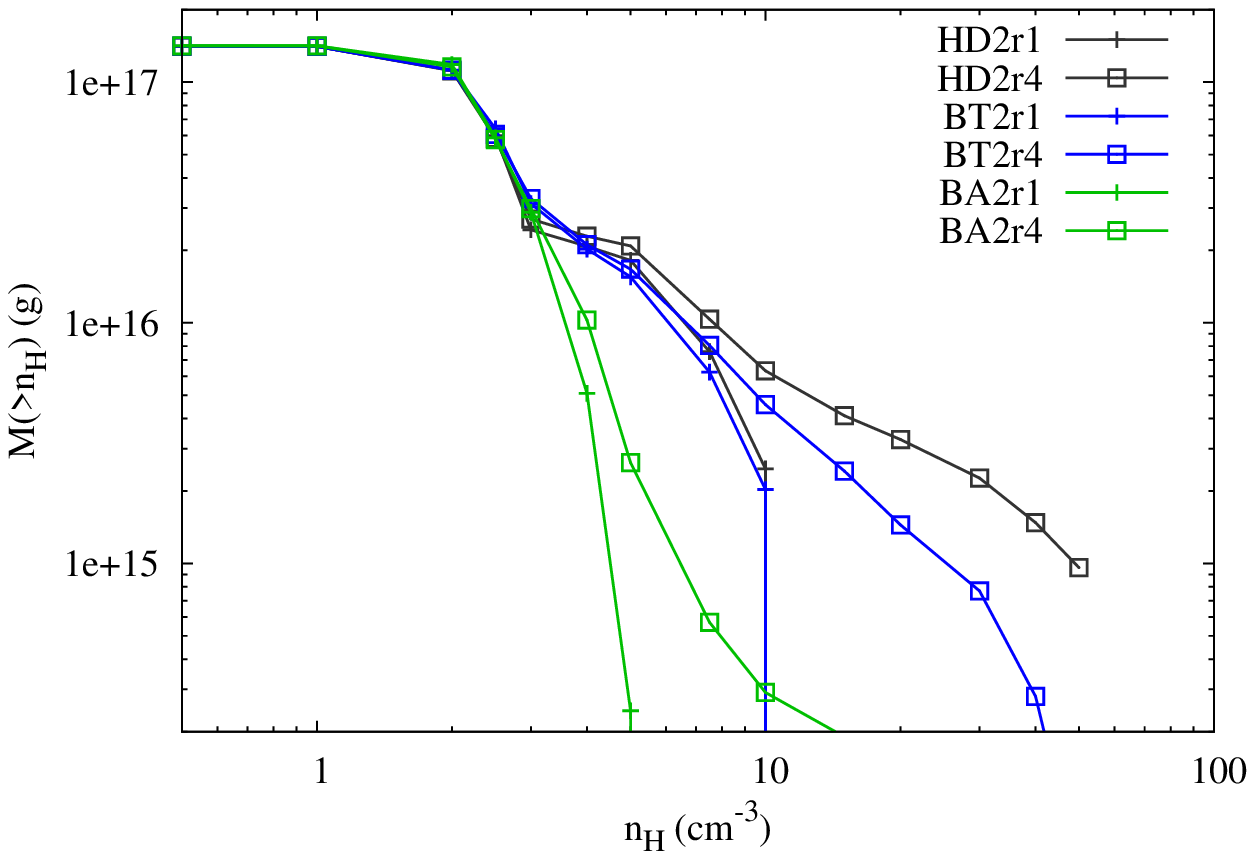}
\caption{
  Density distribution of the ISM in at $t=4\,$Myr, for 2D simulations.
  \label{fig:Density2D}
  }
\end{figure}

\begin{figure}
\centering
\includegraphics[width=0.49\textwidth]{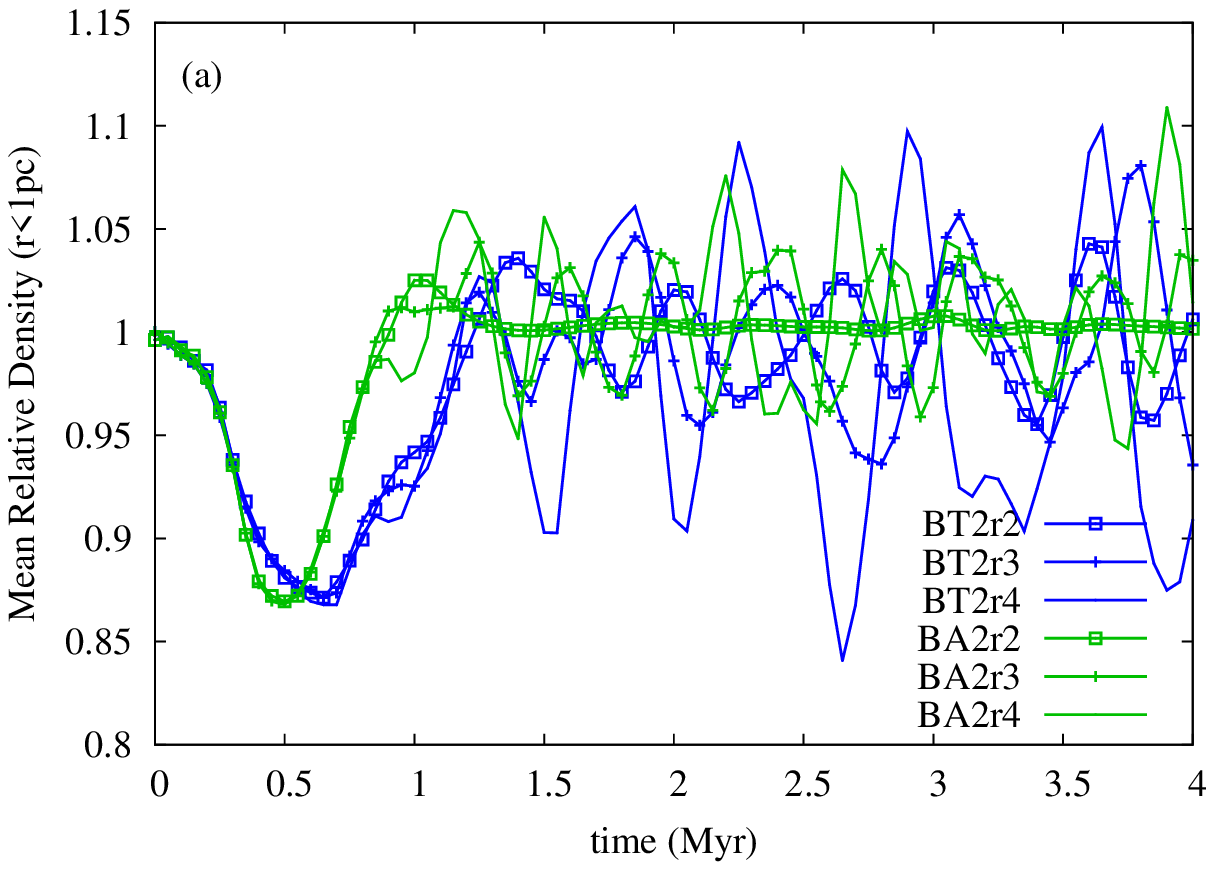}
\includegraphics[width=0.49\textwidth]{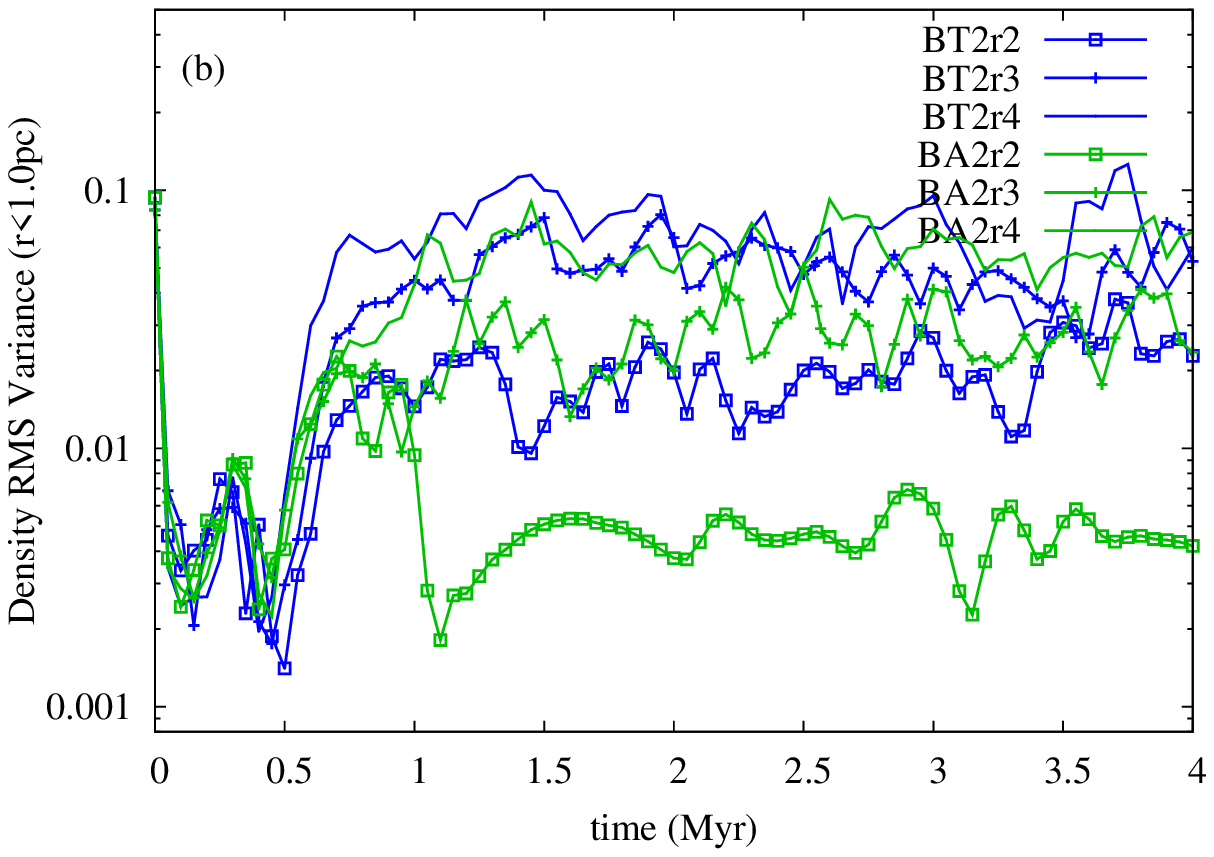}
\includegraphics[width=0.49\textwidth]{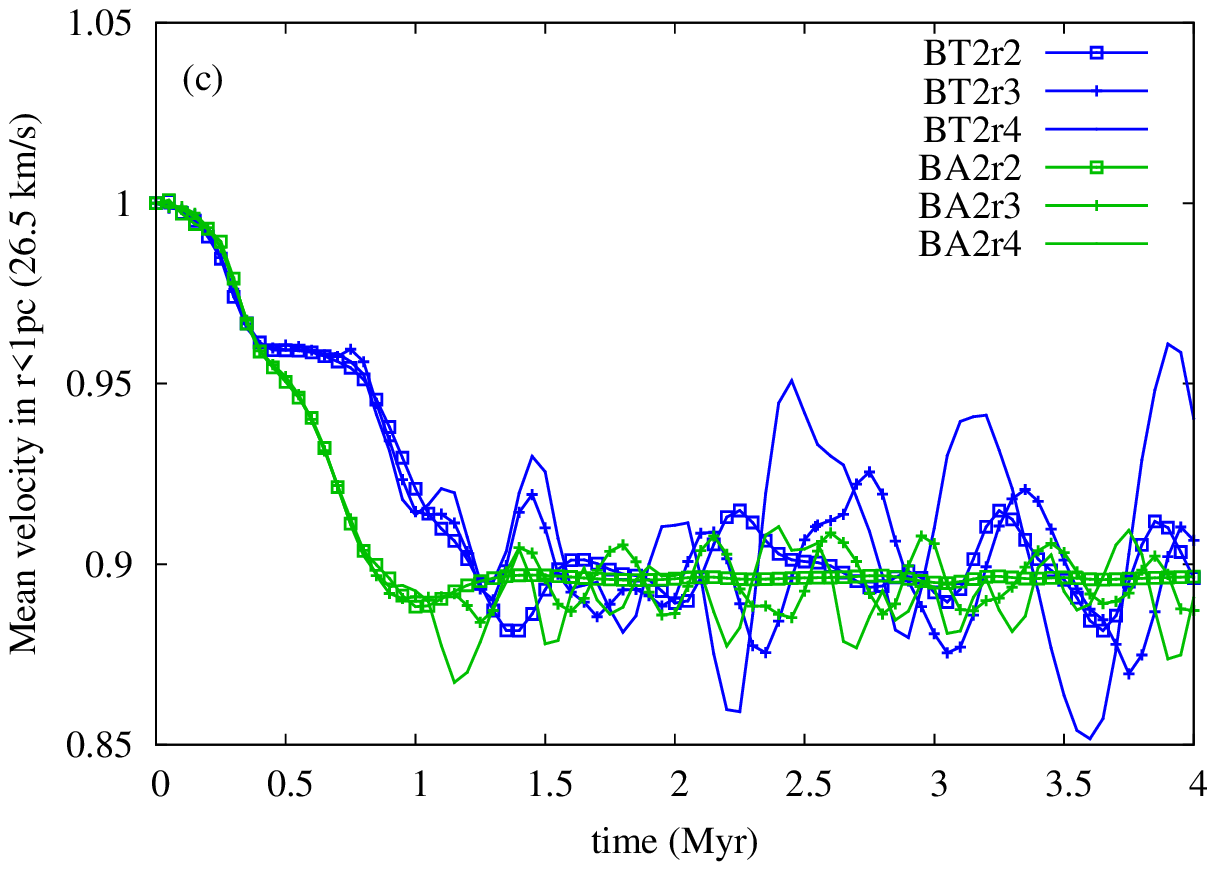}
\caption{
  The mean density (a), RMS density variance (b), and mean gas velocity (c) as a function of time for a volume within 1.0 pc of the star in 2D simulations.
  The density is normalised to the mean ISM input density of $\ensuremath{n_{\textsc{h}}}=2.5\,\ensuremath{\mathrm{cm}^{-3}}$,
  and the velocity to the undisturbed bulk velocity of $v_\star=26.5\,\ensuremath{\mathrm{km}\,\mathrm{s}^{-1}}$.
  \label{fig:NearStar2D}
  }
\end{figure}

A suite of 2D simulations have been run to test the effects of grid resolution on the H\,\textsc{ii} region feedback effects.
They are analogous to the 3D simulations listed in Table~\ref{tab:Sims3D} but with the same and higher resolutions, and are listed in Table~\ref{tab:Sims2D}.
They consist of a 2D domain with $x,y\in[-32.64,18.56]\,$pc with Cartesian (i.e.~infinite slab) geometry.
Fig.~\ref{fig:simBX2r2} shows simulations BT2r2 and BA2r2, having the same resolution and magnetic field configuration as the 3D models BT3r2 and BA3r2 in Section~\ref{sec:3D}.
Hydrogen number density \ensuremath{n_{\textsc{h}}} is plotted in colour, with contours showing the ion fraction $(1-y_n)$, and solid lines crossing the domain showing the magnetic field orientation (in black).
They show most of the same features at BT3r2 and BA3r2.
The model with aligned spatial velocity and magnetic field is stable, but the model where they are perpendicular shows some instability, closer to the hydrodynamic result.
The lateral expanding shell is more compressive in BT2r2 than BA2r2, and the opening angle of the shell downstream is narrower because the shock velocity is lower.
Both models have similar minimum density, although the structure of the downstream wake is different because of the magnetic field orientation.

Results from simulations BT2r3 and BA2r3 are shown in Fig.~\ref{fig:simBX2r3}, showing that with $2\times$ higher resolution both models have unstable I-fronts upstream from the star, and there are many more dense knots than were found in the 3D simulation.
Fig.~\ref{fig:simBX2r4} has still higher resolution models BT2r4 and BA2r4, showing even greater degree of instability in the upstream I-front than BT2r3 and BA2r3, and also stronger compression in clumps and shell.
R-type I-fronts have been shown \citep{NewAxf67,Wil99} to be unstable to weak density perturbations via a shadowing instability.
As long as the magnetic field is not sufficiently strong to make the gas flow entirely one-dimensional, this instability should also be present in magnetised R-type I-fronts.
The stability of the simulation BA2r2, and the similar 3D simulations BA3r1 and BA3r2, is at least partly an artifact of the increased stiffness of magnetic field lines at lower resolution.

\begin{table}
  \centering
  \begin{tabular}{| l | l | c | c |}
    \hline
    ID & $(N_x,N_y,N_z)$ & $\Delta x$ & $\mathitbf{B}$-field ($\mu$G) \\
    \hline
    HD2r1  & $(160,\,160,\,1)$   & 0.32 &  $(0,0,0)$ \\
    HD2r2  & $(320,\,320,\,1)$   & 0.16 &  $(0,0,0)$ \\
    HD2r3  & $(640,\,640,\,1)$   & 0.08 &  $(0,0,0)$ \\
    HD2r4  & $(1280,\,1280,\,1)$ & 0.04 &  $(0,0,0)$ \\
    \hline
    BA2r1 & $(160,\,160,\,1)$   & 0.32 &  $(7,0,0)$ \\
    BA2r2 & $(320,\,320,\,1)$   & 0.16 &  $(7,0,0)$ \\
    BA2r3 & $(640,\,640,\,1)$   & 0.08 &  $(7,0,0)$ \\
    BA2r4 & $(1280,\,1280,\,1)$ & 0.04 &  $(7,0,0)$ \\
    \hline
    BT2r1 & $(160,\,160,\,1)$   & 0.32 &  $(0,7,0)$ \\
    BT2r2 & $(320,\,320,\,1)$   & 0.16 &  $(0,7,0)$ \\
    BT2r3 & $(640,\,640,\,1)$   & 0.08 &  $(0,7,0)$ \\
    BT2r4 & $(1280,\,1280,\,1)$ & 0.04 &  $(0,7,0)$ \\
    \hline
  \end{tabular}
  \caption{
    Simulation properties for two-dimensional calculations in the $(x,y,0)$ plane.
    Columns show, respectively,
    simulation ID,
    number of grid zones in each direction,
    spatial diameter of a grid zone in parsecs,
    magnetic field vector in $\mu$G with respect to the direction of motion of the star.
    Models HD2x are two-dimensional hydrodynamic (no magnetic field), BA2x are two-dimensional MHD simulations with a B-field aligned with the direction of motion, BT2x are two-dimensional MHD with a field transverse (perpendicular) to the direction of motion.
    The number following the B-field designation, e.g.~`r3', represents the resolution.
  }
  \label{tab:Sims2D}
\end{table}

The kinetic energy and momentum in these simulations is plotted in Fig.~\ref{fig:KEMom2D} in the same way as the 3D results in Fig.~\ref{fig:KEMom3D}.
The absolute values of the results are not normalised physically because the slab-symmetry implies a value per unit depth (here cm$^{-1}$), but they are useful as a resolution study.
The results are similar to those from 3D simulations; here the kinetic energy also increases slowly with spatial resolution and is not converging to a maximum value, because the I-front instability gets stronger with higher resolution.

The ISM density distribution function is plotted in Fig.~\ref{fig:Density2D}, where by comparison to Fig.~\ref{fig:Density3D} we see that the higher resolution 2D simulations have stronger compression than their lower resolution counterparts.
We expect a similar trend in 3D, if we could run such high resolution models in 3D.

The properties of the ISM near the star are plotted in Fig.~\ref{fig:NearStar2D}, which again should be compared to the 3D results in Fig.~\ref{fig:NearStar3D}, although symmetry restrictions mean that the results are only indicative.
There is a clear trend of increasing spatial and temporal variability with simulation resolution, also obvious from the simulation snapshots.
The lowest resolution models are not shown because all relax to a steady state with only small variations near the star.
For the next lowest resolution, BA2r2 relaxes to an almost-stationary state, whereas stronger I-front instability in BT2r2 and HD2r2 leads to continuing fluctuations for the duration of the simulation.
These fluctuations are stronger again for models with resolutions r3 and r4.
Only one averaging volume is plotted ($r<1$ pc) but the trends at smaller volumes are the same as for 3D results.

In 2D the mean density is unchanged near the star after the initial evolutionary phase (differing from 3D), and the mean velocity of gas drops by about 10 per cent (similar to 3D).
At high resolution the root-mean-squared (RMS) density fluctuations are also $\approx10$ per cent, and the temporal fluctuations in mean density and velocity are up to 15 per cent.
The H\,\textsc{ii} region does measurably affect gas properties near the star, but it is still at a level that can be treated as a small perturbation.

From the trends with resolution in the 2D simulations, it is clear that many quantities have not fully converged numerically, so we should be cautious in drawing strong conclusions regarding the compression of gas to high densities and the properties of the ISM near the star.
The rate of momentum feedback seems quite robust, being barely changed when the spatial resolution changes by a factor of 8.
The kinetic energy input also only changes by about 20 per cent over the same range of resolutions so, while it has not converged, it is not changing dramatically.

{\changed
\section{Heating and cooling rates for atomic and photoionized gas} \label{app:HeatCool}
The heating and cooling functions used in the rate equation for energy [Eq.~(\ref{eqn:energyrate})] should comprise the main coolants in WNM and photoionized gas.
They are described in more detail here, term by term as they appear in Eq.~(\ref{eqn:energyrate}).
All terms have units erg\,s$^{-1}$ and should be multiplied by $n_\textsc{h}$ to get volumetric rates.

\subsection{Heating processes}
Photoionization heating, $H_{\mathrm{pi}}$ is calculated in a finite-volume way similarly to the photoionization rate $A_{\mathrm{pi}}$ in Eq.~(\ref{eqn:pion_FVrate}) (multifrequency) and Eq.~(\ref{eqn:pion_FVrate_mono}) (monochromatic).
The multifrequency rate is
\begin{equation}
H_{\mathrm{pi}}y_n = \int_{\nu_{\mathrm{th}}}^{\infty}
    \frac{L_\nu \mathrm{e}^{-\tau_\nu}(h\nu-h\nu_\mathrm{th})}{h\nu} 
    \frac{1-\mathrm{e}^{-\Delta\tau_\nu}}{n_{\textsc{h}}V_{\mathrm{shell}}}  d\nu \;,
\label{eqn:pheat_FVrate}
\end{equation}
and the integrals over frequency are again pre-calculated as a function of optical depth following \citet{FraMel94}.

FUV heating, $H_{\mathrm{fuv}}$, from the ionizing star is calculated using eq.~(A3) from \citet{HenArtDeCEA09}:
\begin{equation}
H_{\mathrm{fuv}} = \frac{1.9\times10^{-26}Q_\mathrm{fuv} \mathrm{e}^{-\tau_1}}{4\pi r^2 1.2\times10^{7}}
    \left[1+6.4\left(\frac{Q_\mathrm{fuv}\mathrm{e}^{-\tau_1}}{4\pi r^2 1.2\times10^{7}n_\textsc{h}}\right) \right]^{-1} \;,
\end{equation}
where $Q_{\mathrm{fuv}}$ is the FUV photon luminosity of the star in the range 6-13.6\,eV, and $\tau_1=1.03\times10^{-21} N_\mathrm{H}$ is the dust optical depth ($N_\mathrm{H}$ is the H column density along the ray).
Cosmic ray heating is assumed to be $H_\textrm{cr}=5\times10^{-28}n_\textsc{h}y_n$ \citep{HenArtDeCEA09}.
Absorption of the Galactic FUV interstellar radiation field by PAH grains dominates the gas heating in neutral gas far from the star, and is calculated using eqs.~(19) and (20) from \citet{WolMcKHolEA03}:
\begin{equation}
H_\mathrm{pah} = \frac{1.083\times10^{-25}}{1+9.77\times10^{-3}\left(\sqrt{T}/n_e\right)^{0.73}} \;,
\end{equation}
where we have followed \citet{WolMcKHolEA03} in setting $G_0=1.7$ and $\phi_\textsc{pah}=0.5$ and absorbed these variables into the constant prefactor.
The second term in the \citet{WolMcKHolEA03} expression has been ignored because it is unimportant for $T\ll10^4$\,K.

\subsection{Cooling processes}
Collisional ionization cooling is obtained from a fit to the ionization rate \citep{Vor97},
\begin{equation}
A_{\mathrm{ci}}\psi_\mathrm{H}y_n n_e = 6.34\times10^{-19} y_n n_e \frac{T_p^{0.39}}{0.232+T_p} \mathrm{e}^{-T_p} \;,
\end{equation}
where $T_p\equiv (1.578\times10^5\,\mathrm{K})/T$.

Radiative recombination and bremsstrahlung cooling from H$^+$ are interpolated from the tables for the total cooling coefficient in \citet{Hum94}, although an analytic fit would have been significantly more efficient.
We assume He is singly ionized when H is ionized, so we also add in a bremsstrahlung cooling term for He$^+$ (its recombination cooling is neglected).
These comprise the terms $(C_\mathrm{rr}+C_\mathrm{ff})n_e(1-y_n)$ in Eq.~(\ref{eqn:energyrate}), but neither term is important in photoionized gas at solar metallicity.

Collisional excitation cooling from H$^0$, $C_{\mathrm{cx}}y_n n_e$, is a strong coolant in partially-ionized gas, and its rate is obtained by interpolating data from table 11 in \citet{RagMelLun97}.

Cooling by metals and grains in mostly neutral gas is calculated using eqs.~(21), (C1), (C2), and (C3) in \citet{WolMcKHolEA03}, being the cooling from PAH grains, C$^+$ collisions with H$^0$, C$^+$ collisions with electrons, and O$^0$ collisions with H$^0$, respectively.
These are the main coolants in the diffuse neutral ISM.
The PAH term is
\begin{equation}
 C_\textrm{pah} = 3.02\times10^{-30} T^{0.94} \left(\frac{3.4\sqrt{T}}{n_e}\right)^\beta n_e \;,
\end{equation}
where $\beta\equiv0.74T^{-0.068}$ and we have used the same values for the parameters $G_0$ and $\phi_\textsc{pah}$ as before.
Collisional cooling of C$^+$ with H$^0$ is
\begin{equation}
 C_\textsc{c\,ii}^\textsc{h} = 3.15\times10^{-27} \mathrm{e}^{-92/T} n_\textsc{h}y_n \;,
\end{equation}
and with electrons is
\begin{equation}
 C_\textsc{c\,ii}^\textrm{e} = 1.4\times10^{-23} T^{-0.5} \mathrm{e}^{-92/T} n_e \;.
\end{equation}
Collisional cooling of O$^0$  with H$^0$ is
\begin{equation}
 C_\textsc{o\,i}^\textsc{h} = 3.96\times10^{-28} T^{0.4} \mathrm{e}^{-228/T} n_\textsc{h}y_n \;.
\end{equation}
The term $C_{\mathrm{nt}}$ in Eq.~(\ref{eqn:energyrate}) is $C_{\mathrm{nt}}=C_\textrm{pah}+ C_\textsc{c\,ii}^\textsc{h}+C_\textsc{c\,ii}^\textrm{e}+C_\textsc{o\,i}^\textsc{h}$.

Cooling in photoionized gas is dominated by emission from O and C ions, which have low-energy forbidden transitions that can be collisionally excited in gas with $T\approx5000-10\,000$\,K.
We use an approximate fit to these emission lines, eq.~(A9) from \citet{HenArtDeCEA09},
\begin{equation}
 C_\mathrm{m}n_e(1-y_n) = 1.42\times10^{-22} n_e(1-y_n) \mathrm{e}^{\left[\frac{-33\,610}{T} -\left(\frac{-2180}{T}\right)^2\right]} \;.
\end{equation}
}

\label{lastpage}

\end{document}